\documentclass{aa}  

\usepackage{graphicx}
\usepackage{txfonts}
\usepackage{natbib}
\bibpunct{(}{)}{;}{a}{}{,}
\usepackage{scalerel}
\usepackage{tikz}
\usetikzlibrary{svg.path}

\definecolor{orcidlogocol}{HTML}{A6CE39}
\tikzset{
  orcidlogo/.pic={
    \fill[orcidlogocol] svg{M256,128c0,70.7-57.3,128-128,128C57.3,256,0,198.7,0,128C0,57.3,57.3,0,128,0C198.7,0,256,57.3,256,128z};
    \fill[white] svg{M86.3,186.2H70.9V79.1h15.4v48.4V186.2z}
                 svg{M108.9,79.1h41.6c39.6,0,57,28.3,57,53.6c0,27.5-21.5,53.6-56.8,53.6h-41.8V79.1z M124.3,172.4h24.5c34.9,0,42.9-26.5,42.9-39.7c0-21.5-13.7-39.7-43.7-39.7h-23.7V172.4z}
                 svg{M88.7,56.8c0,5.5-4.5,10.1-10.1,10.1c-5.6,0-10.1-4.6-10.1-10.1c0-5.6,4.5-10.1,10.1-10.1C84.2,46.7,88.7,51.3,88.7,56.8z};
  }
}

\newcommand\orcidicon[1]{\href{https://orcid.org/#1}{\mbox{\scalerel*{
\begin{tikzpicture}[yscale=-1,transform shape]
\pic{orcidlogo};
\end{tikzpicture}
}{|}}}}
\usepackage{comment}
\usepackage[]{hyperref}
\begin{document}

   \title{Multi-wavelength spectroscopic study of shock-driven phenomena in explosive outbursts in symbiotic-like recurrent novae}

   \subtitle{with an emphasis on RS Ophiuchi} 
   \titlerunning{Multi-wavelength phenomenology of Sy-RNe  }
   \authorrunning{Azzollini et al.}

   \author{Alessandra Azzollini\inst{1}\inst{2}\fnmsep\thanks{\href{mailto:}{alessandra.azzollini@uni-wuerzburg.de}}
          \and
          Steven N. Shore\inst{2}\inst{3}\fnmsep\thanks{\href{mailto:}{steven.neil.shore@unipi.it}}
          \and
          N. Paul M. Kuin\inst{4}
          \and
          Kim L. Page\inst{5}
          }

   \institute{Julius-Maximilians-Universität Würzburg, Fakultät für Physik und 
Astronomie, Emil-Fischer-Str. 31, D-97074 Würzburg, Germany
   \and Dipartimento di Fisica "Enrico Fermi", Università di Pisa, Largo Bruno Pontecorvo, 3, 56127 Pisa PI, Italy
              %\email{a.azzollini@outlook.it}
             \and INFN - Sezione di Pisa, largo B. Pontecorvo 3, 56127 Pisa, Italy
             %\email{steven.neil.shore@unipi.it}
             \and
             Mullard Space Science Laboratory, University College London, Holmbury St. Mary, Dorking, Surrey RH5 6NT, UK
             %\email{npkuin@gmail.com}
             \and
             School of Physics \& Astronomy, University of Leicester, LE1 7RH, UK
             %\email{klp5@le.ac.uk}
             }

   \date{Received 12 October, 2022; accepted 13 March, 2023}

% \abstract{}{}{}{}{} 
% 5 {} token are mandatory
 
  \abstract
   {}
   {Our goal is to detail the development of RS Ophiuchi and the other Galactic symbiotic-like recurrent novae throughout their outburst and quiescence, with a particular emphasis on the  propagation of the shock wave during the outburst of the binaries.}
   {The spectral analysis has been performed using archival data according to the features of the individual datasets. Swift grism spectra were reduced and extracted using a combination of the pre-existing {\tt UVOTPY} Python routine and newly written pipelines in Matlab. Other datasets were directly available in reduced form, already corrected for instrumental or background contamination, and calibrated in wavelength and flux or intensity. The work on these was done through pipelines suited for reading the data and elaborating them to extract quantities of interest for the analysis.}
   {We find striking similarities in  different outbursts of the same object and for different novae. For example,  RS Oph 2021 was almost identical to the 2006 outburst, despite having occurred at a different orbital phase with the observations made from a different line of sight through the red giant wind. Despite the intrinsically different properties of the binaries, striking similarities are found for different systems of the same class, for instance, the trend of the electron density over time during outburst appears to follow a general temporal development.}% add? - despite differing XR evolution 
   {}
   

   \maketitle
%
%-------------------------------------------------------------------
\section{Introduction}

Among the classical novae there is a small but important group, the symbiotic-like recurrent novae (hereafter Sy-RNe)\footnote{Although not standard nomenclature, we use this term to avoid confusion with the {\it symbiotic novae}, that constitute a distinct class of explosive  symbiotic binaries (\citet{Nussbaumer1987}).} in which a moderately massive, $\rm{M}\sim1-1.4\;\rm{M_{\odot}}$, white dwarf (WD) orbits within the wind of a red giant (RG) companion  (\citet{anupama3}, \citet{darnley}) and accretes enough gas to emit through an accretion disc. In turn, Sy-RNe are part of a larger taxonomy, Recurrent Novae (RNe), that exhibit more than one observed outburst, with a recurrence timescale $\sim 10-100\;yr$, which is much shorter than that of Classical Novae (CNe) ($\sim10^{4}-10^{5}\;\rm{yr}$ or more).

This study deals with the long period Galactic RNe RS Oph, V745 Sco, V3890 Sgr, T CrB, and all of them have orbital periods of more than a few hundred days and a hot massive WD near the Chandrasekhar limit \citep{1965RNhighmass,Hric,2017Miko,2021Miko}. 
To this group, we add V407 Cyg, which is classified as a D-type symbiotic binary with a long pulsation period Mira variable until, in 2010, it was discovered in outburst in close resemblance to the spectroscopic development of RS Oph (\citet{vcyg1}, \citet{vcyg2}, \citet{iijima}, \citet{munari}). 

The essential property of the Sy-RNe outbursts is that the ejected material is not a radiative or dynamically passive structure, unlike CNe where the ionisation and opacity of the ejecta are governed by irradiation from the WD. In contrast, the ejecta in the Sy-RNe are not freely expanding, even from the first moments after the explosion. The expelled material  supersonically traverses  the dense and extended RG wind. In CNe, the ejecta have constant mass; this never occurs in Sy-RNe. The emission is powered not only by radiation from the central WD, but also from the shocked material crossed by the passing front. The resulting hard continuum is never observed in freely expanding media and lines from high ionisation collisionally excited transitions are observed within the post-shocked gas. The emission strength increases with time and is produced by the highly energetic pulse from the explosion that increasingly ionises the RG wind. Permitted lines are  emitted by the ejecta, from light species (H Balmer lines, He I and He II, for example), and higher ionisation and intercombination lines form in the shocked ejecta as they cool by recombination or collisional de-excitation in extended and less dense ejecta. The presence of a strong continuum, intense emission lines, absorption features due to either the neutral or shocked and ionised giant wind profoundly affects the medium in which the shock expands and produces a peculiar spectroscopic evolution of the outburst that is different from that of other systems.

%--------------------------------------------------------------------
\section{Instruments and datasets}
The latest outbursts of RS Oph, the most widely studied Galactic Sy-RN, occurred in 2006 and 2021. The nova reached optical maximum on 2006 February 12.83 UT (\citet{2006IAUC.8671....2N}) and  on 2021 August 8.93 UT (K. Geary; vsnet-alert 26131). We adopt these times of optical maximum as the start of the respective outburst, T0.

The main focus of this paper is the analysis of spectral data acquired during both events with the ultraviolet (UV) grism mounted on the Ultraviolet/Optical Telescope (UVOT) on board the {\it Neil Gehrels Swift Observatory}. The satellite observed RS Oph in 2006 and 2021, starting observations in the UV-optical at 29.83 and 7.22 days after T0, respectively. The datasets were compared with the 1985 International Ultraviolet Explorer (IUE) sequence and other spectra taken during both quiescence and outbursts of RS Oph itself and remaining Galactic systems of the same class. Table \ref{table:dataset} summarises some of the main features of the data used for the analysis. The corresponding details for the most recent outbursts are provided in Table \ref{table:outburst}. Further information about the complete dataset and the corresponding instruments can be found in Appendix \ref{section:appendix1}.

\begin{table}
\caption{Description of the datasets.}             % title of Table
\label{table:dataset}      % is used to refer this table in the text
\centering                          % used for centering table
\resizebox{\columnwidth}{!}{\begin{tabular}{ccccc}        % centered columns (4 columns)
\hline\hline                 % inserts double horizontal lines
Object & Archive & Event & \# spectra & $\rm{t_{exp}} \left[ \rm{ks} \right]$ \\     % table heading 
\hline                        % inserts single horizontal line
RS Oph & Swift UV grism & 2006 & 70 & 2360 \\
RS Oph & Swift UV grism & 2021 & 54 & 68 \\
RS Oph & IUE low resolution & 1985 & 72 & 197 \\
RS Oph & IUE high resolution & 1985 & 21 & 86 \\
RS Oph & ARAS & quiescence & 81 & 352 \\
RS Oph & ARAS & 2021 & 113 & 384 \\
V745 Sco & IUE low resolution & 1989 & 11 & 56 \\
V745 Sco & SAAO 1.9 m & 2014 & 6 & 4 \\
V745 Sco & CTIO & 1989 & 10 & 9 \\
V745 Sco & SMARTS & 2014 & 5 & 8 \\
V745 Sco & NOT FIES & 2014 & 3 & 5 \\
V745 Sco & ESO UVES & 2014 & 25 & 31 \\
V3890 Sgr & Swift UV grism & 2019 & 26 & 13 \\
V3890 Sgr & IUE low resolution & 1990 & 7 & 46 \\
V3890 Sgr & ARAS & 2019 & 61 & 297 \\
T CrB & IUE low resolution & quiescence & 71 & 177 \\
T CrB & IUE high resolution & quiescence & 8 & 146 \\
V407 Cyg & NOT FIES & 2010 & 18 & 19 \\
V407 Cyg & Ondřejov  & 2010 & 42 & 87 \\
\hline                                   %inserts single line
\end{tabular}}
\end{table}
 
 \begin{table}
\caption{Known outburst events of the Galactic symbiotic-like recurrent novae considered in the present analysis.}             % title of Table
\label{table:outburst}      % is used to refer this table in the text
\centering                          % used for centering table
\begin{tabular}{ccc}        % centered columns (4 columns)
\hline\hline                 % inserts double horizontal lines
Nova & Outburst & T0 $\left[ \rm{MJD} \right]$ \\     % table heading 
\hline                        % inserts single horizontal line
RS Oph & 1985 & 46091 \\
RS Oph & 2006 & 53778 \\
RS Oph & 2021 & 59434 \\
V745 Sco & 1989 & 47737 \\
V745 Sco & 2014 & 56695 \\
V3890 Sgr & 1990 & 48007 \\
V3890 Sgr & 2019 & 58722 \\
V407 Cyg & 2010 & 55265 \\
\hline                                   %inserts single line
\end{tabular}
\end{table}

\section{Data Analysis}
\label{section:analysis}
\begin{figure}[h]
\centering
%\begin{subfigure}{}
%\subfigure
{\includegraphics[width=\hsize,height=0.25\textheight]{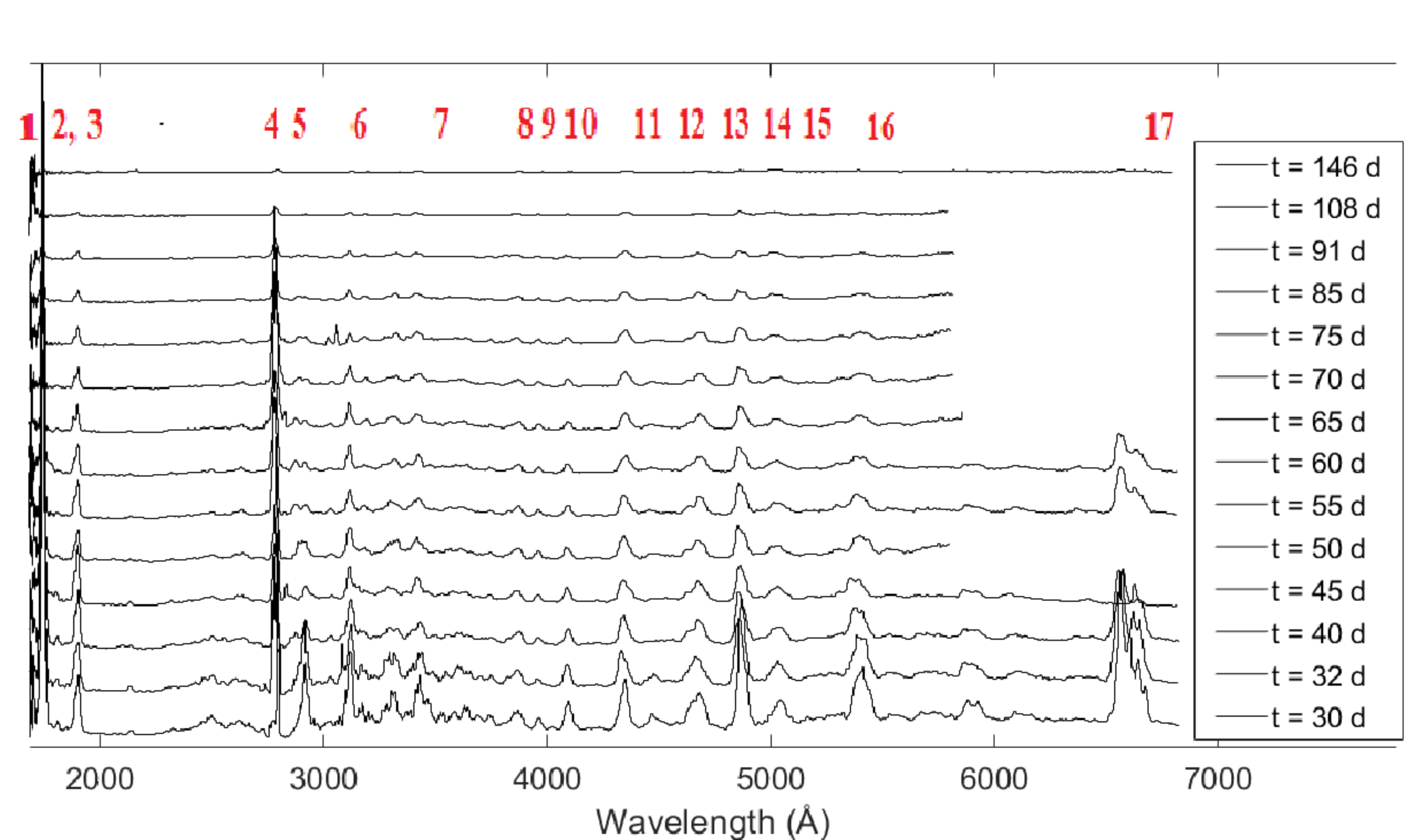}}
\hspace{0.2mm}
%\end{subfigure}
%\subfigure
%\begin{subfigure}{}
{\includegraphics[width=\hsize,height=0.25\textheight]{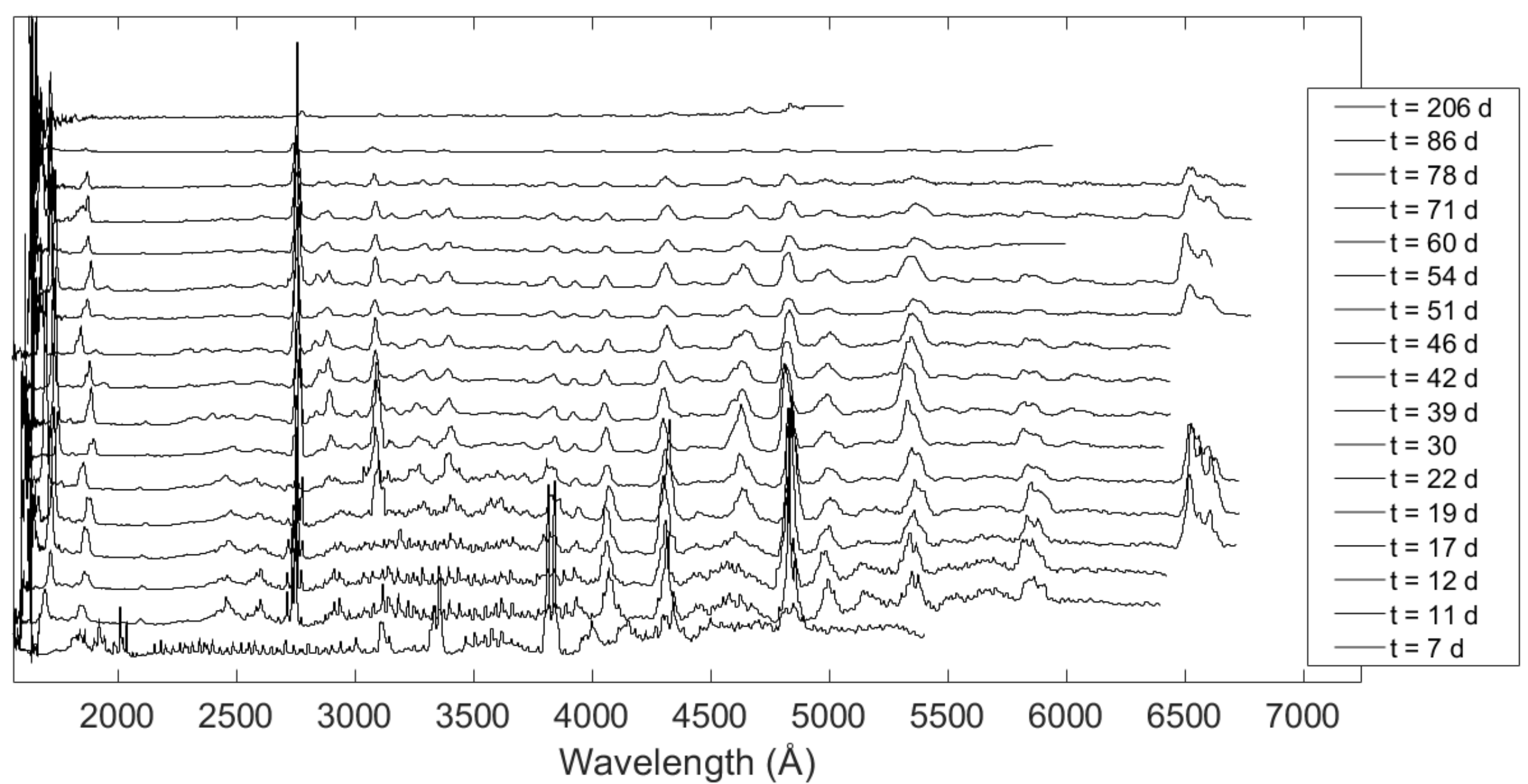}}
%\end{subfigure}
\caption{\label{fig:rsophswift_development} \footnotesize Spectral sequence of RS Ophiuchi at various stages during its 2006 (upper panel) and 2021 (lower panel) outbursts in UVOT data. The flux in units $\left[ \sim 10^{-11} - 10^{-10} \; \rm{erg} \cdot \rm{cm^{-2}} \cdot \rm{s^{-1}} \right]$ of each spectrum was normalised and then plotted with an applied offset of 2. The highlighted emission lines are listed in Table \ref{table:lines}. In the first four 2021 spectra the area of peak sensitivity around 2900 \AA~ was saturated.}
\end{figure}

For our present study, various pipelines have been used to extract and examine spectra from distinct archives. Figure \ref{fig:rsophswift_development} displays various stages of the spectral development of RS Oph during its 2006 and 2021 outbursts. The highlighted emission lines are listed in Table \ref{table:lines}.
 \begin{table}
\caption{Emission lines appearing in the Swift spectra of RS Oph shown in Fig. \ref{fig:rsophswift_development}}             % title of Table
\label{table:lines}      % is used to refer this table in the text
\centering                          % used for centering table
\begin{tabular}{cc}        % centered columns (4 columns)
\hline\hline                 % inserts double horizontal lines
Spectral profile & Line $\left[ \AA \right]$ \\     % table heading 
\hline                        % inserts single horizontal line
1   &   N III] 1750  \\
2   &   Si III] 1892 \\
3   &   C III] 1909  \\
4   &   Mg II 2798  \\
5   &   Mg II 2932    \\
6   &   O III 3133  \\
7   &   [Ne V] 3426 \\
8   &   [Ne III] 3869   \\
9   &   H$\rm{\epsilon}$ 3970    \\
10  &   H$\rm{\delta}$ 4102  \\
11  &   H$\rm{\gamma}$ 4341  \\
12  &   He I 4471   \\
13  &   He II 4686  \\
14  &   H$\rm{\beta}$ 4861   \\
15  &   [O III] 5007    \\
16  &   He I 5016   \\
17  &   H$\rm{\alpha}$ 6563  \\
\hline                                   %inserts single line
\end{tabular}
\end{table}
Because of the limited pointing accuracy of Swift, shifts are present of around $15\;$\AA\ in most of the spectra, greater at lower and higher wavelengths ($\lambda<1900\;$\AA, $\lambda>4000\;$\AA). This is caused by the variation of the non-linear dispersion over the detector resulting in an uncertain wavelength zero point position. In the 2006 campaign, Swift grism observations were done without an initial "slew in place" (SIP)\footnote{https://swift.gsfc.nasa.gov/analysis/uvot\_ugrism.html} and the spectra appeared all over the detector due to pointing accuracy of around 3 arcminutes; with SIP it is $\sim 15$ arcseconds. As shown in Kuin et al. (\citeyear{swiftcal}), the error in the wavelength zero point is therefore less with SIP, and consequently, better in the 2021 data. This zero point offset does not affect the results of the spectral analysis, and therefore it does not need to be rectified. Moreover, an additional blue shift due to physical effects in the source (the presence of the wind and the Strömgren sphere, for example) affects lines in various ways: highest ionisation lines are usually more shifted than low ionisation and permitted resonance lines, with respect to their reference rest wavelength. In the latest spectra of the 2021 dataset, the shift is more noticeable\footnote{The 2021  observing strategy did not always include a SIP, to avoid the hot columns in XRT.}. Every image of the database is affected by this.
Almost all  spectra are flux calibrated, except for the majority of the ARAS archive, NOT, and Ondrejov observations. Consequently, some V3890 Sgr, V745 Sco and V407 Cyg figures are in relative intensity or counts rather than flux. A special issue arises for the ARAS sequences:  the archive is a heterogeneous collection of data acquired by different observers and instruments, and therefore individual resolving power is an important consideration when comparing spectra. For the remaining datasets, the archive is done in such a manner that every image contains a single complete spectrum, each comprising a different wavelength range. 

\subsection{Measurement of the spectral lines}
\label{subsection:measurement}
\emph{Peak line fluxes\/} (they rarely correspond to the centre of the line) were measured, either in intensities or fluxes depending on the specific spectrum. A more reliable quantification for the intensity (or flux) for an emission line is given by the \emph{integrated intensity\/} and \emph{integrated flux\/}. The values were obtained from integration  without fitting an assumed formal profile. For every peak, the velocity shift from the reference zero point was estimated. Some lines, especially in the IUE spectra, show distorted profiles due to saturation and were excluded. The Full Width at Zero Intensity (FWZI) and Full Width at Half Maximum (FWHM) were measured for each line. 

\subsection{Light curves}
\label{subsection:lightcurves}

\begin{figure}[ht!]
\centering
\includegraphics[width=0.5\textwidth]{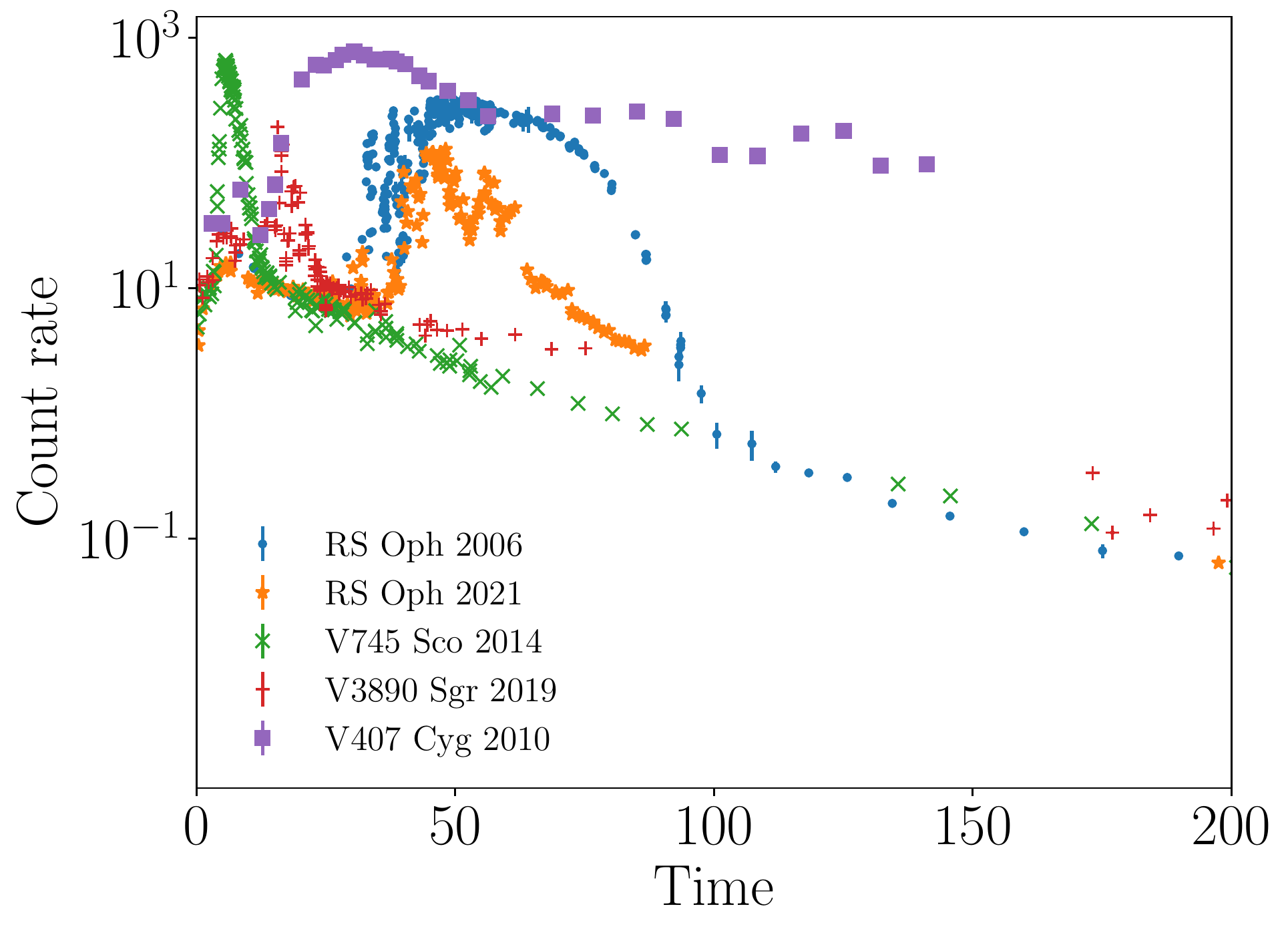}
\caption{\label{fig:lightcurve} \footnotesize XRT (0.2-10 keV) light curves during the events: RS Oph 2006 and 2021 (\citet{rsoph2021xr}), V745 Sco 2014 (\citet{page}), V3890 Sgr 2019 (\citet{10.1093/mnras/staa3083}), V407 Cyg 2010. A constant scaling factor is applied, of 10 for V745 Sco and V3890 Sgr, $3 \times 10^{3}$ to V407 Cyg.}
\end{figure}

Optical photometry was obtained from the American Association of Variable Star Observers (AAVSO) database (\cite{aavso}).  The Swift X-Ray Telescope (XRT) light curves over $0.3-10 \; \rm{keV}$ are displayed in Fig. \ref{fig:lightcurve} %\ref{fig:lightcurve_rsoph}, \ref{fig:lightcurve_v745sco2014}, \ref{fig:lightcurve_v3890sgr2019} and \ref{fig:lightcurve_v407cyg2010}.

In RS Oph, the early decline in XRs is consistent with shock propagation. Around 26 days, the so-called Super Soft Source (SSS) phase begins and increases some orders of magnitudes in just a few days. The XR SSS phase (\citet{2011ApJ...727..124O}) arises when the photosphere of the WD reaches a constant bolometric luminosity due to continued nuclear burning (CNO processing) occurring in its envelope. This phase lasted approximately for the subsequent $30-40\;\rm{days}$, making the WD temperature increase from $\sim 65$ to $\sim 90\;\rm{eV}$ until $t \simeq 60\;\rm{days}$, followed by a decrease, first slowly and then more rapidly, after about 80 days (\citet{2011ApJ...727..124O}). At the latter date, the fading of the SSS signals the presumed end of the shell-hydrogen burning phase on the WD. The appearance of the SSS component is common in most novae. In general, different systems show  similar light curves, although with different timescales. More rapid evolution is usually associated with a more massive WD.

\subsection{Line profiles: an overview}
\label{subsection:profiles}
    Studying single profiles evolution over time for the individual systems reveals the existence of multiple structures at different times or in correspondence with different species. Some examples are shown in Appendix \ref{section:appendix2}. The presence of many inhomogeneities is common in the envelopes and between different regions of classical novae, and the emission is produced at different densities in separate components of the envelope.  This alters line ratios and the form of the line profiles. In SyRNe line profiles usually show two superimposed components. A broad contribution is from the ejecta and shock, a narrow profile comes from the precursor ionisation front which lies ahead of the shock in the RG wind. In addition, absorption components with very narrow cores and broad P Cyg wings appear on top of the most intense emission lines (e.g. H I, He I, Fe II),  especially in the earliest epochs. The features at higher velocity are from the explosion ejecta, while the absorption elements originated in the intervening wind material, seen in projection against the white dwarf and the ejecta (\citet{vaytet2007swift}, \citet{shore1996}, \citet{nussbaumer1995proof}). 
 
\section{Discussion}
\label{section:discussion}
 
\subsection{The Balmer series}
\label{subsection:balmer}

\begin{figure}[h]
\centering
\includegraphics[width=0.5\textwidth]{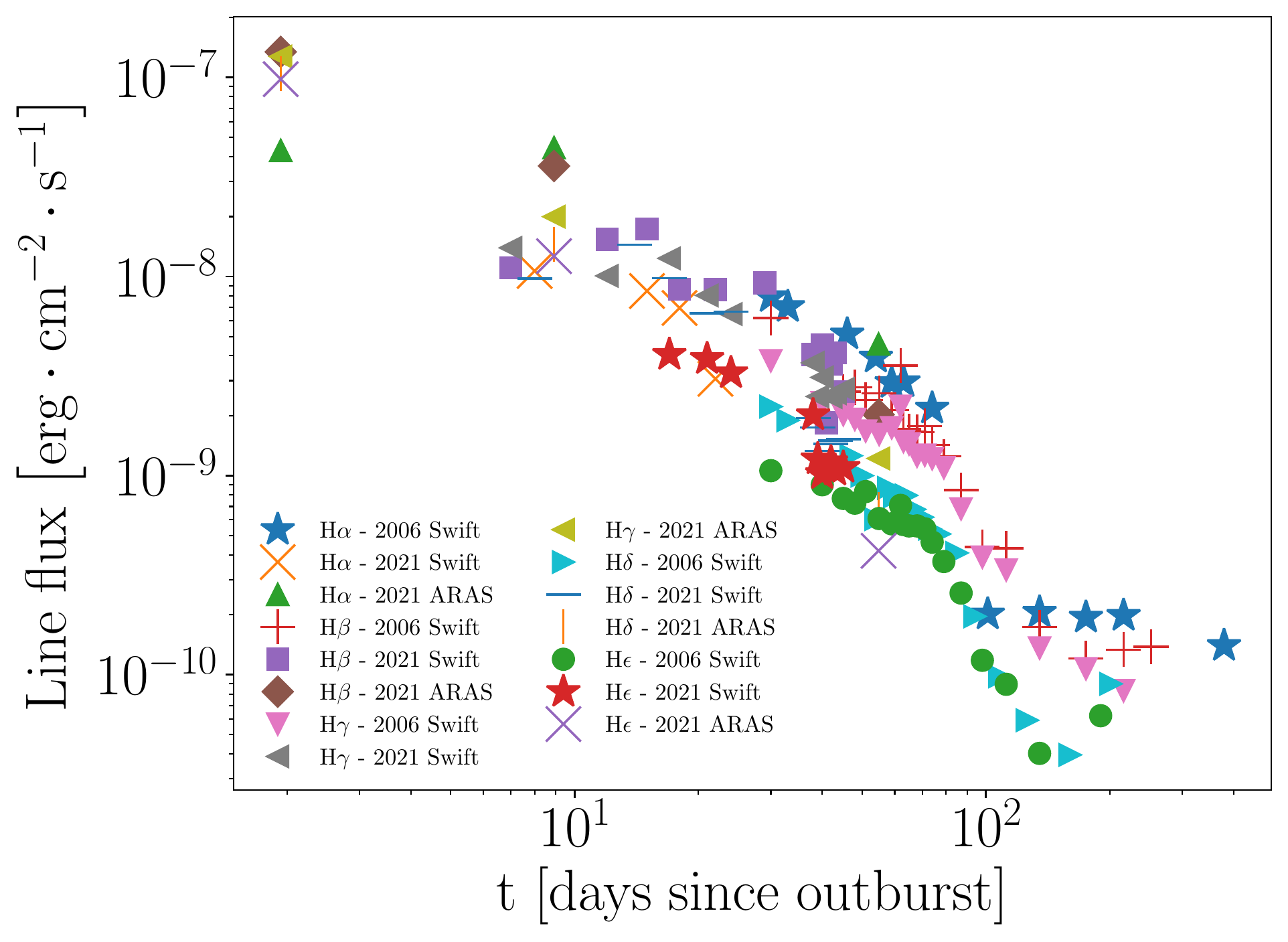}
\caption{\label{fig:balmer_rsoph}  \footnotesize Balmer series lines in RS Oph 2006 and 2021: development during outbursts. The scaling between the two events is a constant factor of 22. Data are binned and averaged over intervals of 3 days. It is important to note the upturn after the end of the SSS day 90.}
\end{figure} 

\begin{figure}[h]
\centering
\includegraphics[width=0.5\textwidth]{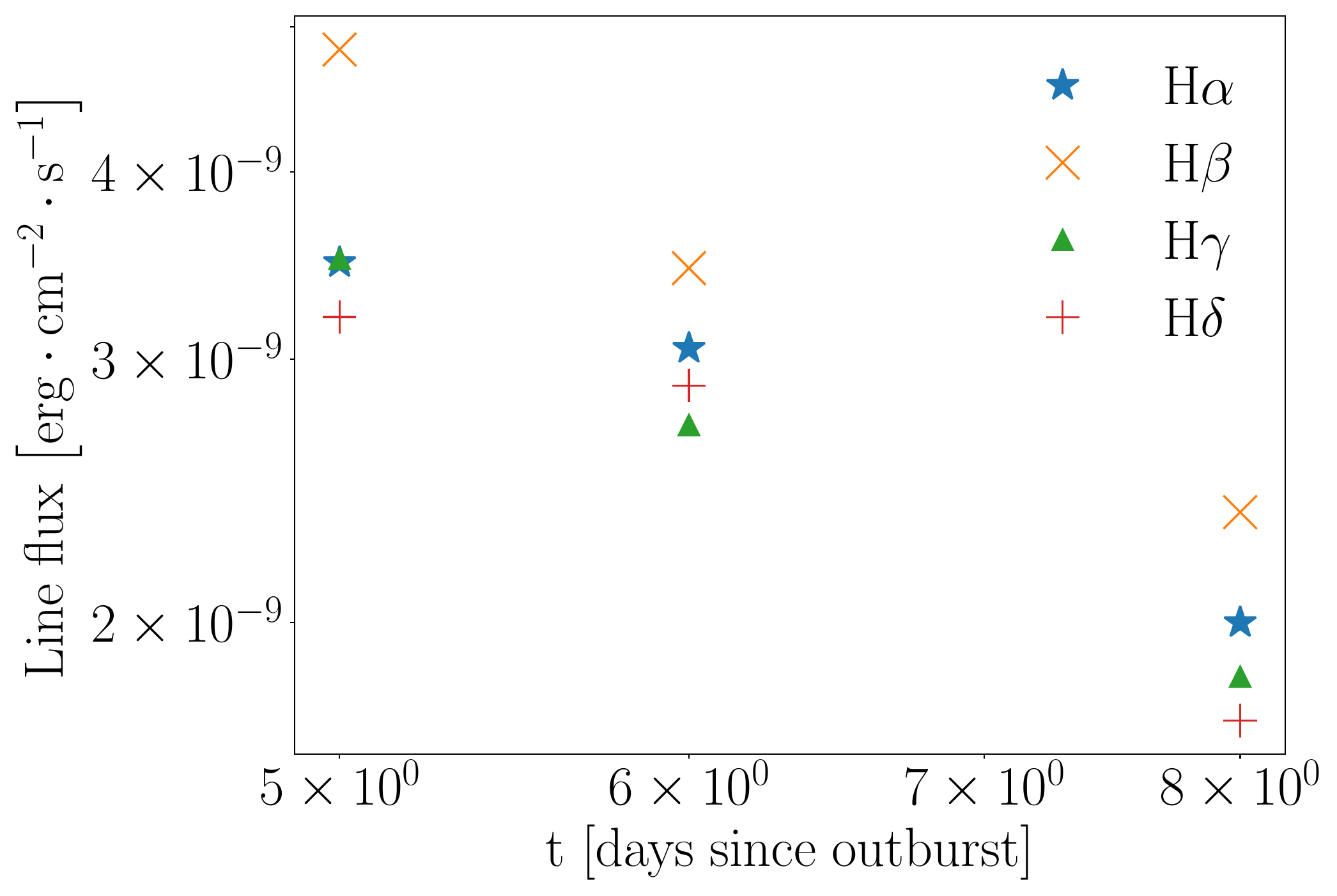}
\caption{\label{fig:balmer_v3890sgr2019swift} \footnotesize Hydrogen transitions in V3890 Sgr 2019: Balmer lines evolution in Swift UVOT UV grism spectra.}
\end{figure}

\begin{figure}[h]
\centering
\includegraphics[width=0.5\textwidth]{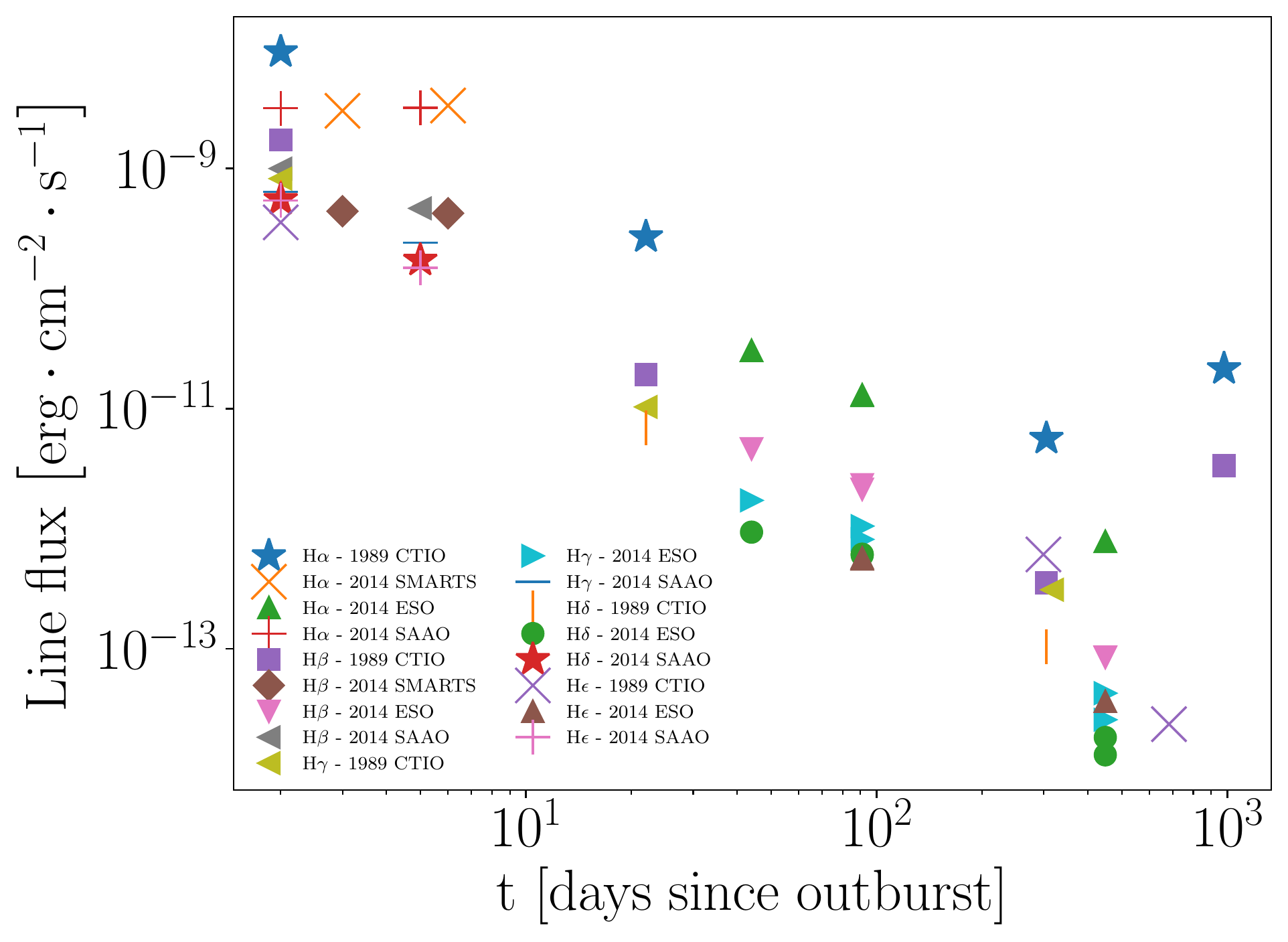}
\caption{\label{fig:balmer_v745sco} \footnotesize Development of Balmer series lines during the 1989 and 2014 outbursts of V745 Sco. All but SMARTS data are binned and averaged over intervals of 3 days.}
\end{figure}

\begin{figure}[h]
\centering
\includegraphics[width=0.5\textwidth]{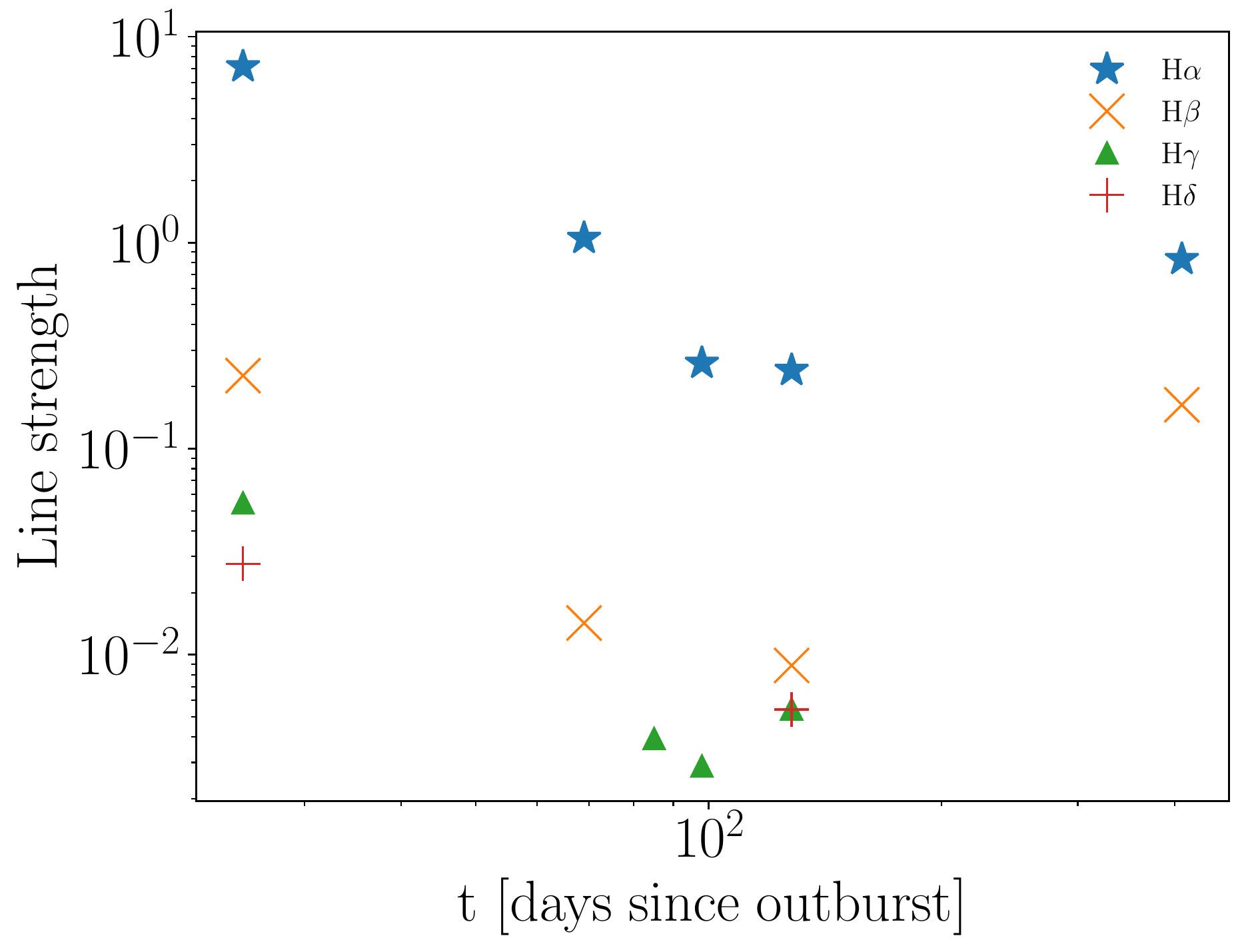}
\caption{\label{fig:balmer_v407cygnot} \footnotesize Hydrogen transitions in V407 Cyg 2010: Balmer lines evolution in NOT spectra. Data are binned and averaged over intervals of 3 days.}
\end{figure}

\begin{figure}[h]
\centering
\includegraphics[width=0.5\textwidth]{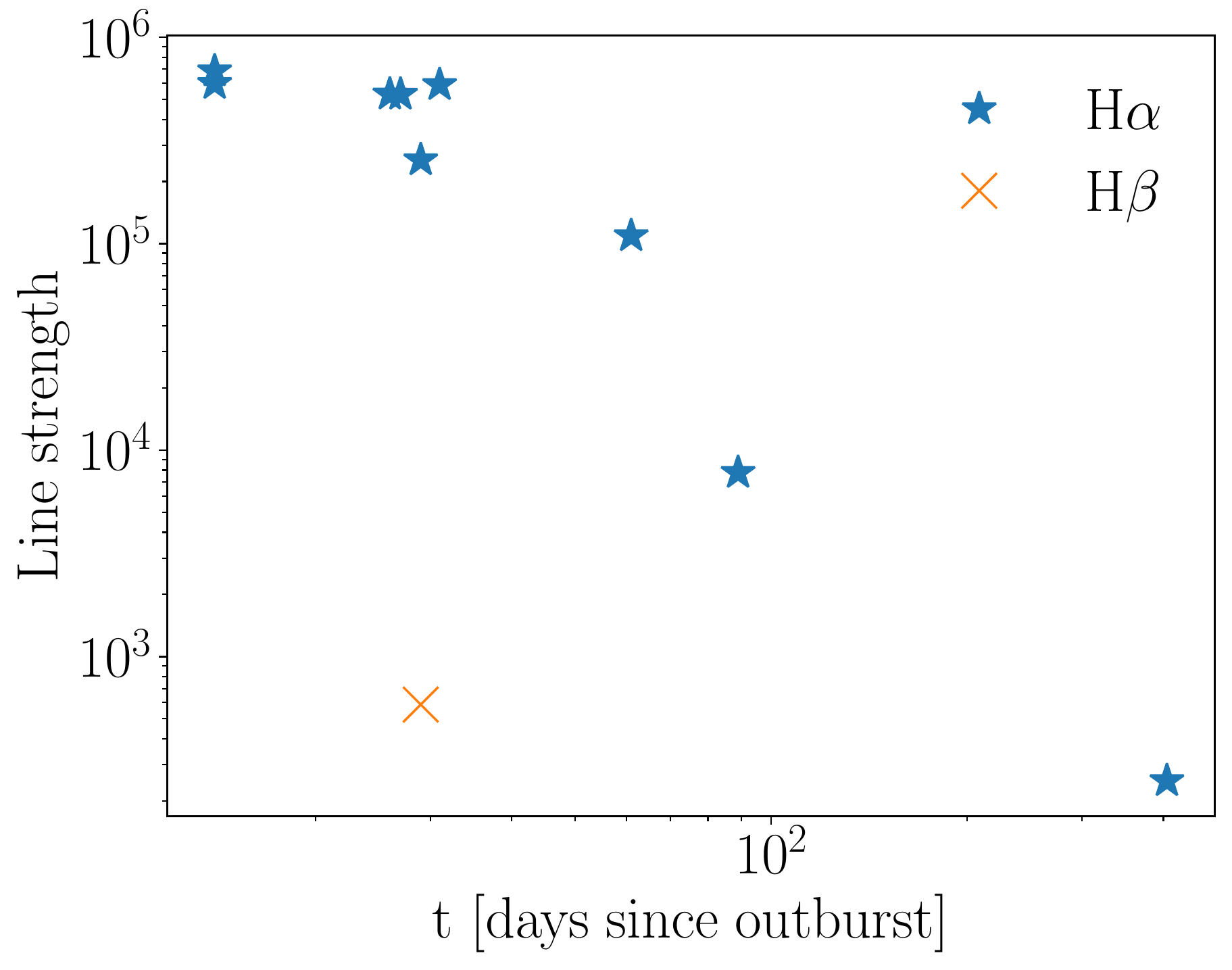}
\caption{\label{fig:balmer_v407cygond} \footnotesize Hydrogen transitions in V407 Cyg 2010: Balmer lines evolution in Ondřejov spectra. Data are binned and averaged over intervals of 3 days.}
\end{figure}

The time series for the Balmer integrated fluxes in our systems are shown in Figs.\ref{fig:balmer_rsoph} - \ref{fig:balmer_v407cygond}.\footnote{The ARAS spectra for RS Oph were limited to those having flux calibration with resolutions between about 850 and 1140.} The two outbursts of RS Oph show similar trends. The outburst progression is essentially the same, but with systematically lower integrated fluxes. A precursor component appears on all of the Balmer lines before the emission was dominated by the expanding ejecta. The initial portion reflects this and it persists during the period when the mass of the ejecta is increasing while the shock remains almost ballistic. The deceleration occurs when sufficient matter has accumulated. The  discrepancy between the two outbursts can be explained by the different orientations of the system relative to the line of sight because of the differential scattering and absorption of the intervening RG wind.

Early phases show composite profiles: extended emission component from the  ejecta with  narrow intense central peaks from the circumstellar environment ionised the shock precursor.  In later stages, the two components become gradually less distinct. In the Swift data of the 2006 outburst of RS Oph, after the decline phase, the integrated flux begins rising again after about 160 days. The enhanced strength of the Balmer lines in the last spectra, particularly H$\delta$ and H$\epsilon$  can be attributed to a newly reformed accretion disc around the WD (\citet{ourpaper}). A similar trend was observed for other transitions, for example He II 4686 \AA\; and He\,I 5048\,\AA. After about 80 days after T0 in RS Oph, the change in the slope corresponds to the point at which the shock reaches wind break-out, which nearly coincides with the end of the supersoft XR phase. 

Despite differences in timescales and sampling, the observed systems all show similar characteristics. Specifically, after eruption Balmer line profiles show a narrower core and very broad wings extending up to $2000\;\rm{km\;s^{-1}}$ (or even $6000\;\rm{km\;s^{-1}}$ in Swift spectra). In Swift spectra, the narrow Balmer lines are unresolved, while the FWZI is between 6000 and 12000 km s$^{-1}$ at the start of the event and  decreasing thereafter to as low as 2000 km s$^{-1}$. In contrast, during quiescence, the FWZI of the Balmer lines never exceeds 900 km s$^{-1}$ (suggesting they form in an accretion disc). The highest values are found for H$\rm{\alpha}$, the other lines of the series are generally narrower and weaker. 

In the dataset used for the analysis, only the spectra of V407 Cyg show strong P Cygni absorption originating from the environment. The disappearance of narrow absorption components on H$\beta$ (and the variations in line profiles of high ionisation species) signals the ionisation of the ambient medium \citep{2011A&A...533L...8S}. In many respects, the 2010 eruption of V407 Cyg was remarkably similar to RS Oph 1985 \citep{vcyg1}.

In V745 Sco, about 22 days after the 1989 outburst the Balmer lines almost completely vanished in the CTIO spectra, especially H$\rm{\gamma}$, and beyond. In the earlier spectra, lines clearly show the double component, with a narrow central peak whose width is $\rm{FWZI} \sim 2500 - 3000\;\rm{km\;s^{-1}}$ and extended asymmetric wings, for a total width $\rm{FWZI} \sim 5000-8000\;\rm{km\;s^{-1}}$ in most profiles. SMARTS data from the 2014 outburst show similar features on the central peak of the narrow component. It is an actual structure and not an artefact, also observed on $H\rm{\alpha}$ in ESO spectrum at $\rm{t}=44\;\rm{days}$ and both H$\rm{\alpha}$ and H$\rm{\beta}$ for the NOT sequence. This may be shadowing from the red giant, as suggested for V407 Cyg in 2010 (Shore et al., \citeyear{vcyg2}), or produced by transient structures in the surrounding wind. As the precursor advances in the circumstellar gas, these regions produce additional complex features on the emission profiles, which are slightly different from pure absorption by the wind. The  ESO spectra also show an absorption component at $\simeq 25\;\rm{km\;s^{-1}}$ on the H$\epsilon$ profiles. Another important feature of Balmer lines in this sequence is the blueshift in the central peak. It increases over time, from $\simeq -65\;\rm{km\;s^{-1}}$ at $\rm{t} = 44$ to $\simeq -135\;\rm{km\;s^{-1}}$ at $\rm{t} = 449$ for H$\rm{\alpha}$ and H$\rm{\beta}$, from $\simeq -115\;\rm{km\;s^{-1}}$ at $\rm{t} = 44$ to $\simeq -165\;\rm{km\;s^{-1}}$ at $\rm{t} = 449$ for H$\rm{\gamma}$, from $\simeq -76\;\rm{km\;s^{-1}}$ at $\rm{t} = 44$ to $\simeq -158\;\rm{km\;s^{-1}}$ at $\rm{t} = 449$ for H$\rm{\delta}$ and $\simeq -170\;\rm{km\;s^{-1}}$ at $\rm{t} = 449$ for H$\rm{\epsilon}$. Again, this is the wind radial velocity probed while the shock travels further out.

As already noted, NOT and Ondřejov spectra of the 2010 outburst of V407 Cyg are the only ones showing clear evidence of line-of-sight absorption from the wind superimposed on emission lines. The V407 Cyg sequence was also the only one showing the wind absorption invert to emission as the precursor reaches the wind peripheral zone (\cite{vcyg1}, \cite{vcyg2}). The absorption is deeper at the beginning of the sequence, and progressively diminishes until it is no longer visible after $\rm{t}\simeq 400\;\rm{days}$. At this point, the ejecta is no longer strongly interacting with the surrounding wind from the giant. The wind component appears in the spectra with the velocities listed in Table \ref{table:windvel}. These are compatible with the typical velocities of RG winds similar to the one in this system.

\begin{table}
\caption{Velocity properties of the wind components in V407 Cyg 2010 spectra.}             % title of Table
\label{table:windvel}      % is used to refer this table in the text
\centering                          % used for centering table
\begin{tabular}{ccc}        % centered columns (4 columns)
\hline\hline                 % inserts double horizontal lines
Dataset & Line & $\rm{v_{rad}}  \; \left[ \rm{km}\;\rm{s^{-1}} \right]$ \\     % table heading 
\hline                        % inserts single horizontal line
NOT & H$\rm{\alpha}$ & $(-80, -73)$ \\
 & H$\rm{\beta}$ & $(-40, -37)$ \\
 & H$\rm{\gamma}$ & $(-100, -92)$ \\
 & H$\rm{\delta}$ & $(-70, -52)$ \\
Ondřejov & H$\rm{\alpha}$ & $(-65, -50)$ \\
 & H$\rm{\beta}$ & $-170$ \\
\hline                                   %inserts single line
\end{tabular}
\end{table}

\subsection{ Helium lines}
\label{subsection:helium}
Permitted He I lines appear a few days after the luminosity peak together with possible hints of He II (although these lines are usually quite weak or blended ). Several days later, He II  appears and reaches  maximum intensity while He I is still visible even if it shows narrower and weaker profiles. The He II evolution is especially interesting for our purposes because this line  traces the shock. He II 1640 \AA\; appears in almost every IUE spectrum of all systems from immediately after the  peak, from day 6 onwards, and remains visible until about day 250, 14 and 145 days for RS Oph, V745 Sco and V3890 Sgr, respectively. He II 4686 \AA\; shows a similar evolution: as with He II 1640, it appears at $\rm{t} \simeq 5\;\rm{days}$ in the ARAS sequence of V3890 Sgr and $\rm{t} \simeq 14\;\rm{days}$ in Swift set of RS Oph 2021, reaching maximum strength between 1 and 2 months later and remaining visible for longer than 100 days in RS Oph and V407 Cyg. It starts fading at around 20 days for V745 Sco and V3890 Sgr. The profiles are broad in each spectrum, with wings extending up to $\gtrsim 2100\;km\;s^{-1}$ in the UV and even $\sim 5000\;\rm{km\;s^{-1}}$ for the optical line in Swift spectra.  The Swift grism PSF is, at worst, $2300\;\rm{km\;s^{-1}}$.

\begin{figure}[h]
\centering
\includegraphics[width=0.5\textwidth]{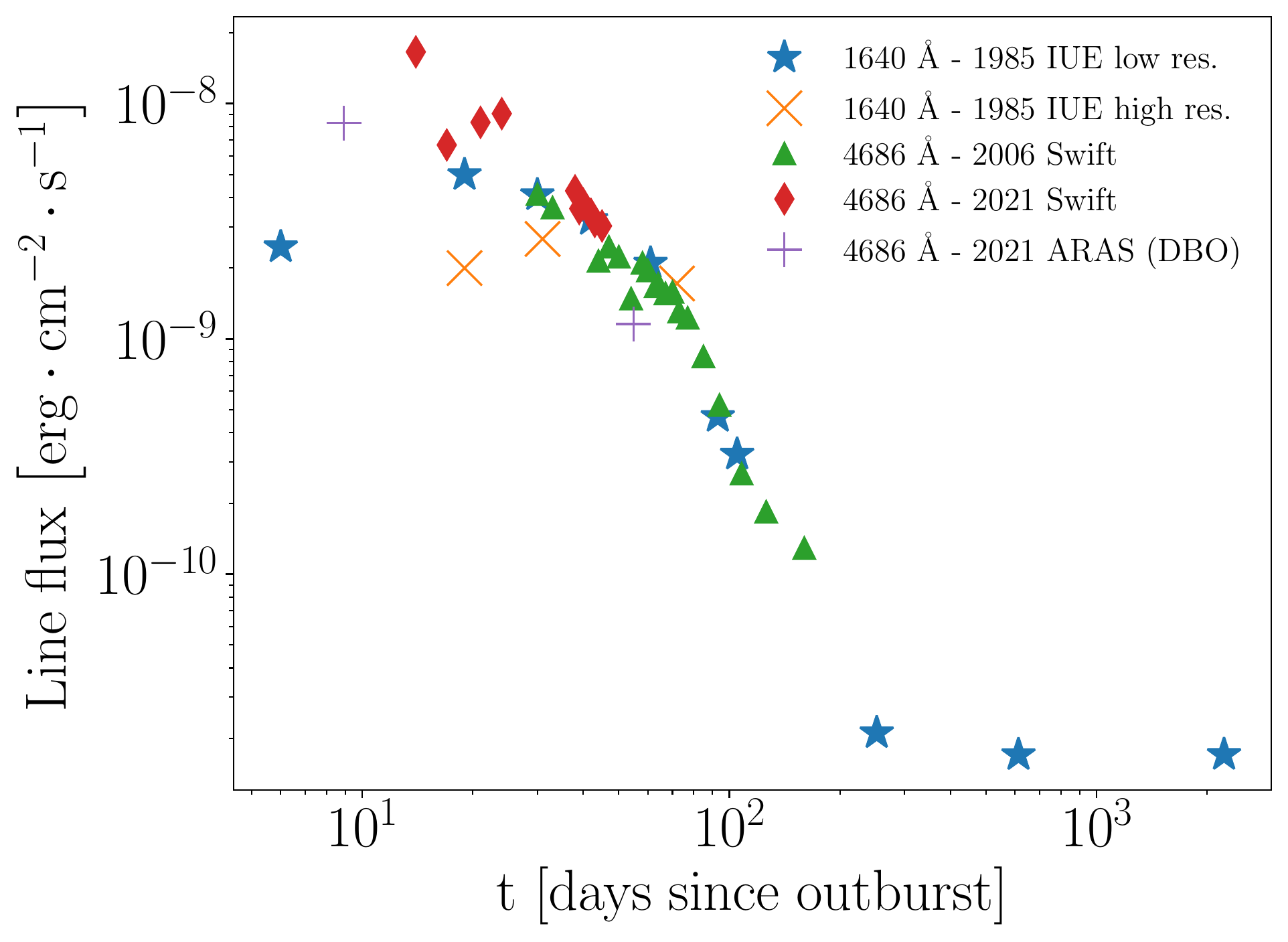}
\caption{\label{fig:HeII_2021vs2006} \footnotesize He II in RS Oph 2006 and 2021: integrated fluxes of lines. The scaling for the 2006 event is a constant factor of 22. Points are binned and averaged over intervals of 3 days.}
\end{figure}

\begin{figure}[h]
\centering
\includegraphics[width=0.5\textwidth]{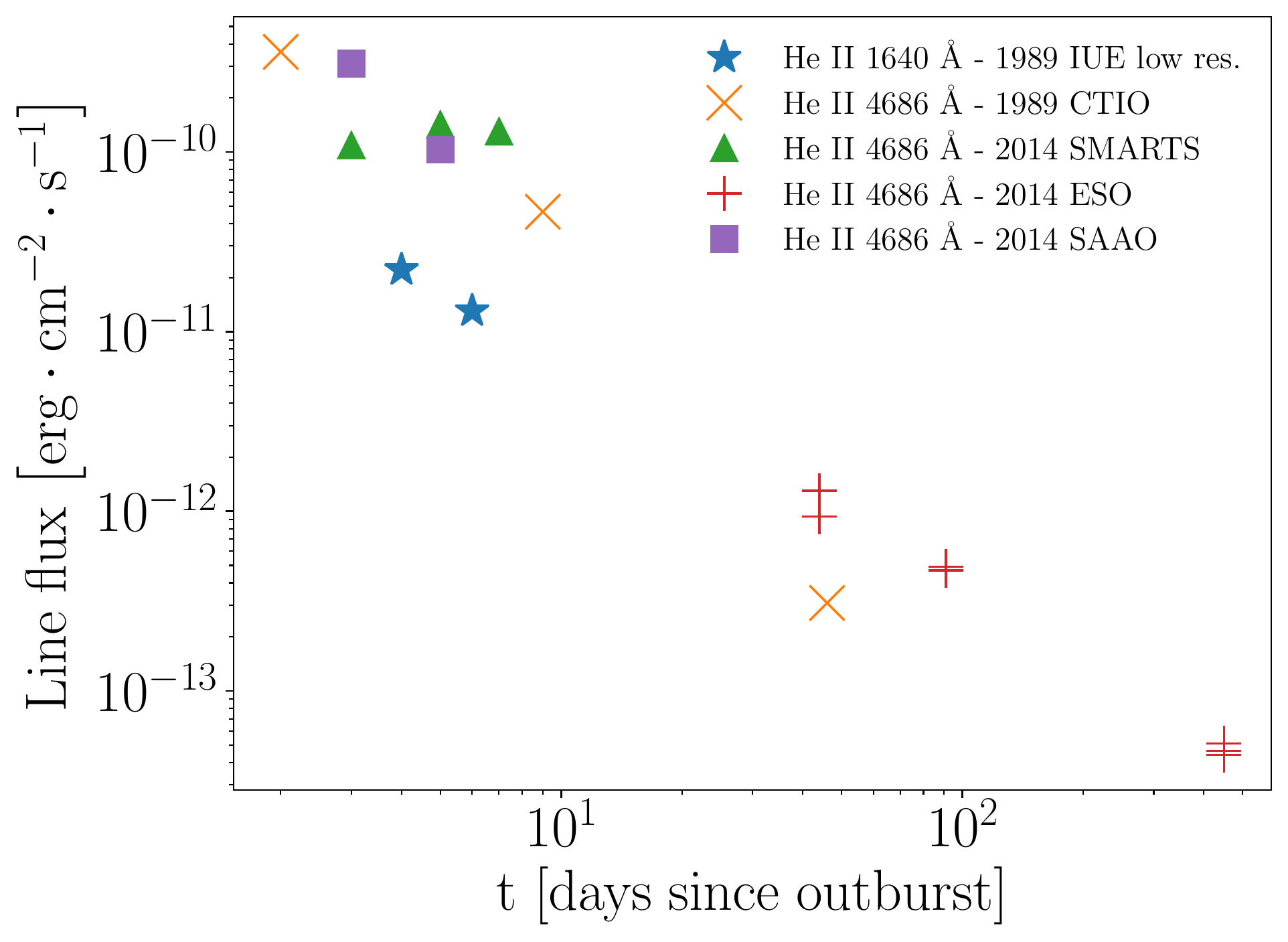}
\caption{\label{fig:HeII_v745sco} \footnotesize He II in V745 Sco 1989 and 2014: lines evolution. Data are binned and averaged over intervals of 3 days.}
\end{figure}

\begin{figure}[h]
\centering
\includegraphics[width=0.5\textwidth]{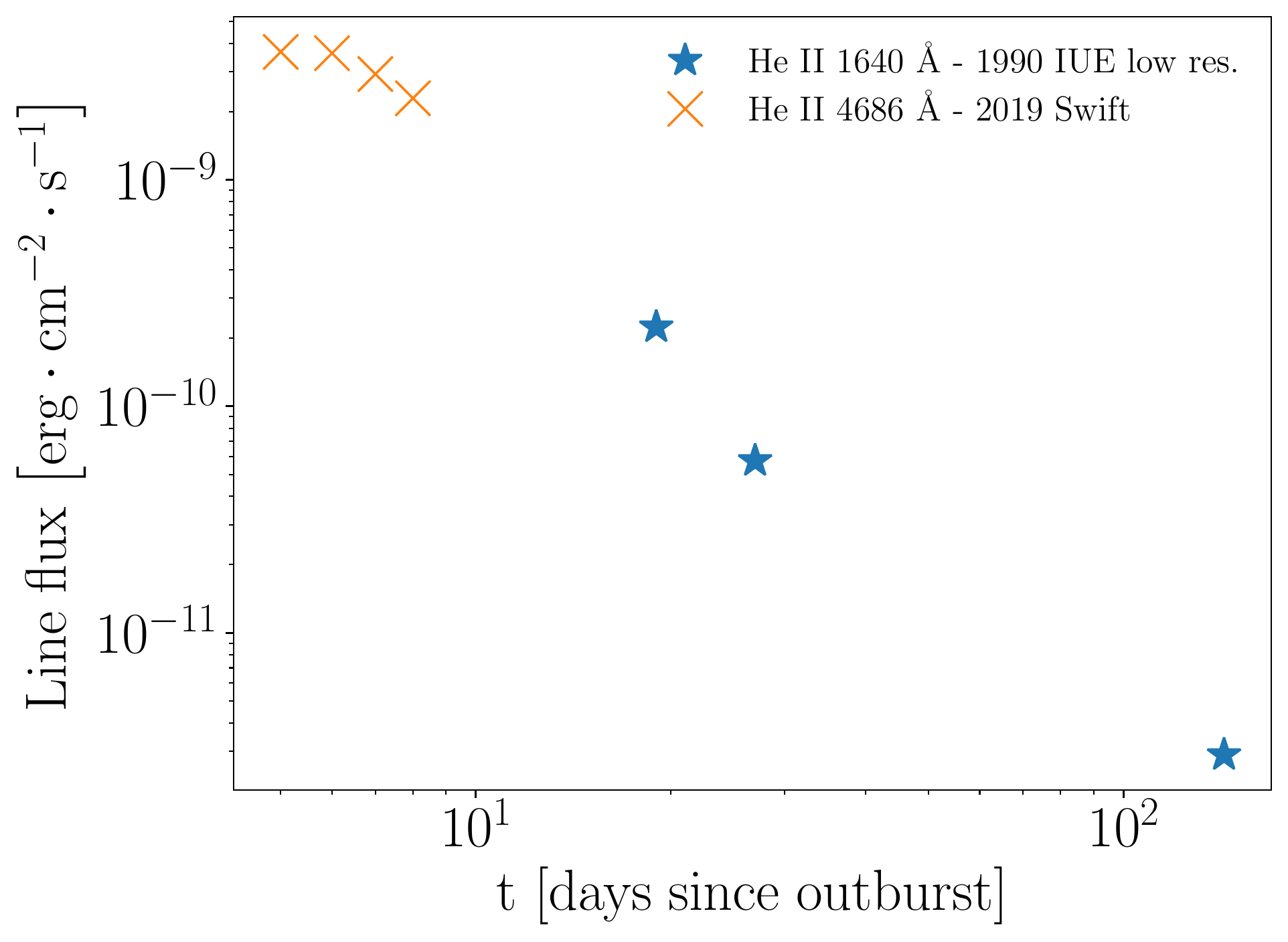}
\caption{\label{fig:HeII_v3890sgr} \footnotesize He II in V3890 Sgr 1990 and 2019: lines evolution. Data are binned and averaged over intervals of 2 days.}
\end{figure}

\begin{figure}[h]
\centering
\includegraphics[width=0.5\textwidth]{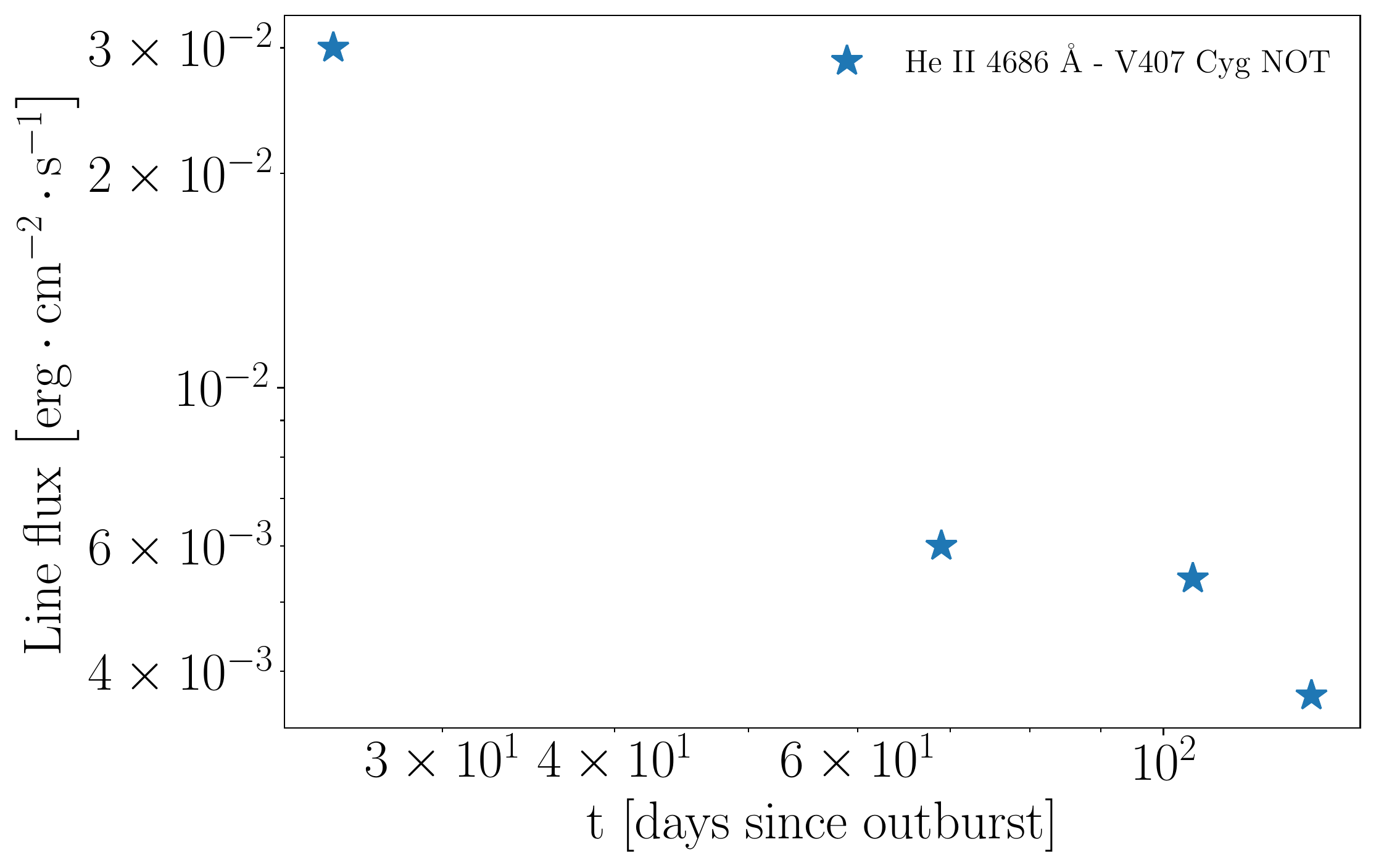}
\caption{\label{fig:HeII_v407cyg} \footnotesize He II in V407 Cyg 2010: line evolution. Data are binned and averaged over intervals of 3 days.}
\end{figure}

\begin{figure}[h]
\centering
\includegraphics[width=0.5\textwidth]{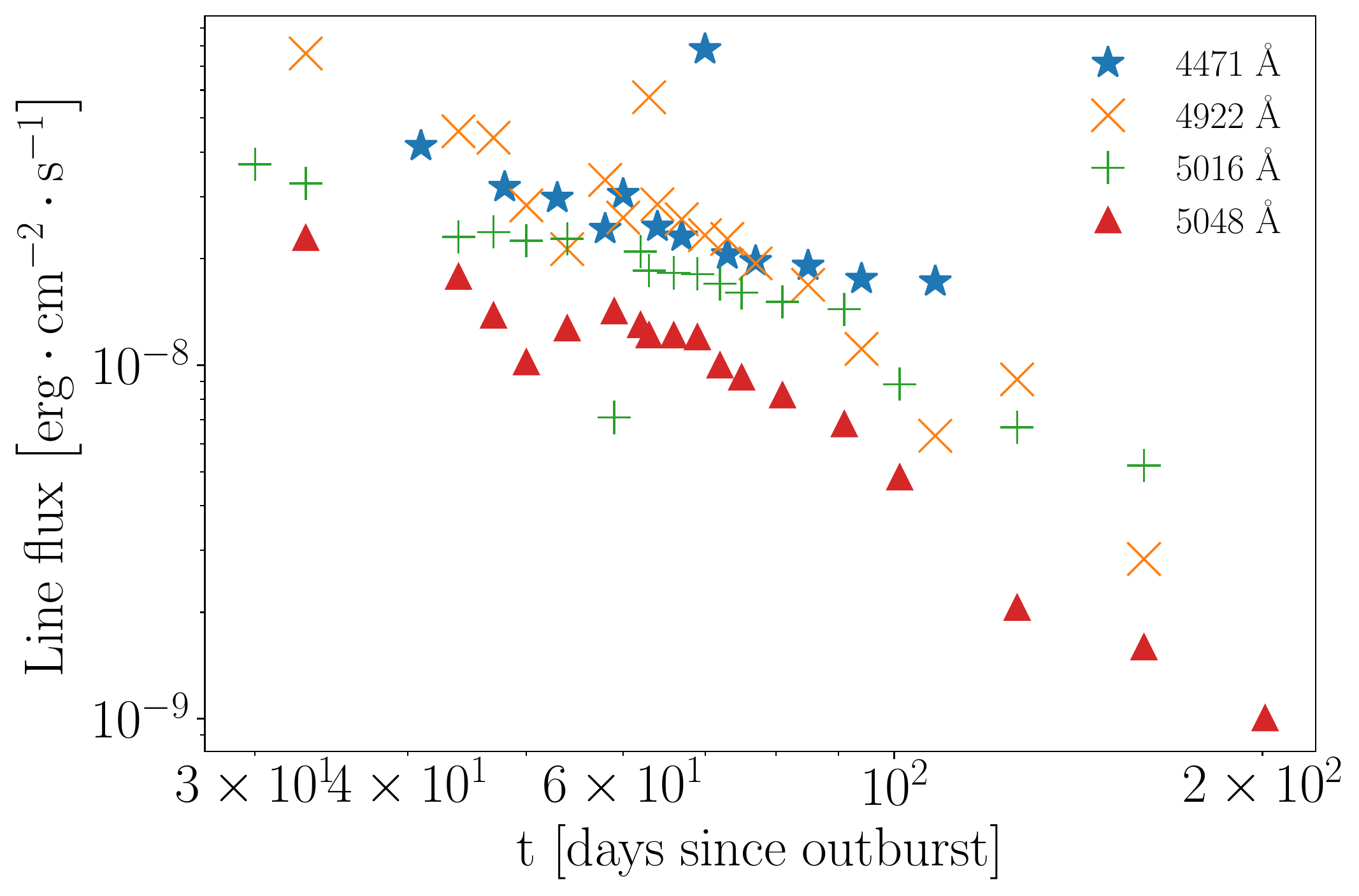}
\caption{\label{fig:HeIintflux} He I in RS Oph 2006: integrated fluxes evolution of lines in Swift spectra. Data are binned and averaged over intervals of 3 days.}
\end{figure}

The He II lines are produced by shocked ejected material during its cooling phase in the first months after outburst, whereas unshocked ejecta are responsible for He I emission, whose origin - and subsequent behaviour - by recombination and not just collisional excitation is similar to H lines.
Figs.\ref{fig:HeII_2021vs2006}-\ref{fig:HeII_v407cyg} show the temporal evolution of He II lines. Another illustrative example is given by He I lines in RS Oph 2006, shown in Fig.\ref{fig:HeIintflux}. These lines are not resolved or distinguishable in the 2021 spectra.

As discussed in \citet{shore1993interpretation}, the line of sight towards the WD changes within the wind of the RG on timescales of months. Hence, the obscuration from wind line absorption and scattering against the compact star depends on phase without redistribution. Any observation of the WD outer envelope or its nebula is affected by this wind extinction. He II 1640 is particularly influenced by such an atmospheric obscuration, and the evidence is that ionisation apparently changes as the obscuration itself changes.

Power-law fits $\rm{F} \propto \rm{t}^{\rm{\alpha}}$, where F is the strength of the line - either flux, intensity or counts depending on the dataset -, and t the stage of the outburst in terms of days since the beginning, were applied to the data. Again, the change in the trend after about 80 days from RS Oph 2006 follows the shock break-out stage. Another feature is the increase in the He II strength after 160 days, in agreement with the observed behaviour of H$\rm{\delta}$ and H$\rm{\epsilon}$ line of the Balmer series and presumably associated with the reformation of the accretion disc around the WD. The other systems show similar sequences but on different timescales, for example, both V745 Sco and V3890 Sgr evolve much more rapidly, and this is clearly seen in the sharp slope variation already around 5-6 days since the beginning. This change is not detected in V407 Cyg, for which the strength of He lines steadily declines during the whole sequence, from 25 days onwards.

\subsection{Si III and C III}
\label{subsection:SiC}
\begin{figure}[h]
\centering
\includegraphics[width=0.5\textwidth]{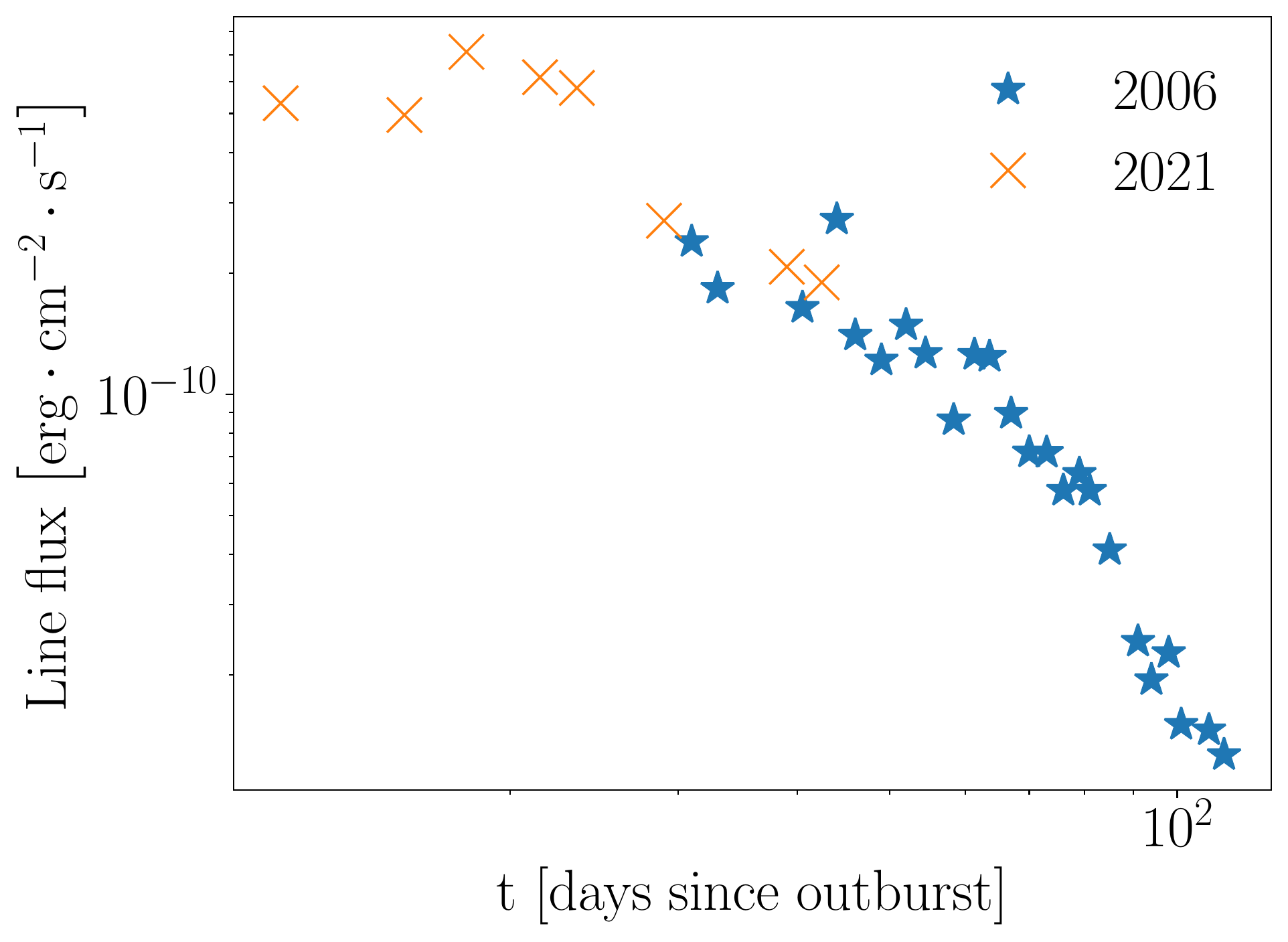}
\caption{\label{fig:Si_2006vs2021} Si III] 1892 \AA\; in RS Oph 2006 and 2021: evolution for emission lines in Swift spectra. Data are binned and averaged over intervals of 3 days. The scaling between the two events is a constant factor of 150.}
\end{figure}

\begin{figure}[h]
\centering
\includegraphics[width=0.5\textwidth]{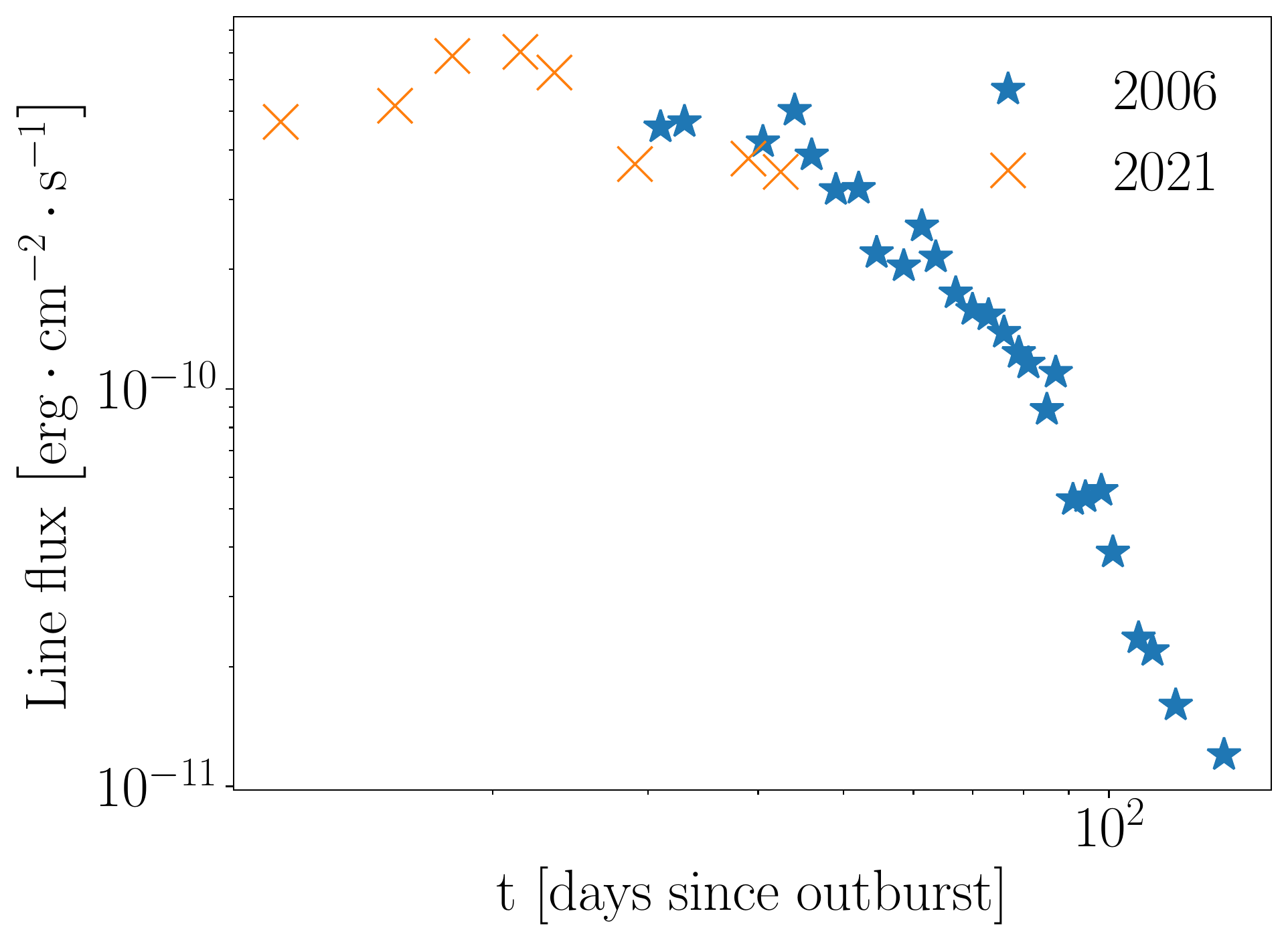}
\caption{\label{fig:C_2006vs2021} C III] 1909 \AA\; in RS Oph 2006 and 2021: evolution for emission lines in Swift spectra. Data are binned and averaged over intervals of 3 days. The scaling between the two events is a constant factor of 150.}
\end{figure}

The Si III] $3 \rm{s}^{2}\; ^1\rm{S}_{0} \rightarrow 3\rm{s}\;3\rm{p}\;^3\rm{P}^{0}_{1}$ $\rm{\lambda}$ 1892 \AA\; and C III] $2\rm{s}^{2}\;^1\rm{S}_{0} \rightarrow 2\rm{s}\;2\rm{p}\;^3\rm{P}^{0}_{1}$ $\rm{\lambda}$ 1909 \AA\; lines are intercombination low excitation transitions formed in the photoionised region. Due to their high critical densities, these lines can be used as diagnostics in a dense environment. The line ratio can be used to obtain the electron density nearly independent of the electron temperature (\citet{Nussbaumer1987}, \citet{keenan1992improved}, \citet{keenan1987si}, \citet{feibelman1987c}, \citet{aoki1998correlation} and \citet{kjeldseth1977emission}). The $\frac{\rm{S}\;III]}{\rm{C}\;III]}$ ratio can be used as a diagnostic from observed fluxes without requiring extinction corrections, unlike He II: it is higher for greater densities. Figs. \ref{fig:Si_2006vs2021} and \ref{fig:C_2006vs2021} show the evolution of integrated fluxes.

\begin{figure*}[h]
\centering
\includegraphics[width=\textwidth,height=0.4\textheight]{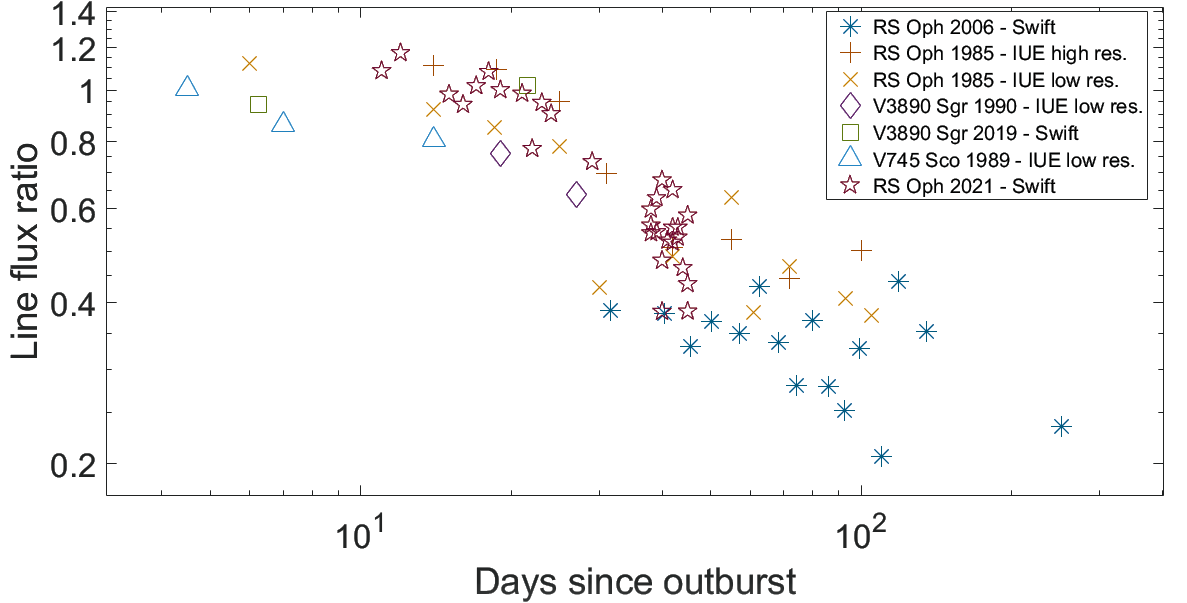}
\caption{\label{fig:ratio_mean} Ratio of Si III] and C III] integrated fluxes for single events as a function of days since the beginning of the outburst. Swift data are averaged over 5 and 3 days for RS Oph 2006 and V3890 Sgr, respectively.}
\end{figure*}

The integrated flux ratio $\rm{r}=\frac{\rm{F}(\rm{\lambda} 1909\;\rm{C \;III})}{\rm{F}(\rm{\lambda} 1892\;\rm{Si\;III})}$ \citep{feibelman1987c} has an average value $\rm{log(r)} \simeq 0.74$ in symbiotic stars. For RS Oph $\rm{log(r)} = 0.68$ in 2006 and $\rm{log(r)} = 0.40$ in 2021 from Swift, $\rm{log(r)} = 0.46$ in IUE, for V745 Sco $\rm{log(r)} = 0.10$ in IUE, for V3890 Sgr $\rm{log(r)} = 0.62$ in IUE. With different timescales but similar behaviour, r increases in the first few days, then decreases until between 60-100 days after the beginning and eventually begins  increasing again at later stages.  This traces the electron density of the medium, see  Fig. \ref{fig:ratio_mean}.

The sort of universality displayed in the electron density evolution traced by the C III]/Si III] ratio cannot be due to the shock dynamics.  The XR light curves, and individual line evolution, show that the systems have different histories and different environments. The shock and precursor are affected by different density gradients and mass loss rates, and different separations between the system components.  Instead, we suggest that Fig. \ref{fig:ratio_mean} shows that the value of $\rm{n_e(t)}$ is biased to a particular time by the recombination process.  Because the recombination rate determines the emissivity, the highest density should weigh in at the earliest time while that in the lower density region will contribute later since the time scales as $\rm{t_{n_e}} \sim \rm{n_e}^{-1}$.  The physics is the same regardless of the environment, although the different systems will show each line following a system-dependent development.

\subsection{Highly ionised species}
\label{subsection:ionised}
Immediately after the eruption, the electron densities of the ambient gas are sufficiently high that only permitted line emission is observed; as the density decreases, the expanding shell begins emitting auroral and then nebular lines, with the highest excitation transitions emerging first because they usually have higher critical density than low excitation transitions. Over time, the density continues to decrease and the forbidden lines  strengthen. Simultaneously, the ionisation level monotonically increases.  High ionisation lines are a fundamental diagnostic of the passage of the shock through the wind. They are formed near the front, hence they probe physical conditions in the region just behind the shock. The role of the precursor is essential since it is strong enough to produce coronal lines in the unshocked wind.

The Mg II $\rm{\lambda\lambda}2798-2802\;$ doublet, forms in the gas immediately behind the ejecta. In contrast, the O III, Ne III and Ne V lines (especially forbidden transitions)  are produced by collisional excitation due to the high velocities and electron temperatures of the precursor and N III arises from the entire structure of the shock region. Ne/H spectra are typical of very fast novae, such that [Ne III] and [Ne V] lines are produced by extremely high-velocity shocks during transition phases when UV becomes transparent and the main nebular lines appear. Profile and strength changes of ionised lines from Mg, N, O and Ne are important to follow the corresponding evolution of shocks.

\begin{figure}[h]
\centering
\includegraphics[width=0.5\textwidth]{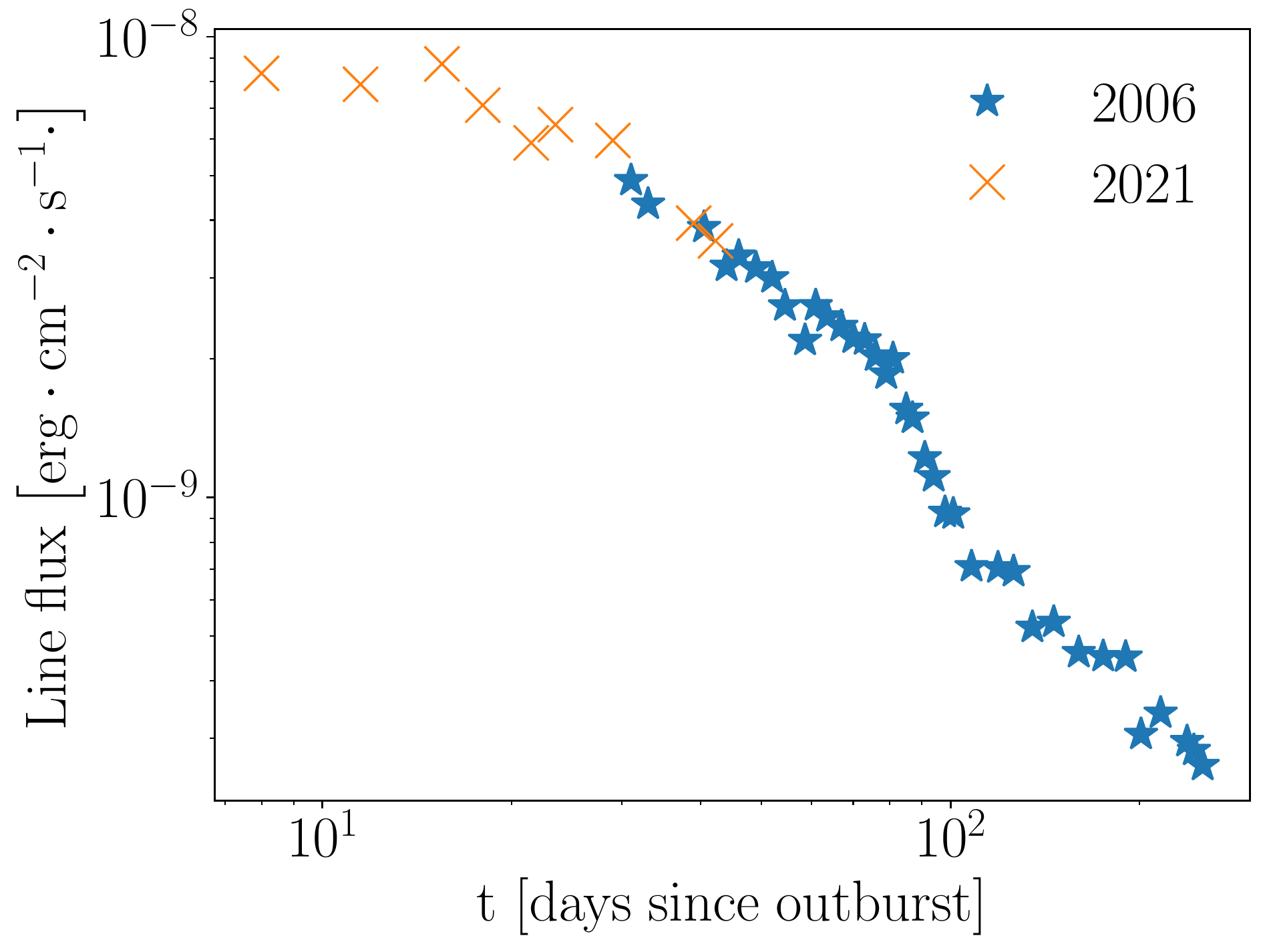}
\caption{\label{fig:OIII5007rsophswift} \footnotesize [O III] $\rm{\lambda}5007\;$ in RS Oph 2006 and 2021: integrated flux as a function of time in Swift UVOT/UV-grism spectra. Data are binned and averaged over intervals of 3 days. The scaling between the two events is a constant factor of 7.}
\end{figure}

The Swift UV grism spectra do not extend below 1700 \AA\  so O III] $\rm{\lambda}$1663 \AA\; and N V $\rm{\lambda}$1240 \AA\ are unobservable  (see, however, \citet{shore1996} for the IUE spectra from the 1985 outburst). Broad profiles from all the other transitions are observed throughout the 2006 and 2021 outbursts.
Fig.\ref{fig:OIII5007rsophswift} shows the integrated flux of [O III] $\rm{\lambda}$5007 \AA\; in days since outburst. 

\begin{figure}[h]
\centering
\includegraphics[width=0.5\textwidth]{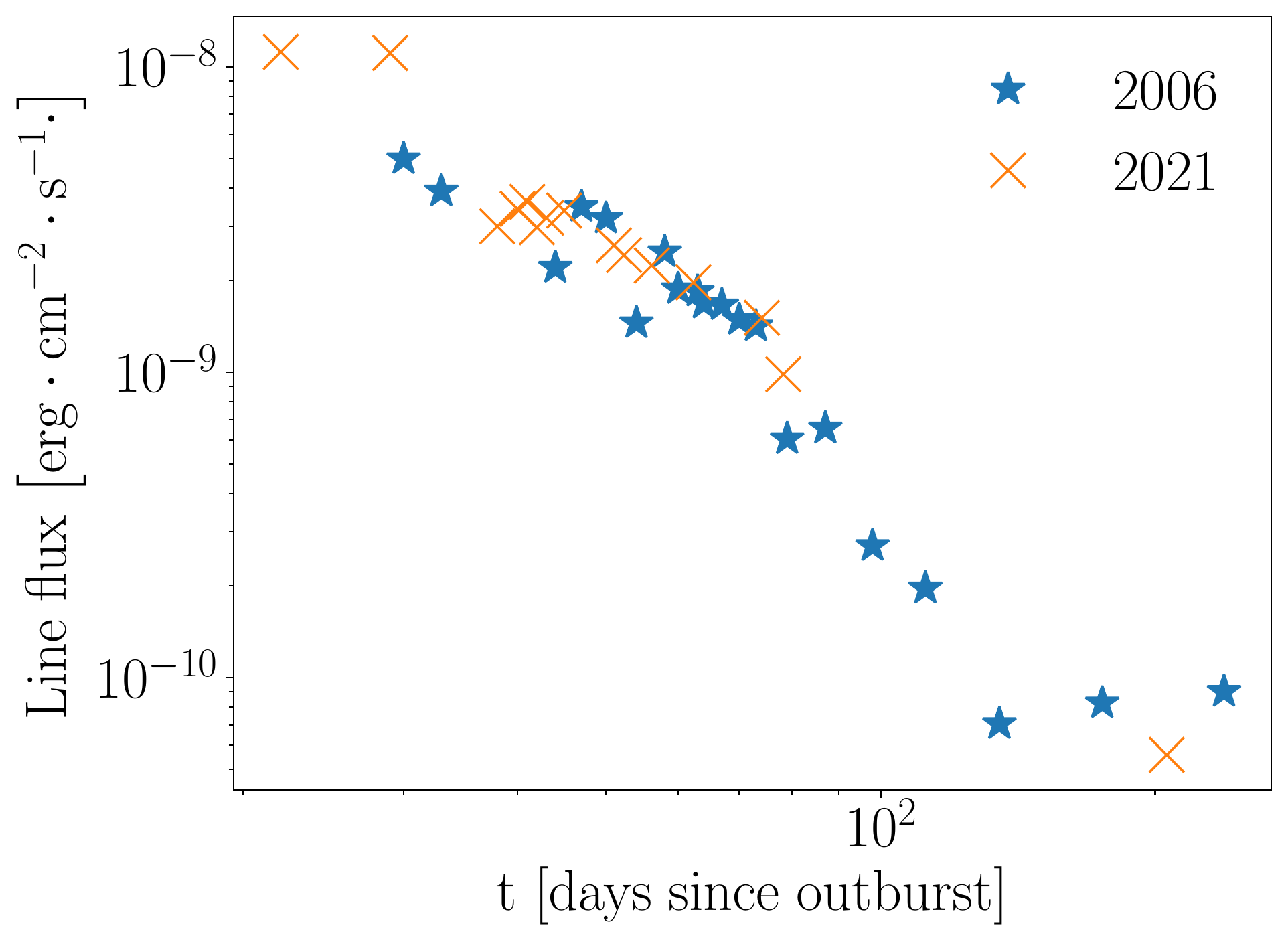}
\caption{\label{fig:OIII3133rsophswift} \footnotesize O III $\rm{\lambda}3133\;$ in RS Oph 2006 and 2021: integrated flux as a function of time in Swift UVOT/UV-grism spectra. Data are binned and averaged over intervals of 3 days.}
\end{figure}

A striking result is that the temporal development is the same in the two events, only the intensity is reduced in 2021.  During outburst, the strength of this line tracks the underlying continuum, which is seen to rise above the zero level between consecutive stages. As a consequence, the actual flux for the profile is different from the peak value. The behaviour over time is consistent, as shown in Fig.\ref{fig:OIII3133rsophswift} for O III $\lambda3133\;$.

\begin{figure}[h]
\centering
\includegraphics[width=0.5\textwidth]{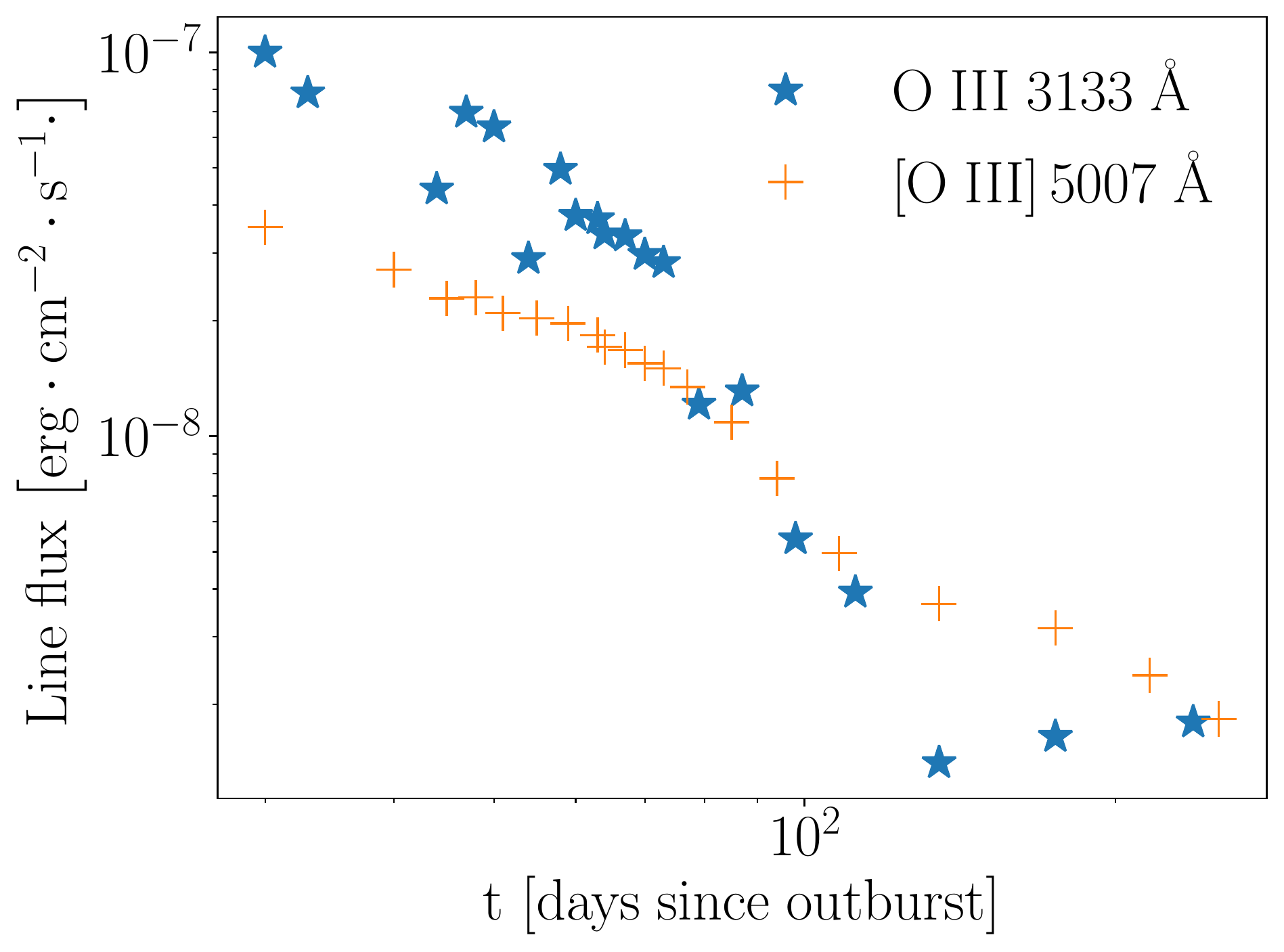}
\caption{\label{fig:OIII_rsoph2006} \footnotesize O III in RS Oph 2006: integrated flux as a function of time for forbidden and permitted lines in Swift UVOT/UV-grism spectra. Data are binned and averaged over intervals of 3 days. The scaling between the two events is a constant factor of 20.}
\end{figure}

\begin{figure}[h]
\centering
\includegraphics[width=0.5\textwidth]{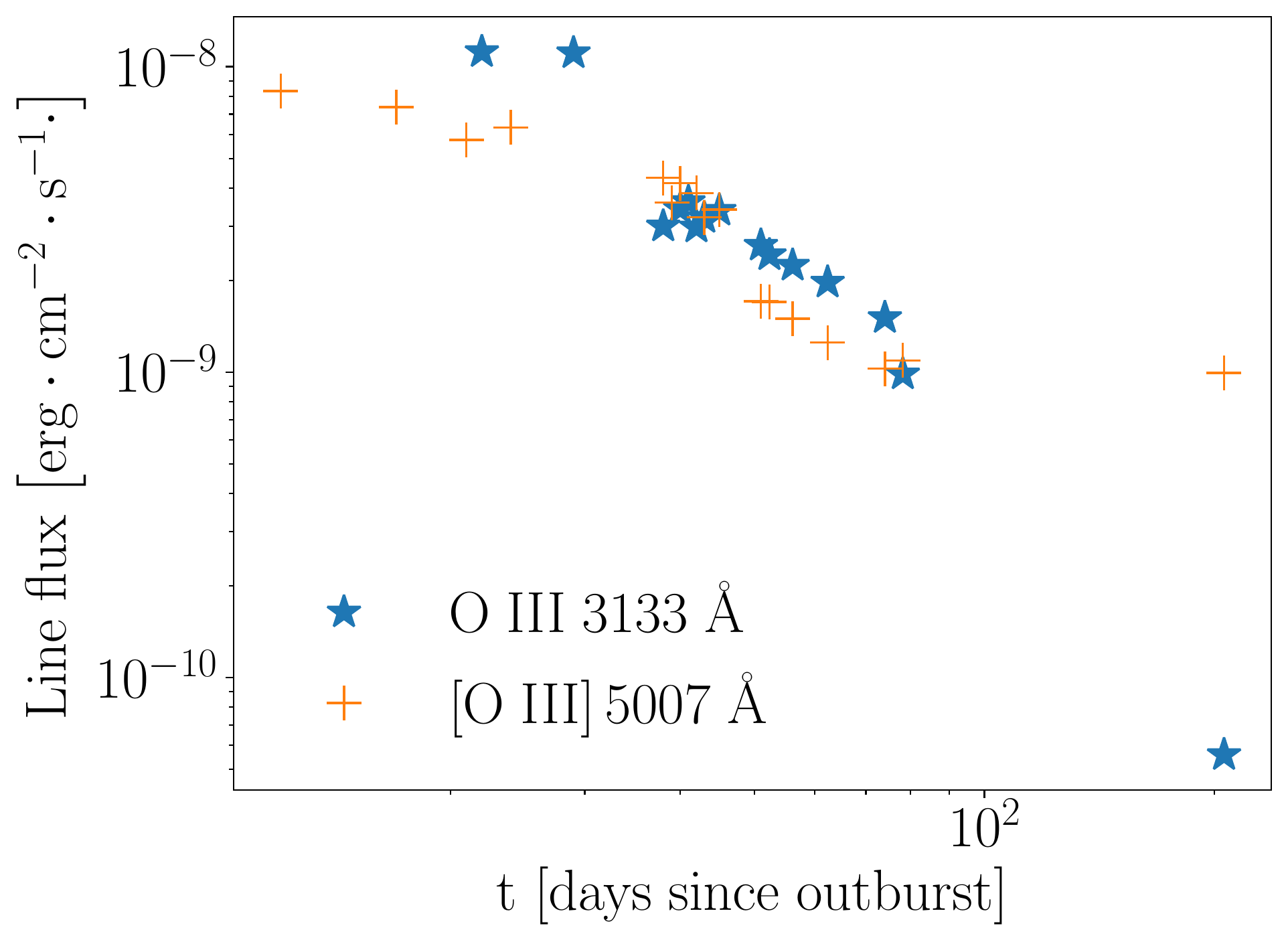}
\caption{\label{fig:OIII_rsoph2021} \footnotesize O III in RS Oph 2021: as in Fig.\ref{fig:OIII_rsoph2006}.}
\end{figure}

\begin{figure}[h]
\centering
\includegraphics[width=0.5\textwidth]{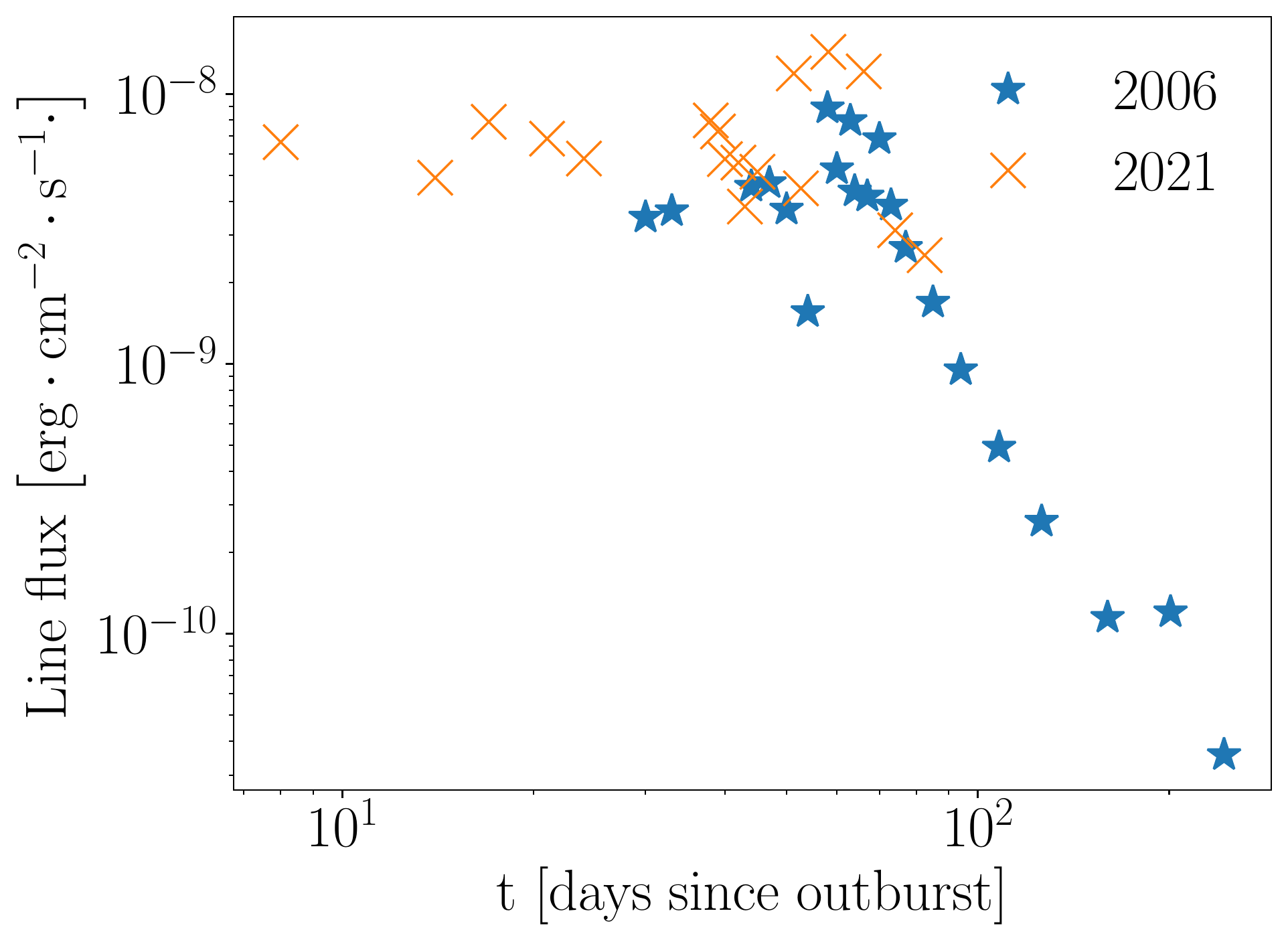}
\caption{\label{fig:MgII} \footnotesize Mg II $\rm{\lambda}2798\;$ in RS Oph 2006 and 2021: integrated flux as a function of time in Swift UVOT/UV-grism spectra. Data are binned and averaged over intervals of 3 days. The scaling between the two events is a constant factor of 22.}
\end{figure}

\begin{figure}[h]
\centering
\includegraphics[width=0.5\textwidth]{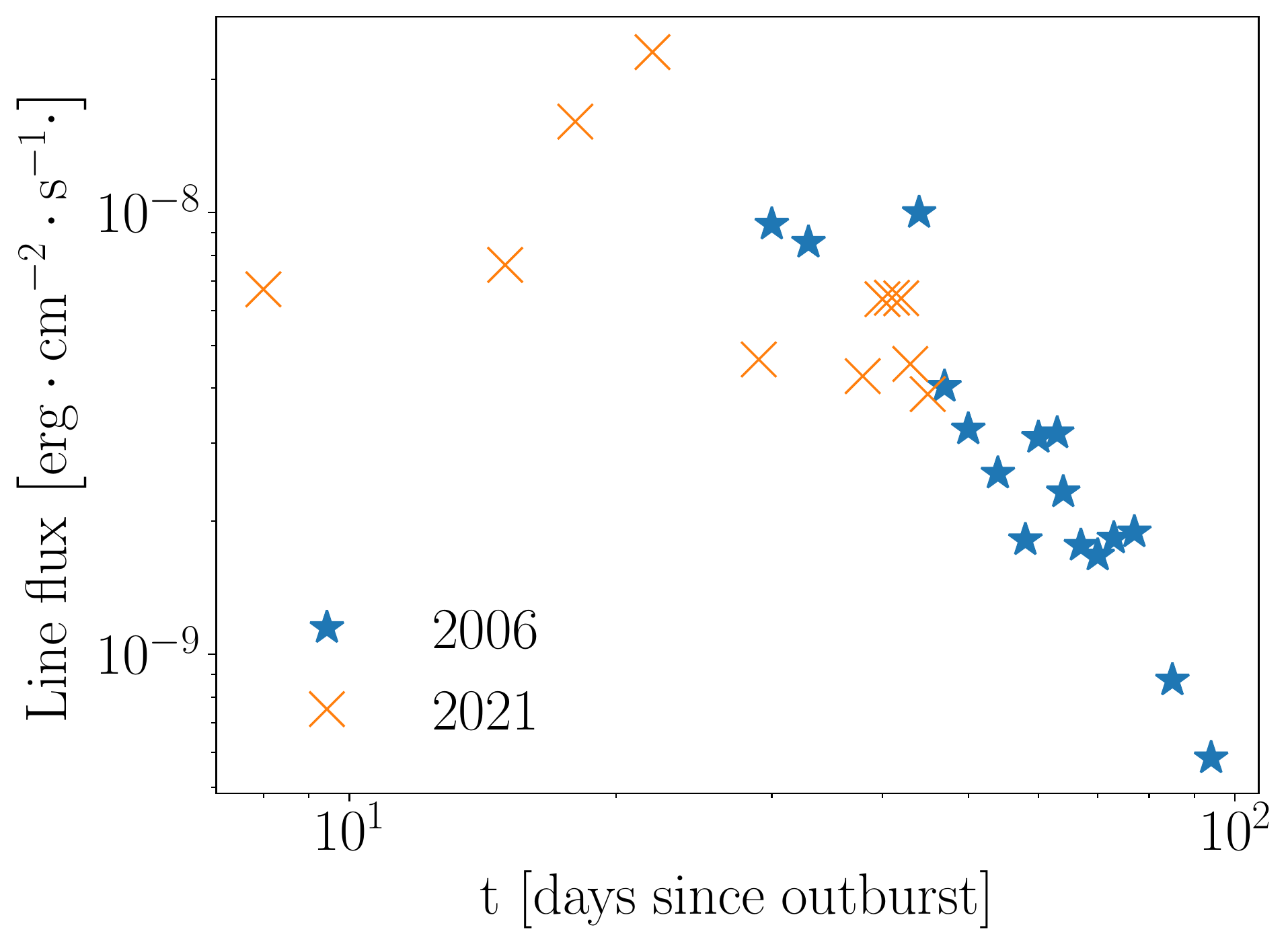}
\caption{\label{fig:NIII} \footnotesize N III] $\lambda1750\;$ in RS Oph 2006 and 2021: integrated flux as a function of time in Swift UVOT/UV-grism spectra. Data are binned and averaged over intervals of 3 days. The scaling between the two events is a constant factor of 50.}
\end{figure}

\begin{figure}[h]
\centering
\includegraphics[width=0.5\textwidth]{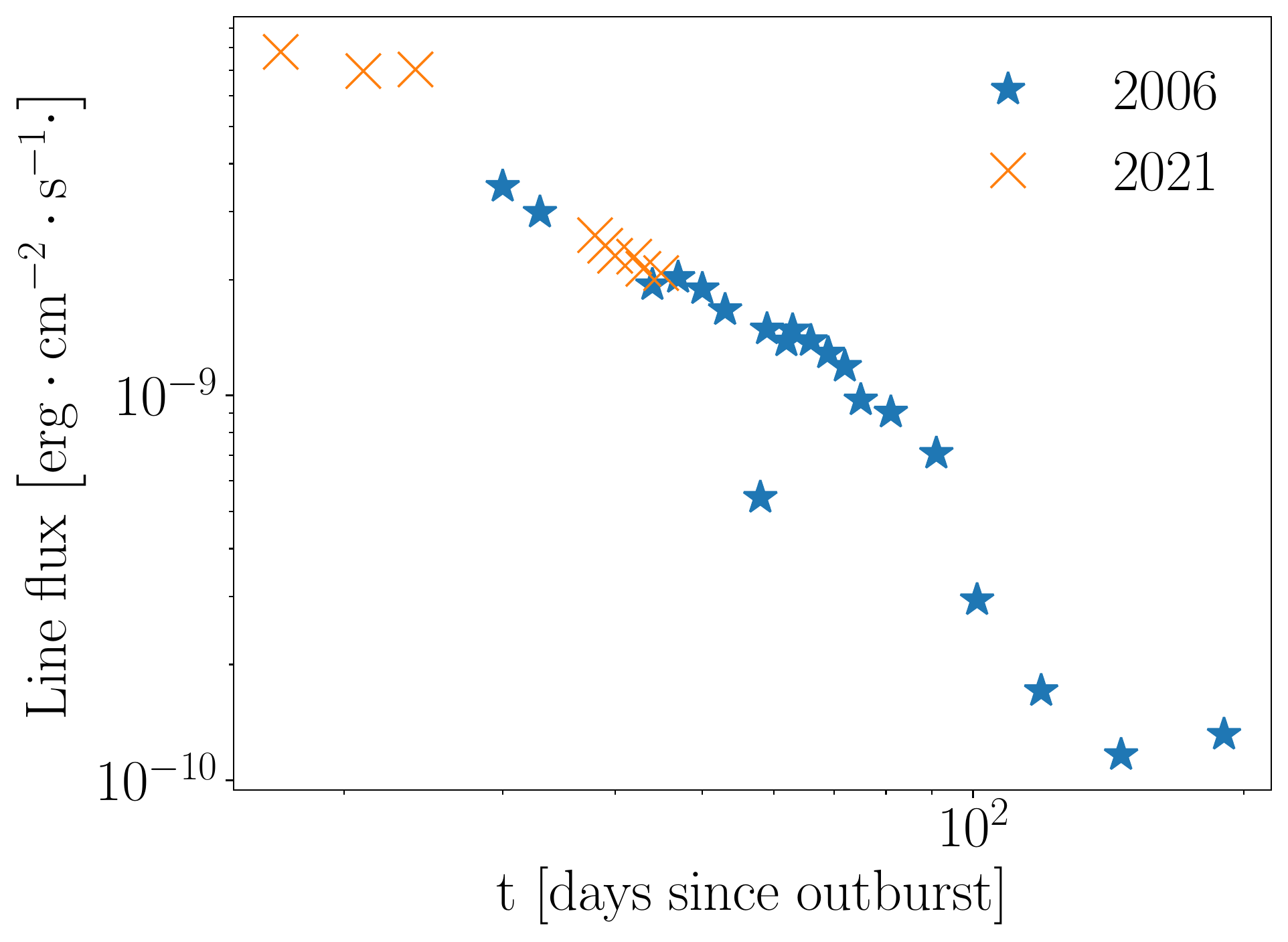}
\caption{\label{fig:NeIII} \footnotesize [Ne III] $\rm{\lambda}3869\;$ in RS Oph 2006 and 2021: integrated flux as a function of time in Swift UVOT/UV-grism spectra. Data are binned and averaged over intervals of 3 days. The scaling between the two events is a constant factor of 13.}
\end{figure}

\begin{figure}[h]
\centering
\includegraphics[width=0.5\textwidth]{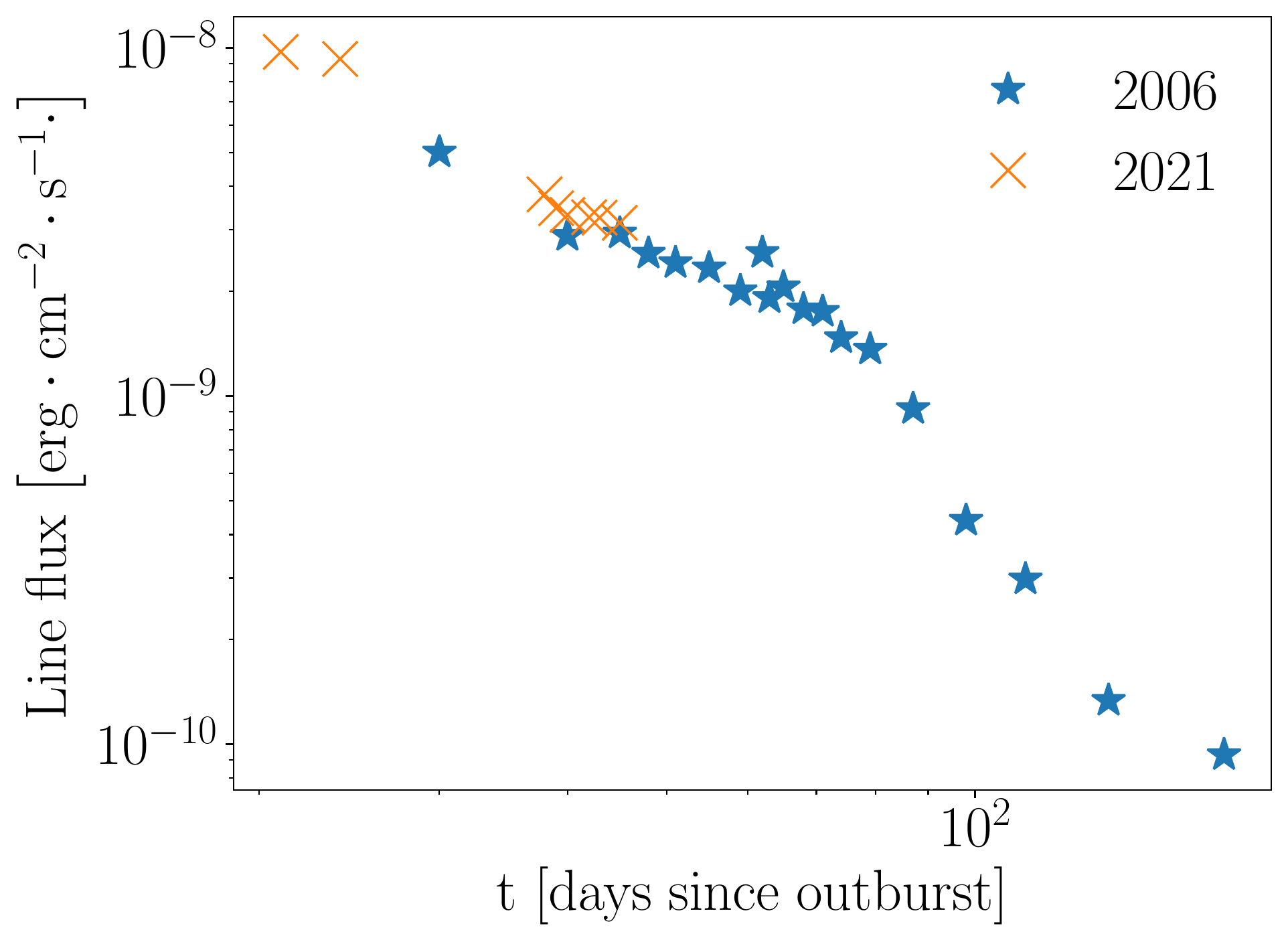}
\caption{\label{fig:NeV} \footnotesize [Ne V] $\rm{\lambda}3426\;$ in RS Oph 2006 and 2021: integrated flux as a function of time in Swift UVOT/UV-grism spectra. Data are binned and averaged over intervals of 3 days. The scaling between the two events is a constant factor of 13.}
\end{figure}

\begin{figure}[h]
\centering
\includegraphics[width=0.5\textwidth]{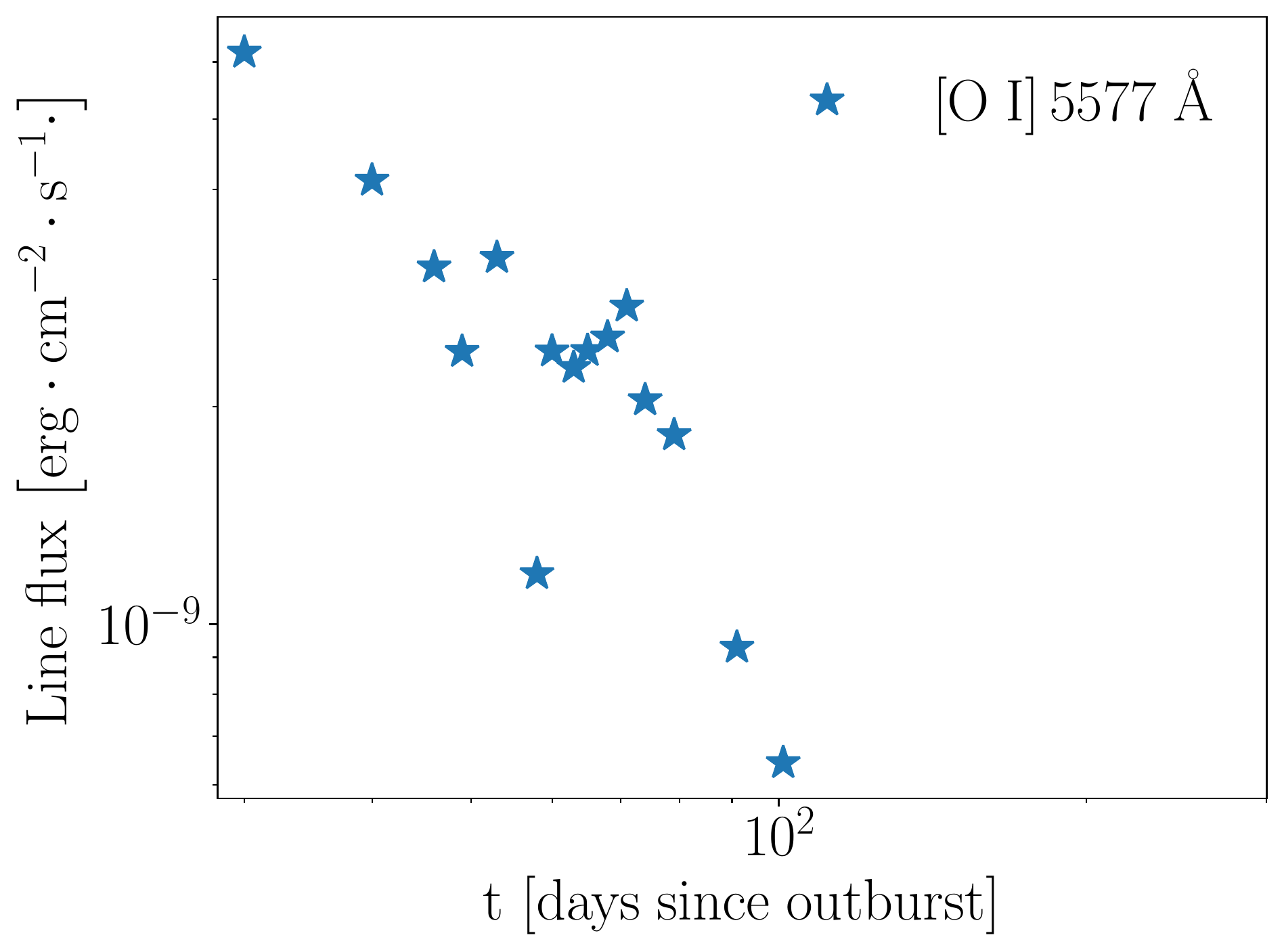}
\caption{\label{fig:ionised2_rsoph2006swift} \footnotesize [O I] in RS Oph 2006: evolution of the line in Swift spectra. Data are binned and averaged over intervals of 3 days.}
\end{figure}

\begin{figure}[h]
\centering
\includegraphics[width=0.5\textwidth]{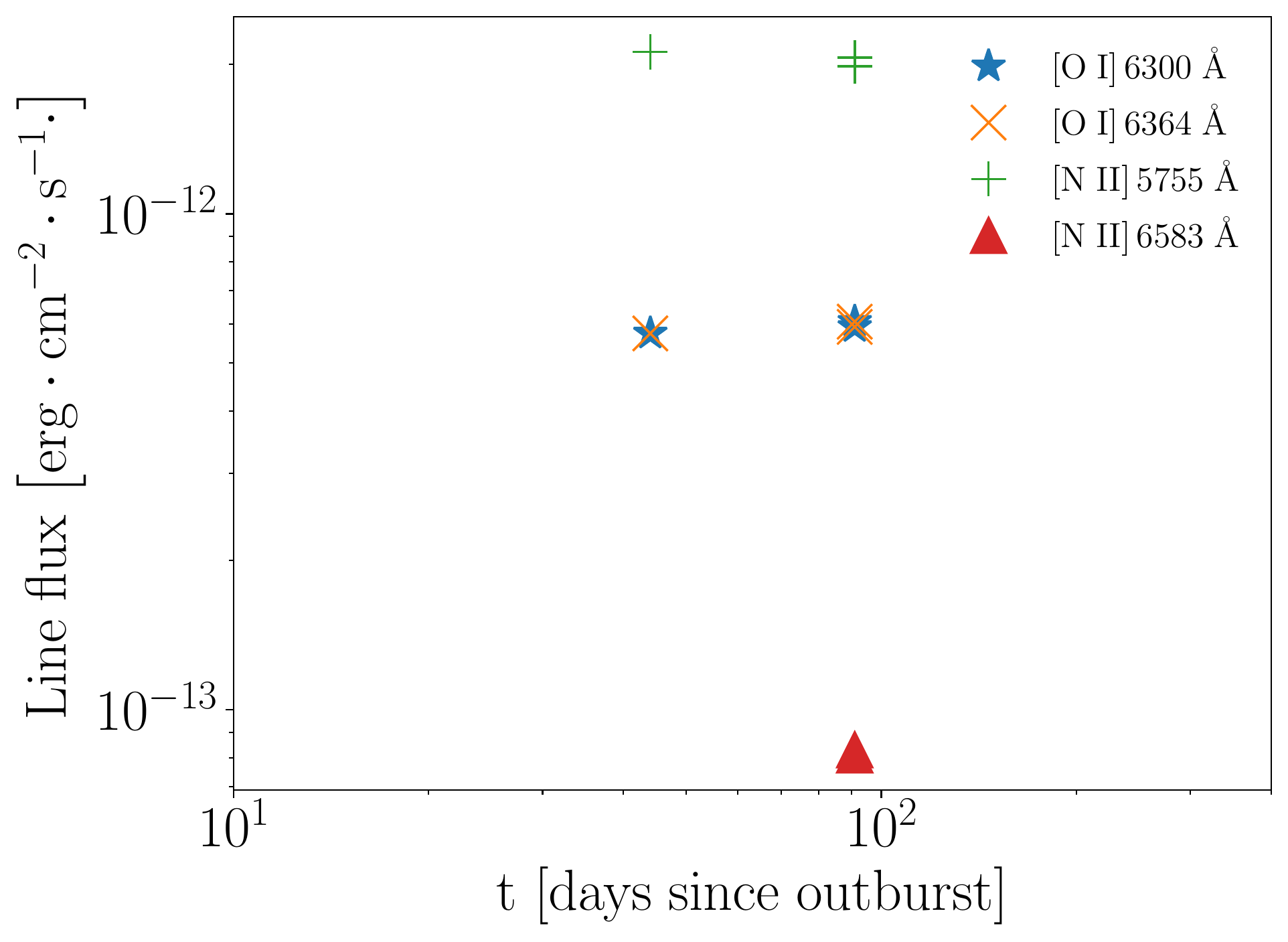}
\caption{\label{fig:ionised2_v745sco2014eso} \footnotesize V745 Sco 2014: neutral and multiple ionised lines in ESO spectra.}
\end{figure}

\begin{figure}[h]
\centering
\includegraphics[width=0.5\textwidth]{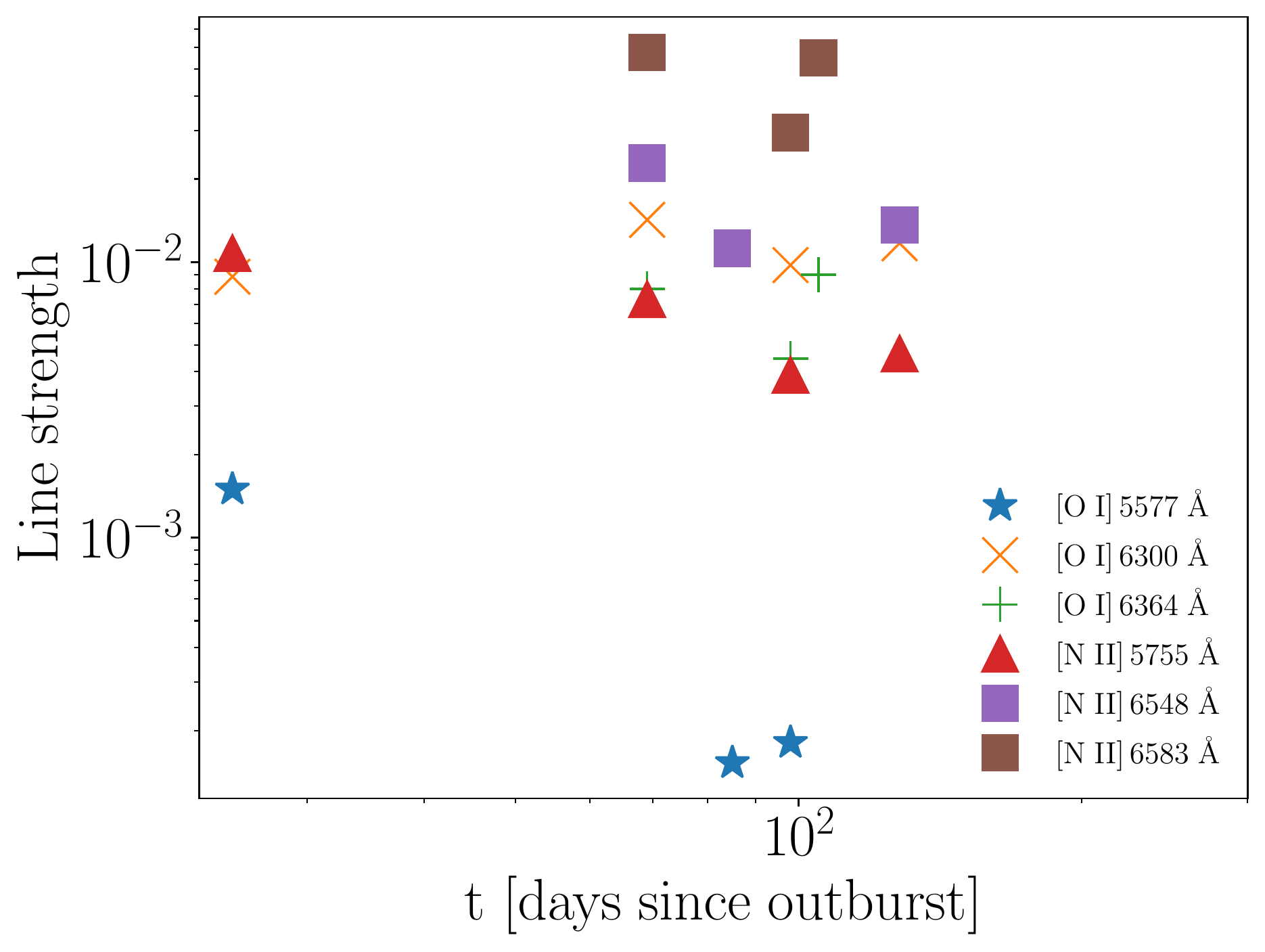}
\caption{\label{fig:ionised2_v407cygnot} \footnotesize V407 Cyg 2010: neutral and multiple ionised lines in NOT spectra. Data are binned and averaged over intervals of 3 days.}
\end{figure}

\begin{figure}[h]
\centering
\includegraphics[width=0.5\textwidth]{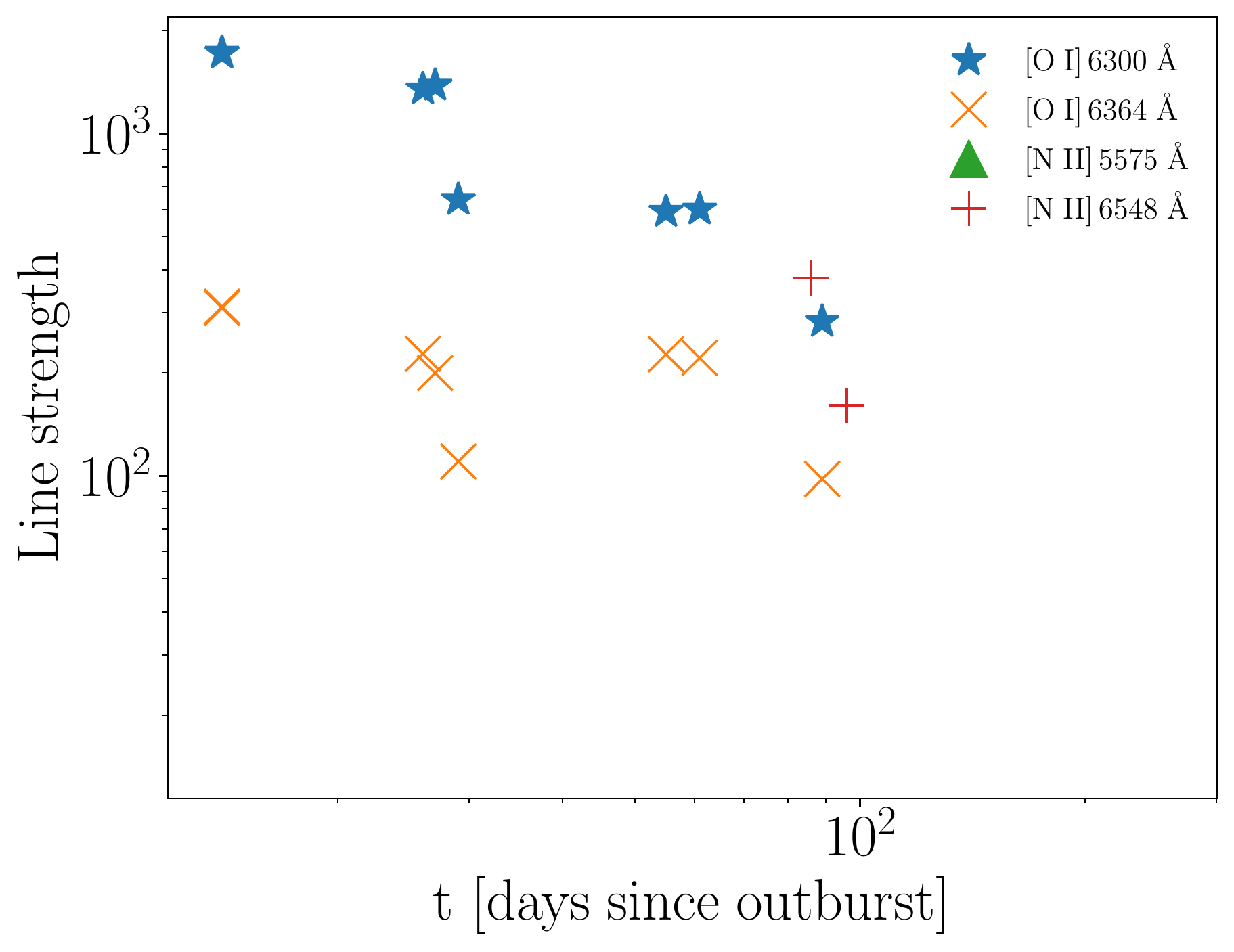}
\caption{\label{fig:ionised2_v407cygond} \footnotesize V407 Cyg 2010: neutral and multiple ionised lines in Ondřejov spectra. Data are binned and averaged over intervals of 3 days.}
\end{figure}

A range of oxygen transitions traces the shock throughout their evolution during the RS Oph event. For example, the  [O III] lines originate close to the shock front. Their evolution during RS Oph outbursts is shown in Figs.\ref{fig:OIII_rsoph2006} and \ref{fig:OIII_rsoph2021}. Applying power law fits, $\rm{F} \propto \rm{t^{\alpha}}$, to the 2006 data for [O III] $\rm{\lambda}5007\;$\AA\  yields   $\rm{\alpha}=-0.95 \pm 0.23$ for $30 \leq \rm{t} \leq 79$ and $\rm{\alpha}=-2.19 \pm 0.40$ for $79 \leq \rm{t} \leq 146$.  For O III $\rm{\lambda}3133\;$\AA\, the results are $\rm{\alpha}=-1.62 \pm 0.39$ for $30 \leq \rm{t} \leq 74$ and $\rm{\alpha}=-4.72 \pm 0.57$ for $74 \leq \rm{t} \leq 146$. Again, the change in the slope of the integrated flux around 80 days corresponds to the break-out and the end of the SSS. Like the H and He lines, the O III flux increases again after 160 days since the beginning of the outburst.  Figs.\ref{fig:MgII}-\ref{fig:ionised2_v407cygond} show how lines from different neutral and ionised species evolve during various outburst events of the Galactic Sy-RNe. An interesting feature is to be mentioned: an additional emission shows up at high-resolution IUE spectra on the red wing of He II 1640 \AA\; line. This is the [O I] multiplet at 1640 \AA, a $2\rm{p}^4\;^3\rm{P}\rightarrow3\rm{s}^3\;\rm{S}^0$ forbidden transition that is pumped by the $2\rm{p}^4\;^1\rm{D}\rightarrow3\rm{s}\;^3\rm{S}^0$ forbidden transition from the upper energy level of the $\rm{\lambda}1302$ \AA\; line and corresponds to the upper state of [O I] 6300 \AA\; (Shore et. al., \citeyear{shore1993interpretation}, \citeyear{vcyg1}). The appearance of this line is connected with high column densities in the neutral wind: O I 1302 \AA\; emission from the ejecta pumps the O I] 1641 \AA\, [O I] 6300, 6364, and 5577\AA\  lines. The 6300 \AA\; doublet is usually observed in symbiotic systems (but the 1641\AA\ line is only observable in high-resolution UV spectra, hence the currently available dataset is limited), and is a useful diagnostic for either cold neutral medium and lower layers of the chromosphere in the  giant. The 6300\AA\ blend initially strengthens until about day 40.  Its intensity gradually decreases, at a progressively faster rate as the ejecta approach break-out. 

As noted earlier, all  the profile sequences show distinct components.  One is broad, due to the expanding ejecta and the shock and of order hundreds to a few thousand km~s$^{-1}$, and a narrow profile from the unshocked wind. These are particularly evident during the outbursts of V745 Sco: for example, the narrow component of [O III] 4363 \AA\; has $\rm{FWZI} = 78, 126, 64 \;\rm{km\;s}^{-1}$ on $\rm{t} = 44, 91, 449\;\rm{days}$ while for the wider one $\rm{FWZI} = 1018, 1290, 188 \;\rm{km\;s}^{-1}$ on $\rm{t} = 44, 91, 449\;\rm{days}$ respectively. In the same sequence, for [Ne III] 3869 \AA\  the two components have $\rm{FWZI} = 104, 141, 89 \;\rm{km\,s}^{-1}$ and $\rm{FWZI} = 708, 1057, 316 \;\rm{km\,s}^{-1}$ on $\rm{t} = 44, 91, 449\;\rm{days}$ respectively and, at $\rm{t} = 449\;\rm{days}$. The N III profiles, instead, display a composite structure similar to the one detected on top of H lines in NOT and SMARTS spectra and are interpreted as the extension of the precursor in the wind. From SAAO data, in [O III] 5007 \AA\; profiles $\rm{FWZI(narrow)} \simeq 1390, 610\;\rm{km\,s}^{-1}$ and $\rm{FWZI(broad)} \simeq 8100, 2600\;\rm{km\,s}^{-1}$ at $\rm{t} = 3, 86\;\rm{days}$. As shown in \citet{vcyg2}, the feature appearing in the V407 Cyg sequence is a signature of the presence of the strong wind emitted by the Mira variable. This almost completely suppresses [O III] $\rm{\lambda}5007\;$ \AA\; emission after about two months. It is not an absorption component, but the shadow zone of the wind. A trace is the anticorrelation of [O I] and [O III] in profile matching and, as a consequence, a correspondence with the Balmer lines is expected with some difference in the profile and the intensity contrasts.

\subsection{Coronal lines}
\label{subsection:coronal}
These transitions are  collisionally excited atomic transitions predominantly from highly ionised  atoms of the iron group. Forbidden lines, such as [Fe XI], [Fe XII], [Fe XIV], [Mg V], [Ar X], appear in post-maximum light spectra of SyRN outbursts.  They are strongest in the earliest stages and arise from two separate regions, the precursor and ionised wind (where $\rm{T_e} \sim 10^4$ K) and the post-shocked ejecta and wind ($\rm{T_e} > 10^6$ K); the profile contribution   from the wind is narrow and that of the post-shock gas is broad, e.g., a few thousand km s$^{-1}$ for the permitted lines. The particular conditions required to produce the coronal emissions occur in the earliest post-maximum stages, typically after the start of the narrow-line phase   when the shock temperature has dropped the initial value $\rm{T}\simeq10^{7}\;\rm{K}$ to $\sim 1.2\times10^{6}\;\rm{K}$. For example, these lines appeared in RS Oph 2006 at 20-30 days, then  began a rapid decline until about day 120  and completely disappeared at $\rm{t}\simeq200\;\rm{days}$, when $\rm{T_{e}}\leq6\times10^{5}$. This signalled the weakening of the shock and is separate from the break-out phenomenon that drives the light curve and permitted lines (Gorbatskij, \citeyear{gorbatskii1972formation} and \citeyear{gorbatskii1973formation}). Throughout the spectral evolution, both the intensity and width of coronal lines decreased over time due to the deceleration of the ejecta.

\begin{figure}[h]
\centering
\includegraphics[width=0.5\textwidth]{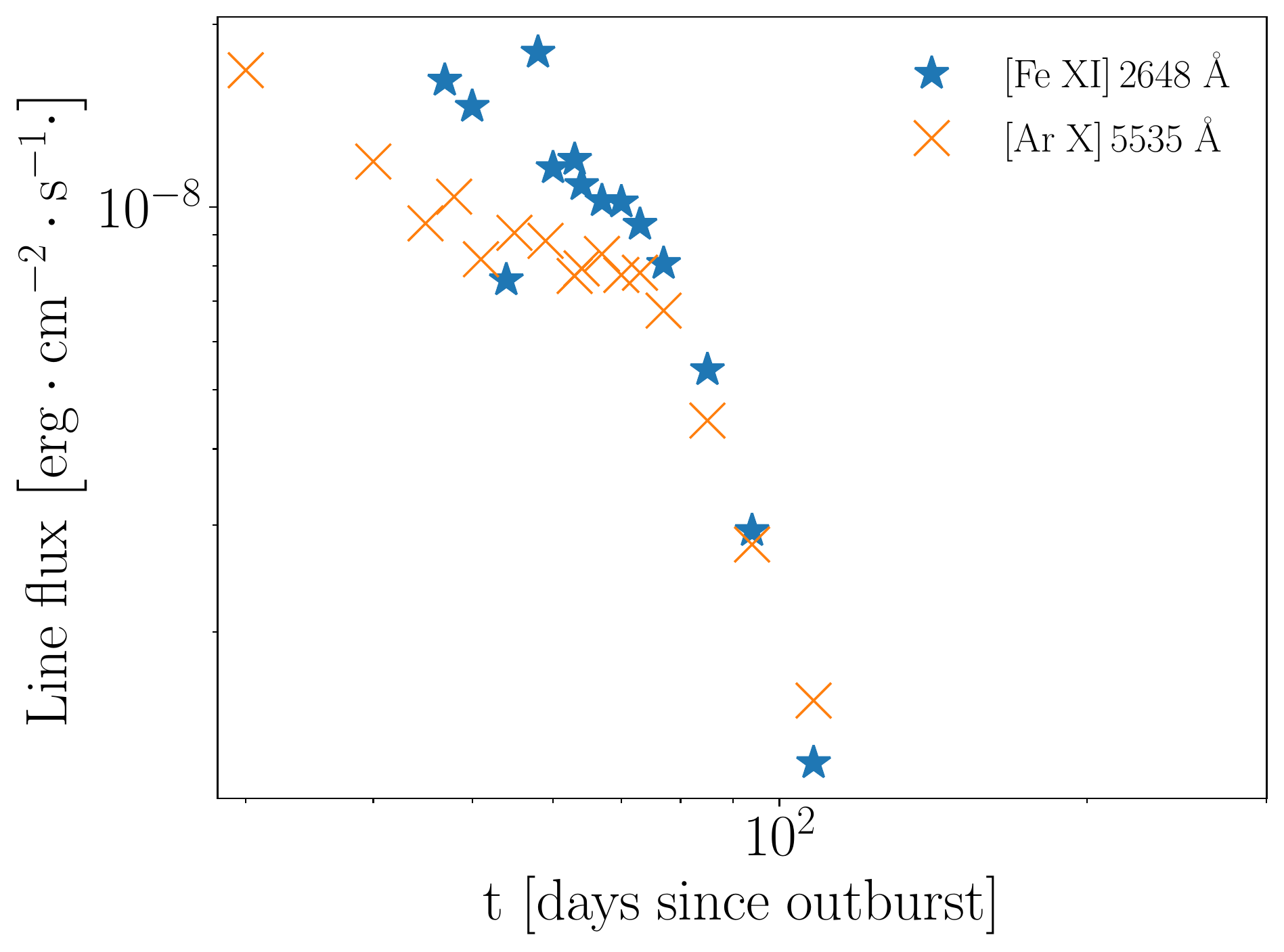}
\caption{\label{fig:coronal_rsoph2006swift} Coronal species in RS Oph 2006: integrated line strength as a function of days since the optical maximum in Swift spectra. Data are binned and averaged over intervals of 3 days.}
\end{figure}

\begin{figure}[h]
\centering
\includegraphics[width=0.5\textwidth]{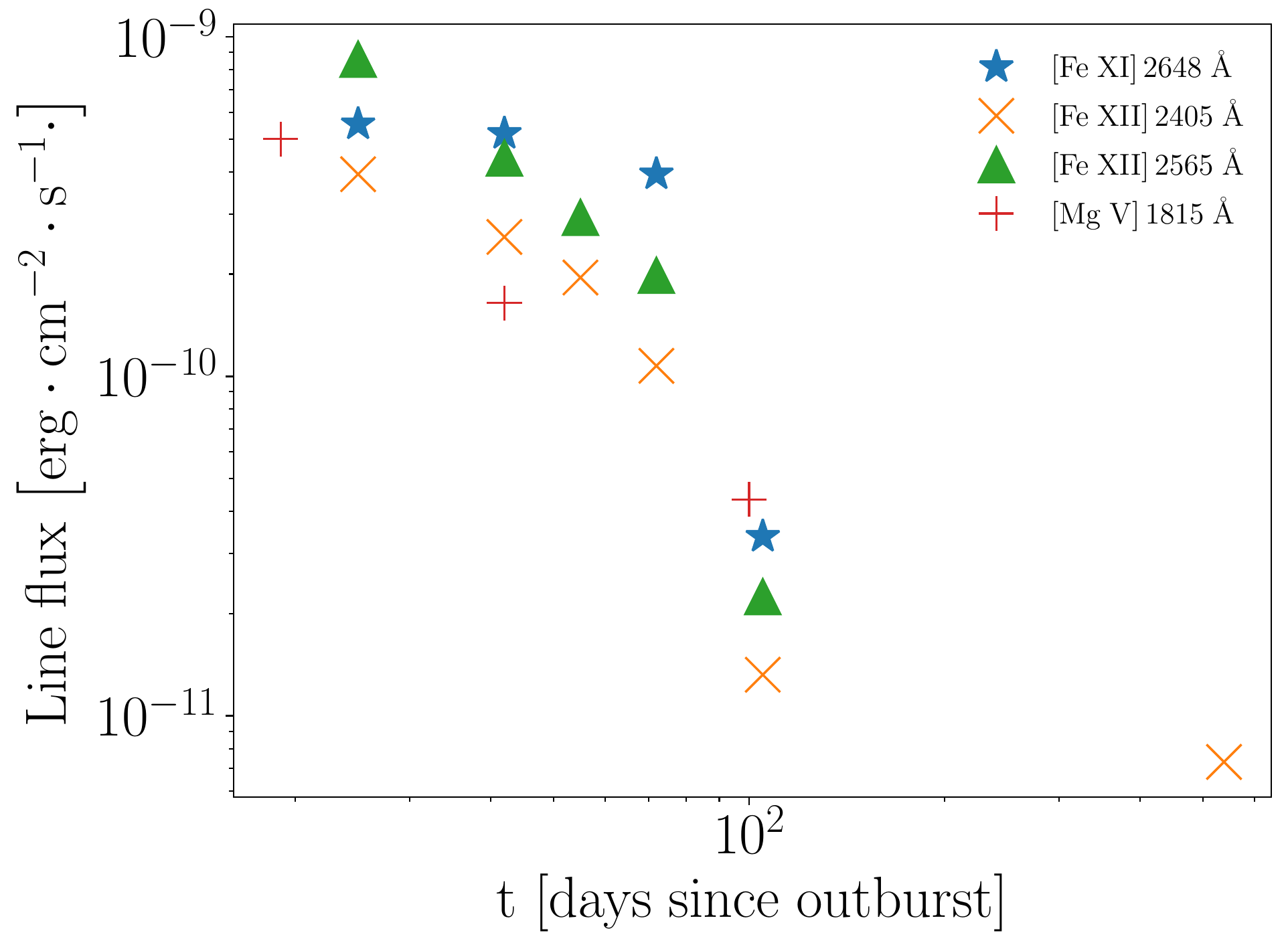}
\caption{\label{fig:coronal_rsoph1985iuelow} Coronal species in RS Oph 1985: integrated line strength as a function of days since the optical maximum in IUE low-resolution spectra. Data are binned and averaged over intervals of 3 days.}
\end{figure}

\begin{figure}[h]
\centering
\includegraphics[width=0.5\textwidth]{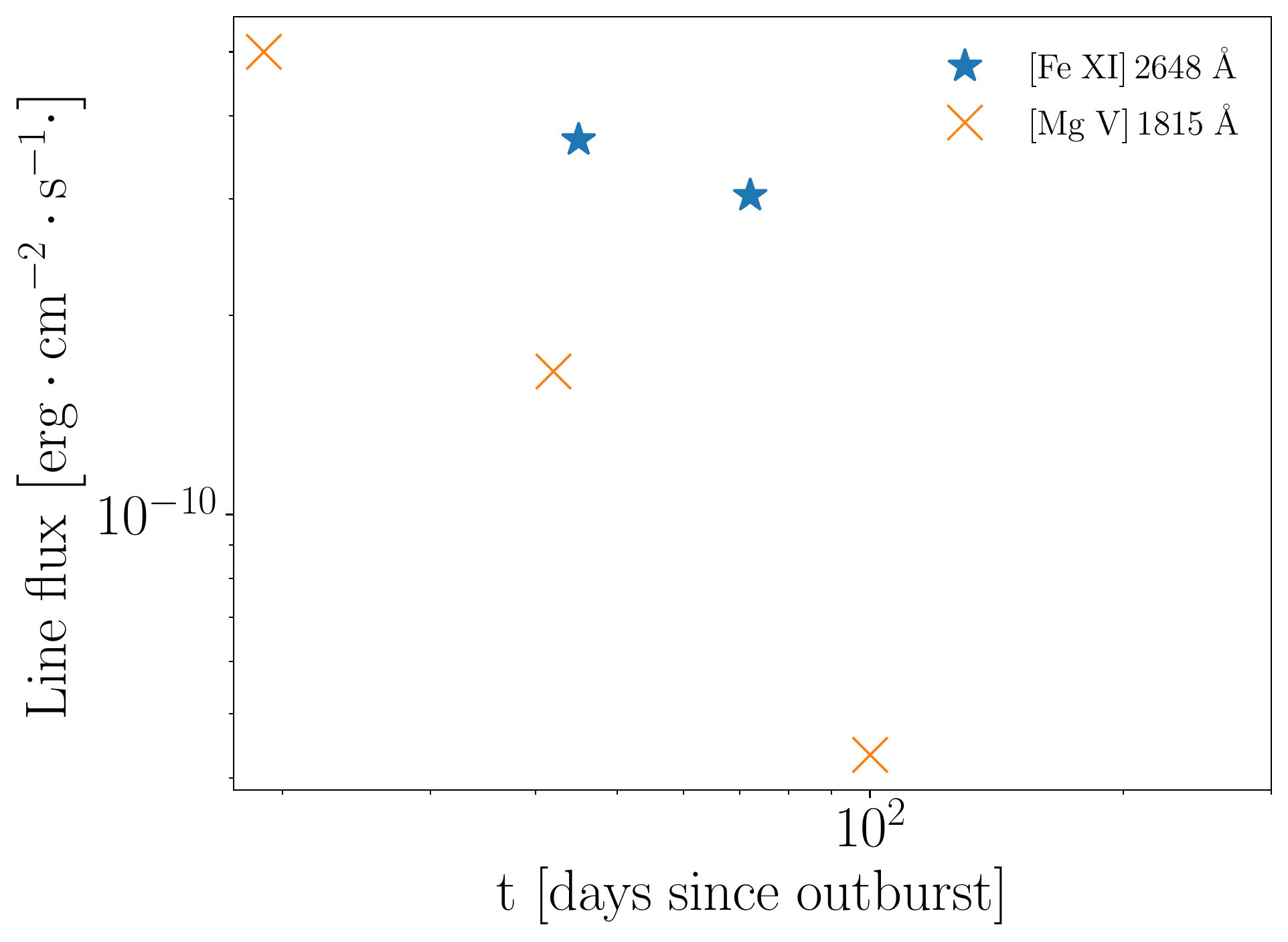}
\caption{\label{fig:coronal_rsoph1985iuehi} Coronal species in RS Oph 1985: integrated line strength as a function of days since the optical maximum in IUE high-resolution spectra. Data are binned and averaged over intervals of 3 days)}
\end{figure}

\begin{figure}[h]
\centering
\includegraphics[width=0.5\textwidth]{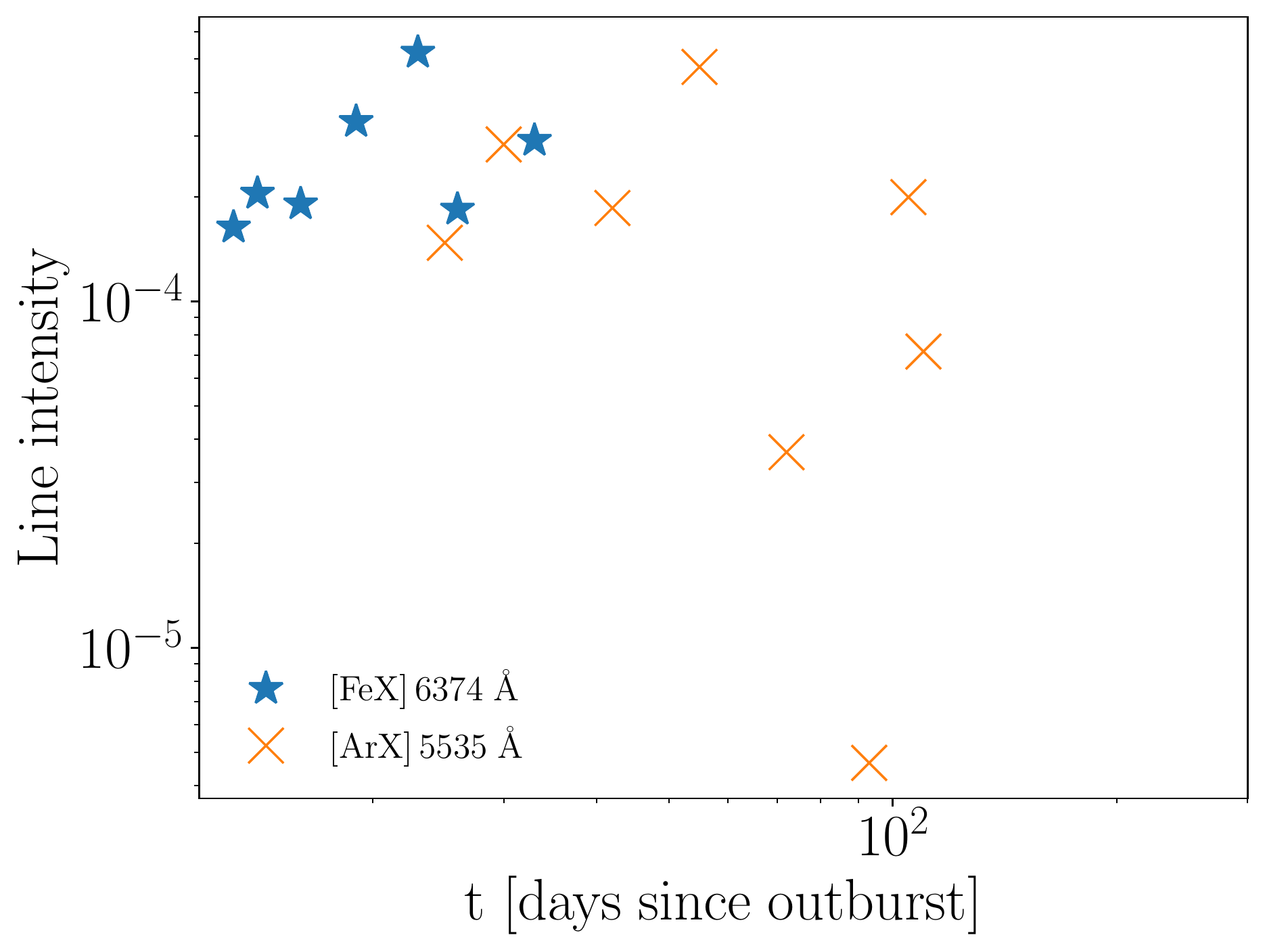}
\caption{\label{fig:coronal_v3890sgr2019aras} Coronal species in V3890 Sgr 2019: integrated line strength as a function of days since the optical maximum in ARAS spectra. Data are binned and averaged over intervals of 3 days.}
\end{figure}

\begin{figure}[h]
\centering
\includegraphics[width=0.5\textwidth]{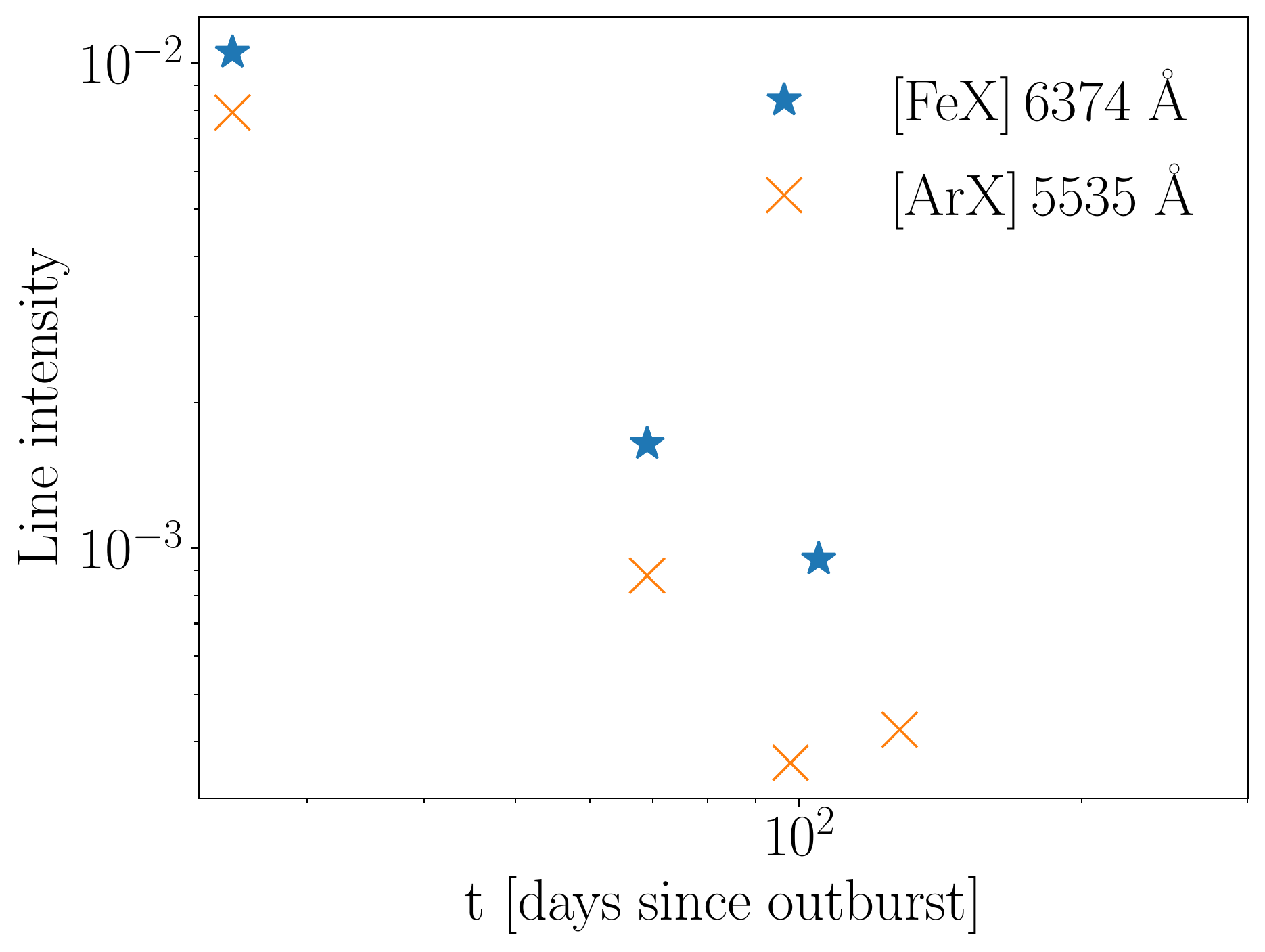}
\caption{\label{fig:coronal_v407cyg2010not} Coronal species in V407 Cyg 2010: integrated line strength as a function of days since the optical maximum in NOT spectra. Data are binned and averaged over intervals of 3 days.}
\end{figure}

Only a few coronal lines are clearly visible and fairly intense in the available datasets, in most sequences they are either blended or too weak  to be measured. The most prominent are [Fe XI] $\rm{\lambda}2648\;$\AA\; and [Ar X] $\rm{\lambda}5535\;$, which show up in almost all optical spectra with occasional occurrence of [Fe XIV] $\rm{\lambda}5303\;$\AA\; and [Fe X] $\rm{\lambda}6374\;$\AA, while in UV the most frequently observed are [Mg V] $\rm{\lambda}1815\;$\AA\; and [Fe XII] $\rm{\lambda}2405\;$\AA. The identification of other lines is uncertain. A blend of [Fe VII] and [Ca V] 6087 \AA\; is probably present in some Swift spectra of RS Oph, but the profiles are unresolved and the identification is uncertain. Figs.\ref{fig:coronal_rsoph2006swift}-\ref{fig:coronal_v407cyg2010not} and Table \ref{table:coronal} display the evolution of emission line intensity over time for several systems.

\begin{table}
\caption{Coronal lines in V745 Sco, V3890 Sgr and V407 Cyg spectra. CTIO and Ondřejov data are binned and averaged over intervals of 3 days, SMARTS and SAAO of 2. Units are $\rm{erg\;cm^{-2}\;s^{-1}}$ for CTIO, SMARTS, SAAO and IUE, counts for Ondřejov.}             % title of Table
\label{table:coronal}      % is used to refer this table in the text
\centering                          % used for centering table
\resizebox{\columnwidth}{!}{\begin{tabular}{ccccc}        % centered columns (4 columns)
\hline\hline                 % inserts double horizontal lines
Event & Dataset & Line & t $\left[ \rm{days} \right]$ & Line strength \\     % table heading 
\hline                        % inserts single horizontal line
V745 Sco 1989 & CTIO & [Fe X] 6374 \r{A} & 5 & $9.71\times10^{-11}$ \\
 & & & 9 & $3.93\times10^{-11}$ \\
 & & & 22 & $4.37\times10^{-12}$ \\
 & & [Ar X] 5535 \r{A} & 5 & $8.42\times10^{-11}$ \\
 & & & 9 & $2.51\times10^{-11}$ \\
 & & & 22 & $2.70\times10^{-12}$ \\
 & & & 317 & $1.77\times10^{-13}$ \\
 V745 Sco 2014 & SMARTS & [Fe X] 6374 \r{A} & 3.5 & $1.11\times10^{-11}$ \\
 & & & 6 & $1.12\times10^{-11}$ \\
 & & [Fe XI] 7892 \r{A} & 3.5 & $6.39\times10^{-12}$ \\
 & & & 6 & $6.29\times10^{-12}$ \\
 & & [Fe XIV] 5303 \r{A} & 3.5 & $7.24\times10^{-12}$ \\
 & & & 6 & $7.32\times10^{-12}$ \\
 & & [Ar X] 5535 \r{A} & 3.5 & $3.65\times10^{-12}$ \\
 & & & 6 & $3.53\times10^{-12}$ \\
 V745 Sco 2014 & SAAO & [Fe X] 6374 \r{A} & 4.5 & $1.09\times10^{-10}$ \\
 & & & 6 & $2.99\times10^{-12}$ \\
 & & & 86 & $2.09\times10^{-12}$ \\
 & & [Fe XIV] 5303 \r{A} & 86 & $9.24\times10^{-13}$ \\
 & & [Ar X] 5535 \r{A} & 3 & $8.74\times10^{-11}$ \\
V3890 Sgr 1990 & IUE low res. & [Fe XI] 2648 \r{A} & 19 & $1.14\times10^{-10}$ \\
 & & & 27 & $4.53\times10^{-11}$ \\
\hline                                   %inserts single line
\end{tabular}}
\end{table}

As in RS Oph in 1985 (\citet{shore1996}), the coronal lines follow the XR light curves, appearing and peaking between 51-130 days. A similar development is displayed by V407 Cyg (\citet{vcyg1}, \citet{vcyg2}): at first the lines strength increases, then declines after the maximum XR luminosity. After the peak in XRs, the coronal lines continue weakening, while their profiles become progressively more asymmetric, similar to the He II lines. The change in emissivity shows how sensitive the coronal lines are to the ionisation state of the gas: a drop in the shock powering leads to rapid recombination for such high charge states, on timescales of days. The rapid decline following the break-out is expected from the short recombination times for these high ionisation states rather than cooling.  These lines also appear in the spectra from CNe, where they reach a freeze-out as the recombination rate becomes throttled by the density decrease due to the expansion rate of the ejecta (\citet{1998AAS...19310105V}).  In SyRNe, two competing effects rapidly neutralise the wind. The ejecta are not in free expansion, and the decreased shock photoionisation input is competing with the advective transport of ionised gas from the wind. Turning the source off, neutral gas issuing from the RG refills the ionised cavity and further decreases the coronal line emission. Turn off of the SSS part reduces the photo-ionisation.  Different timescales characterise the evolution of these lines in various systems: emission arises and then declines from about 20 days in RS Ophiuchi, V3890 Sagittarii and V407 Cygni, whereas V745 Sco shows a faster evolution, with coronal lines appearing and weakening even around 5 days after the beginning of the outburst.

As expected for forbidden transitions, the profiles are systematically narrower than for permitted or intercombination lines, but they are almost always blended with other lines and it is difficult to accurately determine their  width. Another reason for the narrow appearance of coronal lines could be the combined effects of enhanced ionisation produced by the expanding Str{\"o}mgren sphere in the outer part of the red giant wind and continuous ionisation by radiation emitted by the wind-shocked ejecta, as already argued in \citet{shore1996} for 1985 IUE spectra of RS Oph. A comparison is not possible with Swift spectra because the grism resolution is no better than $\rm{\Delta v} \simeq 500\;\rm{km\;s^{-1}}$ and we cannot resolve the features that the high-resolution IUE spectra displayed.

\section{Results and conclusions}
\label{section:conclusion}

The time-dependent evolution of dynamical structures was investigated by searching for certain morphological and physical elements. Clues for the presence of shock fronts are, for example, optical spectra showing multiple velocity components, optical and UV emission significantly enhanced by the absorption of the shock in the dense surrounding, light curves evolving with an early rapid rise towards the peak and subsequent spikes at progressively lower wavelengths. For the profiles, the blue side of an individual line is altered by the shock penetrating in the wind, while the red wing comes from the more rarefied periphery (\citet{vcyg1}). Thin overlapping profiles are due to emission and absorption in the wind of the RG ahead of the forward shock. Furthermore, shocks will also affect chemical species responsible for the lines, in fact instabilities of this kind are usually paired with the propagation of ionisation fronts. When the interaction of the shock with the environment dominates the scenario, it will be likely to see more highly ionised species; as the strength of the front diminishes, larger amounts of lines from neutral elements appear in the spectrum. In light of the development of shock waves in the system, the evolution of spectral line profiles can be understood as a change in the nature of the front itself: the thermonuclear runaway produces an early radiative wave that is largely absorbed by the ejecta, then outer layers of the WD envelope are accelerated at high velocities creating a dense slab; as this layer progressively cools, the shock gradually turns into an adiabatic front at later stages. Over time, lines change from asymmetric profiles with extended bluer wings to more symmetric shapes in later phases; the main repercussion of the shock on the shape of the profiles is seen in the vast and extended wings, much as the wind and the atmosphere of the red giant affect the large component of the peak and the P Cyg absorption. 

\subsection{Density determination}
\label{subsection:density}

The ratio of the Si III] and C III] emission line fluxes was used to estimate the electron density of the medium and its variation over time. The theoretical approach adopted for the evaluation is the same as in \citet{shore1996}, \citet{Nussbaumer1987}, \citet{1986A&A...155..205N}, \citet{1979ApJ...229.1163C}, \citet{1979A&A....75L..17N}, \citet{keenan1987si}, \citet{tapestry}, \citet{osterbrock}. More details about the calculation and the corresponding results can be found in Appendix \ref{section:appendix3}.

\begin{figure}[h]
\centering
\includegraphics[width=0.5\textwidth]{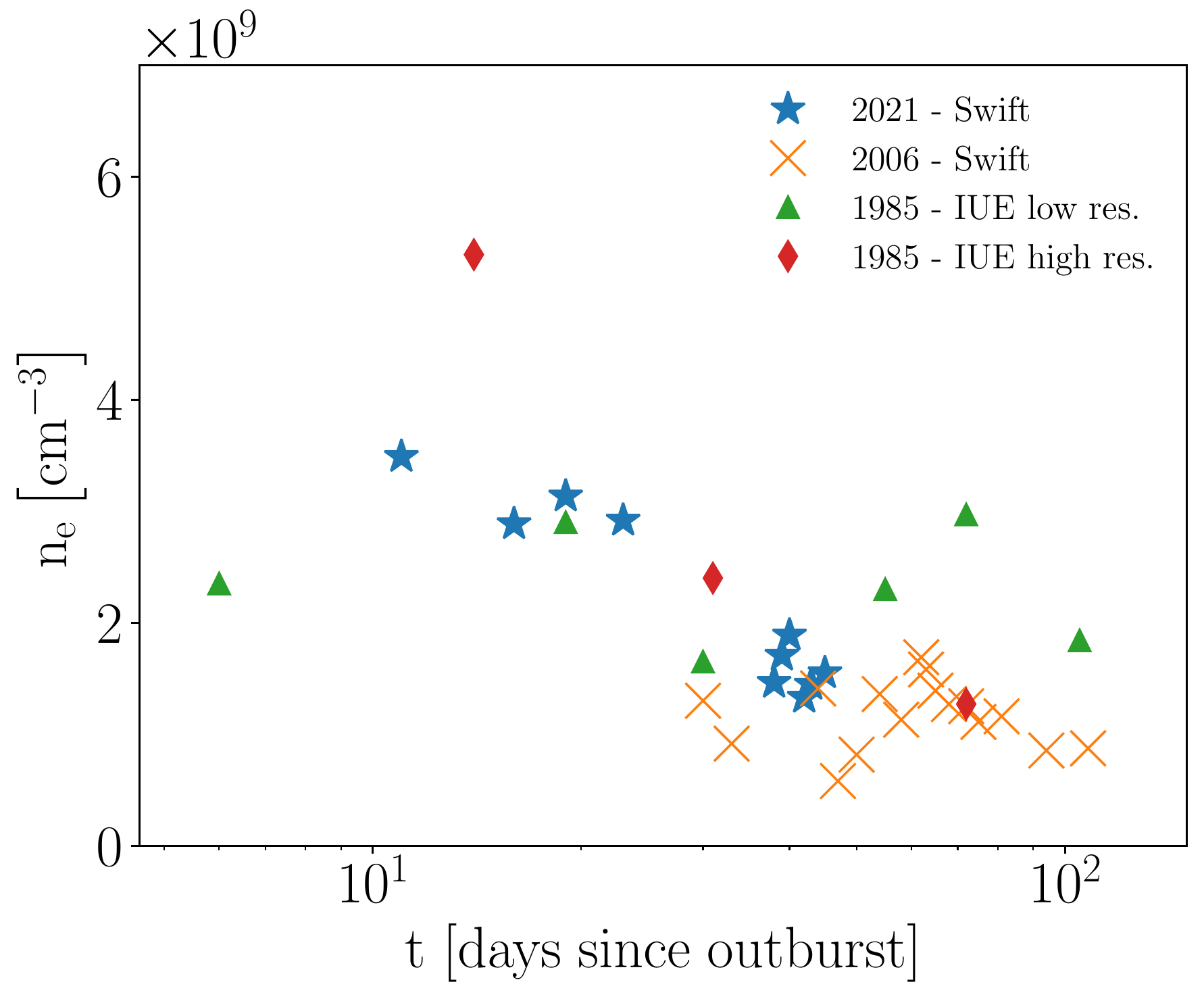}
\caption{\label{fig:Ne_rsoph} Electron density in RS Oph 2021, 2006 and 1985: temporal evolution of $n_{e}$ determined from the ratio of Si III] and C III] emission line fluxes. Data are binned and averaged over intervals of 3 days.}
\end{figure}

The electron density time variation is displayed in Fig. \ref{fig:Ne_rsoph} and Table \ref{table:electrondensity_v745sco} for the three sources. Values from all the eruptions of RS Oph and the 2019 V3890 Sgr outburst are averaged on bins of 3 days. The mean value of $\rm{n_e}$ is $\sim 5$ orders of magnitude lower in later than earlier stages.
  
\begin{table}
\caption{Electron density estimations. Data are binned and averaged over intervals of 2 days for V745 Sco and 3 for V3890 Sgr.}             % title of Table
\label{table:electrondensity_v745sco}      % is used to refer this table in the text
\centering                          % used for centering table
\begin{tabular}{cccc}        % centered columns (4 columns)
\hline\hline                 % inserts double horizontal lines
Event & Dataset & t $\left[ \rm{days} \right]$ & $\rm{n_{e}} \left[ \times10^{9}\;\rm{cm^{-3}} \right]$ \\
\hline                        % inserts single horizontal line
V745 Sco 1989 & IUE low res. & 4.5 & 2.42 \\
 & & 7 & 2.71 \\
 & & 14 & 3.63 \\
V3890 Sgr 1990 & IUE low res. & 19 & 1.74 \\
 & & 27 & 1.63 \\
V3890 Sgr 2019 & Swift & 5.75 & 2.97 \\
 & & 8 & 0.96 \\
 & & 21.5 & 3.21 \\
\hline                                   %inserts single line
\end{tabular}
\end{table}

In the Swift spectra, $n_e$ first declines until 40 days for RS Oph and about 10 days in V3890 Sgr, then rises for the next 10 days and decreases again at later stages. This is consistent with the results of previous analysis, of the 1985 outburst of RS Oph in particular. The trend of electron density is the same noticed by \citet{cassatella1985ultraviolet} and \citet{shore1993interpretation}, \citet{shore1996}, \citet{1990MNRAS.246...78S}, \citet{gonzalez19921990} for the different systems. Measured values are also in fairly good agreement with the calculations of \citet{anupama3} and \citet{1994MNRAS.269..105A}, using the Balmer and He I lines instead.

IUE and Swift data of RS Oph show consistent values at similar stages of the two events. In 2021, $\rm{N_{e}}\sim 3.5 \times 10^{9}\; \rm{cm^{-3}}$ at 11 days, the maximum is $\rm{N_{e}}\sim 3.9 \times 10^{9}\; \rm{cm}^{-3}$ at 12 days and the minimum $N_{e}\sim 9.0 \times 10^{8}\; \rm{cm^{-3}}$ at 40 days. In 2006, in the first detections the electron density is $\sim 1.3 \times 10^{9}\; \rm{cm^{-3}}$, then $\rm{N_{e}}$ has a maximum value of $1.69\times 10^{9}\; \rm{cm^{-3}}$ at 62 days and a minimum $\rm{N_{e}} = 4.34 \times 10^{8}\; \rm{cm^{-3}}$ at 91 days. The high-resolution spectra of the 1985 explosion peak at $t = 14\;days$ with $\rm{N_{e}} = 5.30 \times 10^{9}\; \rm{cm^{-3}}$, then the value decreases and reaches a minimum of $\rm{N_{e}} = 6.00 \times 10^{8}\; \rm{cm^{-3}}$ 42 days after the beginning, in excellent agreement with $\rm{N_{e}} = 6.59 \times 10^{8}\; \rm{cm^{-3}}$ at 41 days in 2006. In 2021 instead, at 42 days $\rm{N_{e}} \sim 1.3 \times 10^{9}\; \rm{cm^{-3}}$. This demonstrates that successive outbursts of this SyRN are amazingly similar. 

\subsection{Ionisation of the ambient medium}
\label{subsection:ionisation}

\begin{figure}[h]
\centering
\includegraphics[width=0.5\textwidth]{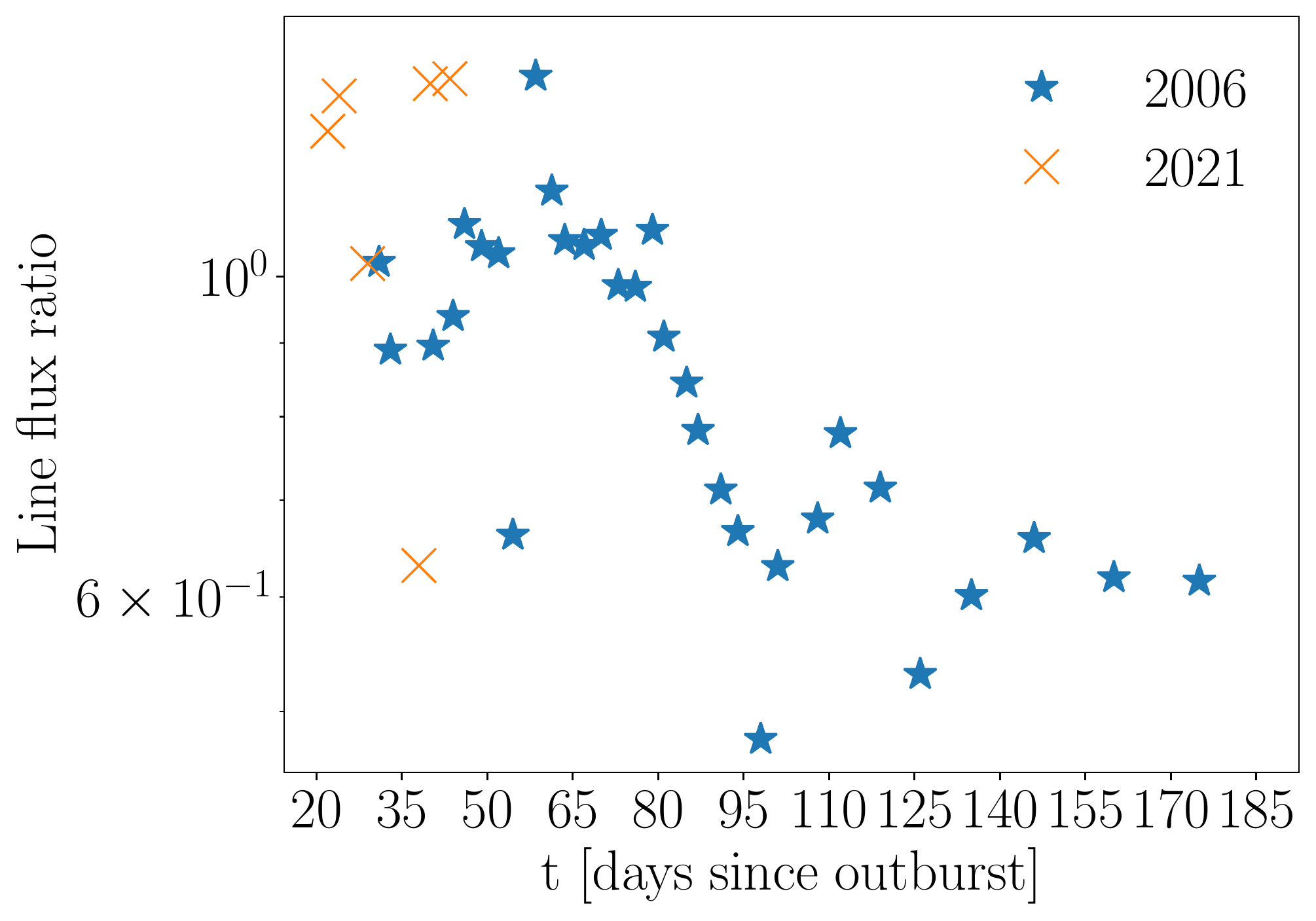}
\caption{\label{fig:NeVNeIIIratio_rsoph} [Ne V] $\lambda$3426 \AA\; to [Ne III] $\lambda$3869 \AA\; flux ratio in RS Oph 2006 and 2021: temporal evolution of the ratio between integrated fluxes of the emission lines. Data are binned and averaged over intervals of 3 days.}
\end{figure}

Neutral neon and oxygen transitions have similar ionisation potentials and occur co-spatially (\citet{10.1093/mnras/238.4.1395}, \citet{1975MNRAS.170..475S}). The shock resulting from the explosion is the main source of ionisation for the ambient medium, and strong high ionisation emission lines are produced by the combined effect of the precursor and the shock (\citet{dopita}). As stressed by \citet{10.1093/mnras/275.1.195} for the 1985 outburst of RS Oph, the ratio between these lines depends on the ions stratification downstream and the velocity of the shock front. In particular, [Ne V] and [Ne III] emission lines are useful diagnostics because their ratio depends on the \emph{ionisation parameter\/}, U, the number density of ionising photons divided by that of H atoms and dependent on the luminosity of the ionising source, the distance from the source and the hydrogen number density. It can be employed as a measurement of the variation in the ionisation structure, and $\rm{r} \equiv \frac{\rm{F}\left([\rm{Ne\; V}]\; \rm{\lambda} 3426\right)}{\rm{F}\left([\rm{Ne\; III}]\; \rm{\lambda} 3869\right)}$ changes as $\rm{U}^{2}$ (\citet{Gorjian_2007}, \citet{Abel_2008}). Unfortunately, $U$  is obtainable in only a limited set of data, since the [Ne III] and [Ne V] emission lines are simultaneously observable only in Swift spectra (which extends coverage beyond 3200 \AA). Fig.\ref{fig:NeVNeIIIratio_rsoph} and Table \ref{table:NeVNeIIIratio_v3890sgr} show the evolution of the ratio. The trend of forbidden Ne lines resembles that of [O III], which confirms that emission from these species arises from similar regions and origin. In particular, as already noted by \citet{10.1093/mnras/238.4.1395}, in Sy-RNe during outburst forbidden lines of this type are excited by fast shocks  and are often stronger than permitted transitions, such as those producing Balmer lines.

\begin{table}
\caption{V3890 Sgr 2019: ratio between integrated fluxes of [Ne V] $\lambda$3426 \AA\; and [Ne III] $\lambda$3869 \AA\; emission lines in Swift spectra. Data are binned and averaged over intervals of 3 days.}             % title of Table
\label{table:NeVNeIIIratio_v3890sgr}      % is used to refer this table in the text
\centering                          % used for centering table
\begin{tabular}{cc}        % centered columns (4 columns)
\hline\hline                 % inserts double horizontal lines
t $\left[ \rm{days} \right]$ & Ratio \\     % table heading 
\hline                        % inserts single horizontal line
5 & 0.97 \\
6 & 1.04 \\
7.5 & 1.32 \\
\hline                                   %inserts single line
\end{tabular}
\end{table}

\begin{figure}[h]
\centering
\includegraphics[width=0.5\textwidth]{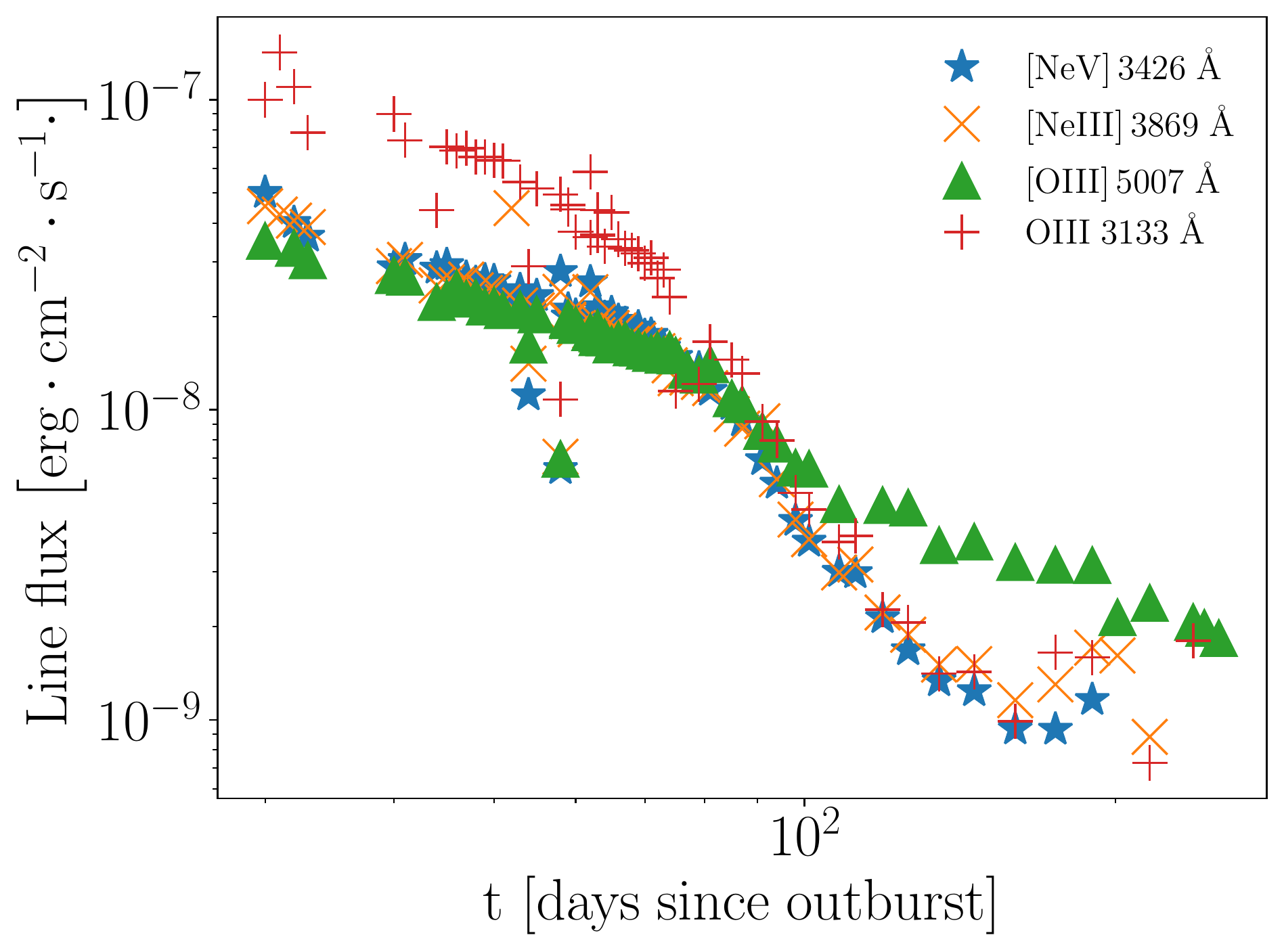}
\caption{\label{fig:NeVNeIIIOIII_rsoph2006} Neon and oxygen in RS Oph 2006: evolution over time of spectral lines in Swift spectra. Data are binned and averaged over intervals of 3 days.}
\end{figure}

\begin{figure}[h]
\centering
\includegraphics[width=0.5\textwidth]{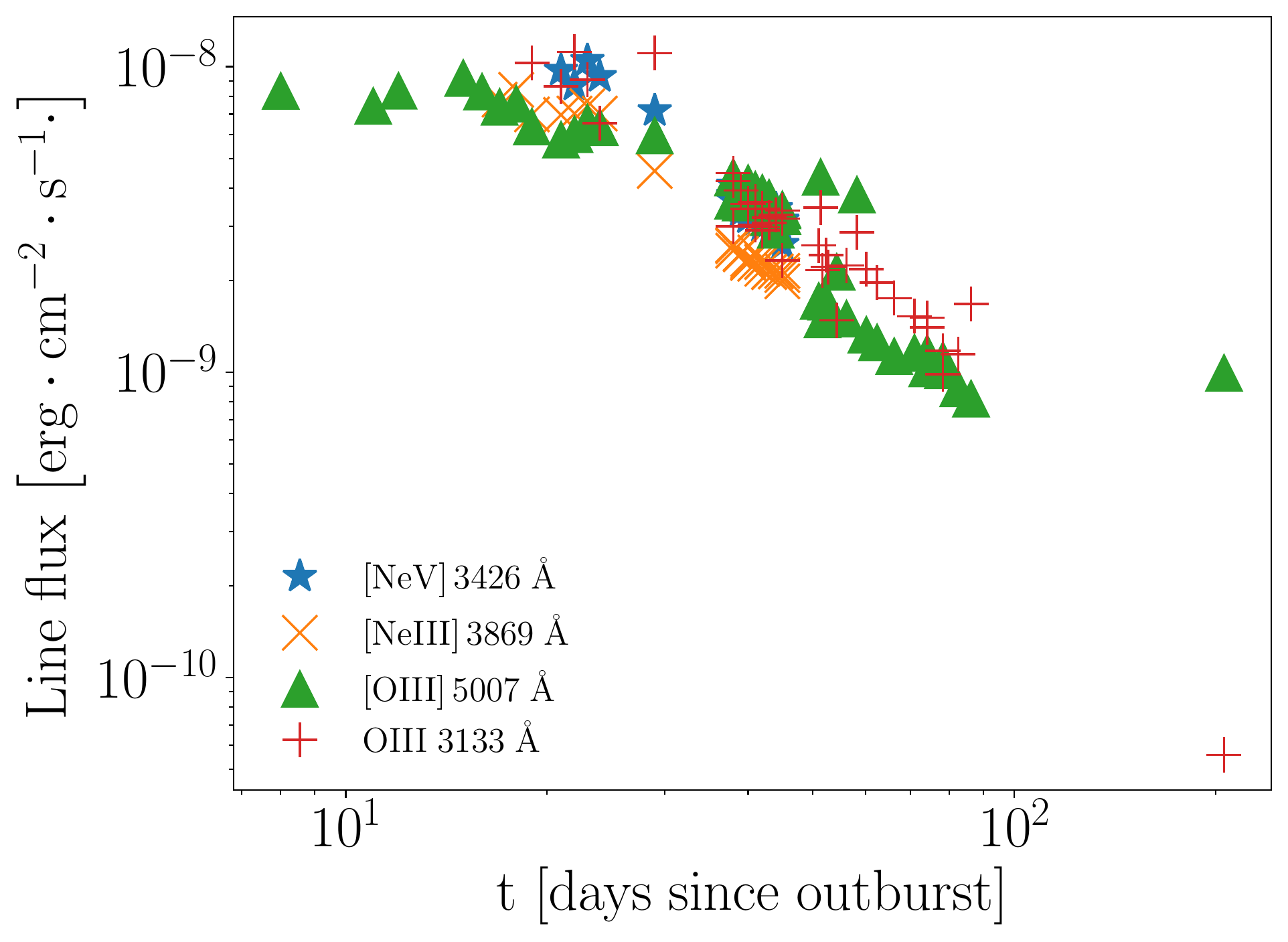}
\caption{\label{fig:NeVNeIIIOIII_rsoph2021} Neon and oxygen in RS Oph 2021: evolution over time of spectral lines in Swift spectra. Data are binned and averaged over intervals of 3 days.}
\end{figure}

The variation of the $r$ reflects how ionisation proceeds at the various epochs (\citet{10.1093/mnras/275.1.195}).  Another important diagnostic  for this evolution would be the ratio of [O III] emission lines. Unfortunately, the plasma diagnostic for  [O III] 4959 \AA\; and 4363 \AA\ is not available since the 4363 \AA\ line is always blended with H$\gamma$ in the Swift spectra. The only possible comparison between these species is the trend of the flux as shown in Figs.\ref{fig:NeVNeIIIOIII_rsoph2006} and \ref{fig:NeVNeIIIOIII_rsoph2021} for RS Oph. The Ne and O lines show  similar variations. In the early phase until break-out around 80 days, the different evolution is caused by the fact that the transition at 3133 \AA\; is permitted, whereas the others are forbidden and the O III is additionally fluorescently excited by the Lyman series. The  evident difference is the [O III] $\lambda$5007 \AA\ behaviour relative to the other lines after about 100 days. These low ionisation forbidden lines persist for far longer, driven in part by the continuing UV illumination by the remnant WD.  The other lines, instead, decline more rapidly and the corresponding curve shows a clear dip.

The time series traces the shock development. As the front propagates through the circumstellar medium, the gas is heated at high temperatures and emits high ionisation lines and a hard continuum, and therefore ionisation increases in the earlier phases. Then, as the front progressively decelerates the gas cools and  recombines and ionisation becomes less efficient. This stage is represented by the declining phase in the evolution of $r$. The ionisation turns off in the later phases, but the RG wind continues to   advect the previously ionised gas. At the same time, density is decreasing but, since the timescale for this decline is about the same as for recombination, ionisation freezes out at this point and the rate of weakening of lines decreases. As in \citet{2008ASPC..401...19S} for RS Oph 1985, the rate of decline after break-out becomes the same for all  species.

In RS Oph 2006, after about 160 days the [Ne III], [Ne V], and O III fluxes increase, similar to some Balmer and He I lines. This signals the outset of a new source for ionising photons, arguably renewed accretion from the giant wind (\citet{ourpaper}). For at least some of the UV lines, a change in the observer's line of sight through the wind could have obscured part of the emitted radiation, which is absorbed by the RG wind. In this case, comparing such spectroscopic results with X-ray photometry is useful because it can help to understand which process is more likely responsible for the observed features. The trend of XRs light curve is marked by a similar slight climb in the latest points. The resemblance between the two types of data signals that a restructured accretion disc is probably responsible for the increase in emission.

\section{Summary}
The present work focused on the five Galactic SyRNe, exploring their behaviour during outburst and quiescent stages and collecting information about shock-driven phenomena from spectroscopic observations in the optical and ultraviolet. The dynamics of the environment were analysed by looking at emission line profiles, temporal evolution and other useful diagnostics.  In particular, the Swift grism detection of the two latest explosions of RS Ophiuchi provided the unique opportunity to investigate diagnostic spectral features that were unattainable in previous surveys or compare earlier results with \emph{new\/} ones. This added interesting material to the pre-existing knowledge of both the individual object and general description. The main result from the comparison is that the SyRNe evolve in strikingly similar ways. The observed discrepancies, the reduced intensity above all, are due to the different orbital periods and orientations of the system along the line of sight to the WD. From 2021 November 4th, Swift observations of RS Oph were no longer possible due to the nearby Sun which lasted until the first days of February 2022. In addition, for some time, the spacecraft was subject to hardware issues and this prevented further observations of the system. However, the available data are enough to properly follow the evolution.  The latest outburst of this nova and concomitant observations in multiple wavelengths allowed us to deepen our insight into the dynamics behind these explosive events and the nature of the sources.

\begin{acknowledgements}

We acknowledge the use of public data from the {\it Swift} data archive and the IUE archive at MAST. Spectra taken by David Boyd and published in the ARAS archive have been used and we thank him, Francois Teyssier, Olivier Garde, Christian Buil and the ARAS group of observers for their selfless contribution to the community.  
Based on observations collected at the European Organisation for Astronomical Research in the Southern Hemisphere; 
%under ESO programme(s) PPP.C-NNNN(R).
some of the observations used in this paper were obtained with the Southern African Large Telescope (SALT); some data were obtained at Ondrejov 
%Ond\UTF{00C5}\UTF{0099}ejov 
Observatory and some  with the SMARTS  telescope which is operated as part of the SMARTS consortium.  
We also acknowledge the variable star observations from the AAVSO International Database contributed by observers worldwide and used in this research.
Based on observations made with the Nordic Optical Telescope, owned in collaboration by the University of Turku and Aarhus University, and operated jointly by Aarhus University, the University of Turku and the University of Oslo, representing Denmark, Finland and Norway, the University of Iceland and Stockholm University at the Observatorio del Roque de los Muchachos, La Palma, Spain, of the Instituto de Astrofisica de Canarias.  NPMK and KLP acknowledge UKSA support.  We thank Jordi Jos\'e, Elena Mason, and Glenn Wahlgren for discussions. Also, we thank the anonymous referee for the valuable suggestions.

\end{acknowledgements}

%\section{Bibliography}
\nocite{*}
\bibliographystyle{aa}

\begin{appendix} %First appendix

\section{Instruments and datasets}
\label{section:appendix1}

\subsection{Swift Ultraviolet/Optical Telescope}
 \label{subsection:swift}

The Ultraviolet and Optical Telescope (UVOT) (\citet{uvot}) on board of the \emph{Neil Gehrels Swift Observatory\/} (\citet{swift}) is a 30-cm diameter modified Ritchey-Chrétien telescope, which allows rapid observations of optical and UV counterparts. UVOT is also able to provide spectral data in two grisms (one optical, one UV). The UVOT UV-grism provides images with spectra of modest signal-to-noise ratio in the wavelength and magnitude ranges $1700-5200\;$\AA, $9-15\; \rm{mag}$, with a  resolution $R\approx75$. The main limit to UVOT spectroscopic observations is the second-order overlap beyond 2900\,\AA. However, offsetting the source spectrum on the detector permits order separation. In this regard, wavelength calibration of the spectra is difficult because of the non-linear grism dispersion and the uncertainty in the wavelength zero point. 
The  sensitivity is also a function of the position of the source on the detector and  the point spread function (PSF) broadens towards the red end of the UV grism range (2900 - 6700 \r{AA}). 
The reddest part of Swift grism data is most affected by sensitivity drops of the instrument. Besides the TiO bands, this is due to overlapping lines from the second and third order in the grism image.

The Swift UVOT UV grism archive contains spectra acquired during the most recent outbursts of RS Oph and the 2019 event of V3890 Sgr. The archival set consists of grism images corrected for the default detector distortion from which the spectrum of interest must be extracted. For the extraction and calibration of the spectra, we use the {\tt UVOTPY} UVOT/grism data reduction software (Kuin, \citeyear{uvotpy}).

The  RS Oph dataset from the 2006 and 2021 outbursts consists of 124 grism images.  There are  26 for V3890 Sgr 2019 corresponding to virtually every phase of the evolution of the event. The coverage for RS Oph\footnote{The Swift non-UV-optical observations started earlier.} is from around day 30 (March 14th 2006, MJD 53807.8), to quiescence one year later, May 13th 2007 (MJD 54170.1) in 2006 and from day 7 (August 15th 2021, MJD 59441.22) to day 206 (March 1st 2022, MJD 59639.97) in 2021. For V3890 Sgr, coverage extends from  day 5 (September 2nd 2019, MJD 58728.1) to day 21 (September 21st 2019, MJD 58747.9). 
        
\subsection{International Ultraviolet Explorer}
\label{subsection:iue}
The International Ultraviolet Explorer (IUE) satellite (Boggess et. al., \citeyear{1978Natur.275..372B} and \citeyear{1978Natur.275..377B}) provided spectrophotometry at  high ($0.1-0.3\;$\AA) and low ($6-7\;$\AA) resolution in the ultraviolet  from 1150 to 3200 \AA. The long wavelength spectrographs operated in the range $1850-3300\;$\AA, the short wavelength one between $1150-2000\;$\AA\; and each had two dispersion modes: high-resolution using an echelle and a cross-disperser grating giving roughly $0.2\;$\AA\; resolution, and low resolution that employed the cross-disperser alone with a resulting resolution around $6\;$\AA. The dataset used for this analysis was taken from the IUE archive (MAST) and contains both long and short-wavelength spectra of RS Ophiuchi, V745 Scorpii, V3890 Sagittarii and T Corona Borealis. The RS Oph and T CrB sets contain both low and high-resolution spectra (see Table \ref{table:dataset}).

The RS Oph spectra cover nearly the entire 1985 outburst, with low-resolution data obtained from day 6 (1985 Feb 2, MJD 46098.8) with T0 at 1985 Jan 27, MJD 46092) throughout the later stages and into quiescence with the latest images acquired on 1995 Sept 15, MJD 49975.7) and 1991 Mar 2, MJD 48317.5).  The high-resolution dataset includes 21 spectra taken in 59 long and short spectral intervals and covers the outburst interval from February 14th 1985 (MJD 46110.4, 18 days after time zero) to May 17th 1985 (MJD 46202.1), 3 months after the beginning. The wavelength ranges covered are $1100-2000\;$\AA\; and $1800-3300\;$\AA. The eleven V745 Sco spectra cover the 1989 outburst from day 4 (1989 Aug 3, MJD 47741.3) to day 14 (1989 Aug 19, MJD 47757.7) at low resolution with both small and large apertures. V3890 Sgr was observed in outburst between 1990 Apr 25-27 and IUE observations of this system cover day 19 (1990 May 15, MJD 48026.2), to day 143 (1990 Sep 16, MJD 48150.8) at low resolution in both wavelength intervals. The data for T CrB have been used only for comparisons between the line profiles during outburst and quiescence, however, they cover long evolution periods of this currently "quiet" nova with spectra at both low and high resolution.

\subsection{European Southern Observatory}
\label{subsection:eso}
The ESO dataset is from  Ultraviolet and Visual Echelle Spectrograph (UVES) (\citet{2000SPIE.4008..534D}), a high-resolution optical spectrograph operating with high efficiency from 300 to 1100 nm. The instrument splits the incoming light in two arms - UV ($\rm{\lambda\lambda} 3000-5000\;$\AA, the Blue arm), and Visual ($\rm{\lambda\lambda} 4200-11000\;$\AA, the Red arm). The resolving power is about 40,000, reaching a maximum two-pixel resolution of 80,000 in the Blue arm and 110,000 in the Red.

The archival data for the 2014 outburst of V745 Sco comprise a set of 25 flux-calibrated high-resolution ($\rm{R} > 40000$) spectra obtained during the 2014 outburst of V745 Sco from 2014 March 22 (MJD 56738.3) throughout the explosive stages to quiescence on 2015 May 1  (MJD 57143.4) in two  wavelength ranges ($3600-5000\;$\AA\; and $5500-9500\;$\AA).

\subsection{South African Astronomical Observatory}
\label{subsection:saao} 
The dispersion and the resolving power of the Spectrograph Upgrade: Newly Improved Cassegrain (SpUpNIC) (\citet{crause2016spupnic}) on the 1.9 m Radcliffe telescope of the South African Astronomical Observatory (SAAO) vary depending on the grating, but in general this instrument has a dispersion between $30-210\;$\AA\;$\rm{mm^{-1}}$ and a resolution of $0.5-5\;$\AA. Instead, the resolving power goes from 700 for the grating G7 to 6500 for G5 (\citet{crause2016spupnic}).   V745 Sco spectra were taken with G7 during its 2014 outburst from 2014 February 8  (MJD 56696.1, 2 days after the beginning of the event) to May 3  (MJD 56780.9). The data are all flux calibrated and taken in two wavelength regions, $3500-7000\;$\AA\; and $6000-6800\;$\AA. 

\subsection{Cerro Tololo Inter-American Observatory}
\label{subsection:ctio} 
The CHIRON \citep{chiron} spectrograph (\citet{2004AAS...20516501B}) covers a spectral range $4100-8700\;$\AA\ wih resolution of $\rm{R} = 80,000$ for the image slicer in normal or iodine mode, $\rm{R} = 25,000$ in fibre mode, $\rm{R} = 140,000$ when the narrow slit is used for bright stars (\citet{chiron}). It was used to observe the V745 Sco early evolution from February 9  to February 13  (MJD 56697.4 - 56701.4), from 3 to 12 days after the explosion. The previous outburst of this binary system was followed (\citet{williams}) at low resolution, from 1989 August 1 (MJD 47740.0) to quiescence on 1992 April 4 (MJD 48716.2).

\subsection{Nordic Optical Telescope}
\label{subsection:not} 
The high-resolution FIbre-fed Echelle Spectrograph (FIES) \citep{2014AN....335...41T} mounted on the Nordic Optical Telescope (NOT) \citep{1985VA.....28..561A} has maximum spectral resolution $\rm{R}=67,000$ and covers the spectral range from 3000 \r{A} up to 9000 \r{A} without gaps. In 2010 and 2014, FIES registered the activity of the symbiotics V407 Cyg and V745 Sco. For V407 Cyg, spectral acquisition began on April 2nd (MJD 55288.2), 23 days after the start of the event, and ended on July 16  of the same year (MJD 55393.0), while V745 Sco was observed on MJD 56700.3 and MJD 56707.3. The resulting spectra are not absolutely flux calibrated in the wavelength region 3500-7500 \r{A} and can be used for comparison with other datasets, especially for line profiles (Shore et al. \citeyear{vcyg1}, \citeyear{vcyg2}).

\subsection{Ondrejov Observatory}
\label{subsection:ond} 
Observations of V407 Cyg were obtained from March 24th 2010 to April 22nd 2011 (MJD 55279.1 - 55674.0). The Ondrejov spectra taken with the SITe005 800 × 2000 chip (\citet{ondrejov}) were obtained in the vicinity of H$\rm{\alpha}$, supplemented by spectra at both bluer (4000–5000 \AA) and redder (8000–9000 \AA) intervals with a dispersion of 0.24\AA/px and a spectral coverage of around 500 \AA\ (\citet{vcyg2}).

\subsection{Astronomical Ring for Access to Spectroscopy}
\label{subsection:aras} 
The Astronomical Ring for Access to Spectroscopy (ARAS) archive is a web open-access database (\cite{teyssier2019eruptive}, \citet{aras}) containing spectra collected by volunteers. It contains several spectra from both outburst and quiescence of symbiotic-like recurrent novae in the visible wavelength range. In particular, spectra were obtained during the 2019 outburst of V3890 Sgr from August 28th to October 3rd (MJD 58723.8 - 58759.1), during the quiescence of RS Oph in various intervals between 2012 and 2020 and then during its 2021 outburst from day 1.82 (August 9th 2021, MJD 59435.82) to 76.75 (October 23th 2021, MJD 59510.75).

\section{Spectral profiles}
\label{section:appendix2}

\begin{figure}[ht!]
\centering
\includegraphics[width=0.5\textwidth]{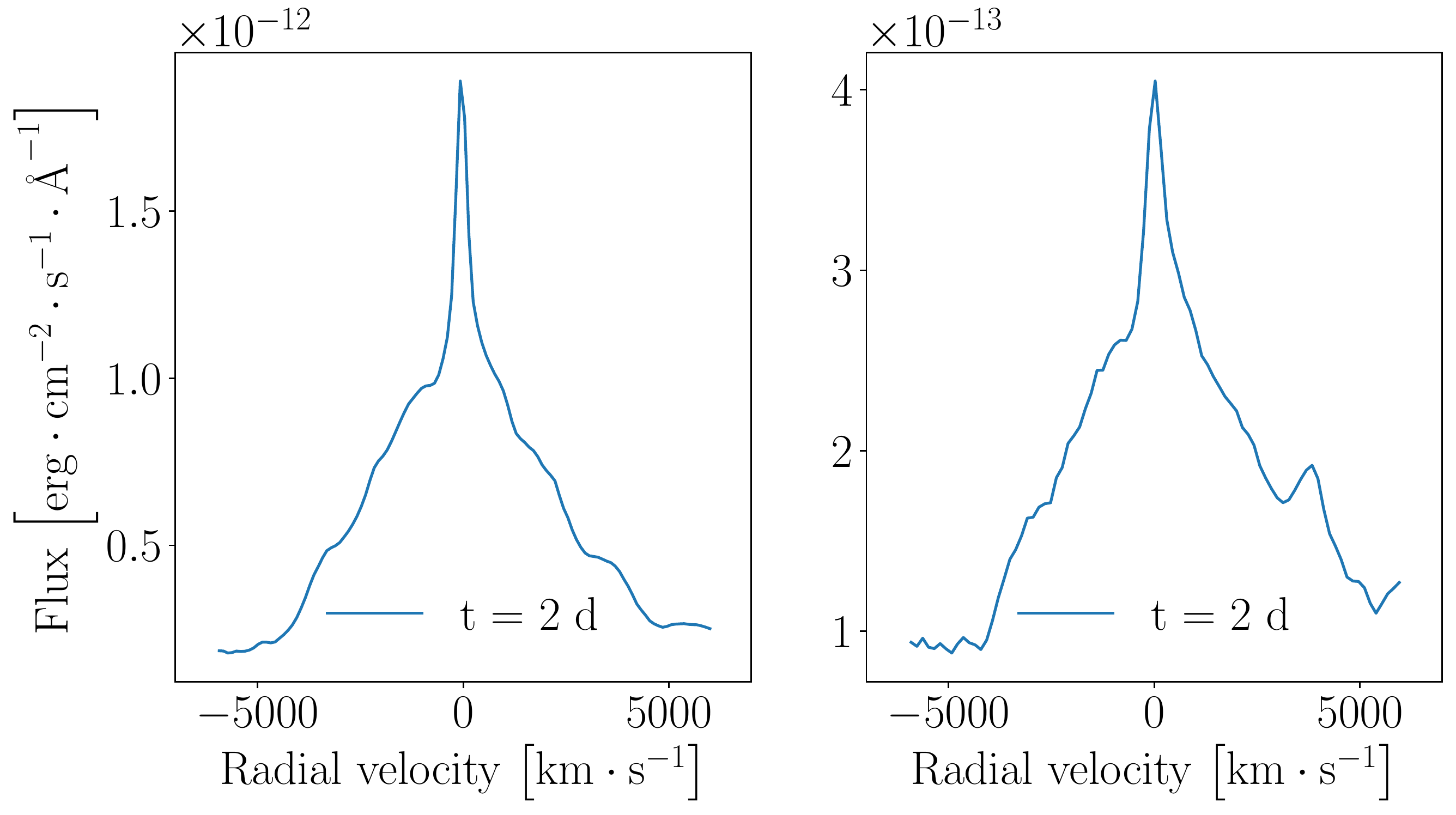}
\caption{\label{fig:v745scosaao} \footnotesize H$\rm{\alpha}$ and H$\rm{\beta}$ in V745 Sco 2014: Balmer lines in SAAO spectra on MJD 56696.1.}
\end{figure}

\begin{figure}[ht!]
\centering
\includegraphics[width=0.5\textwidth]{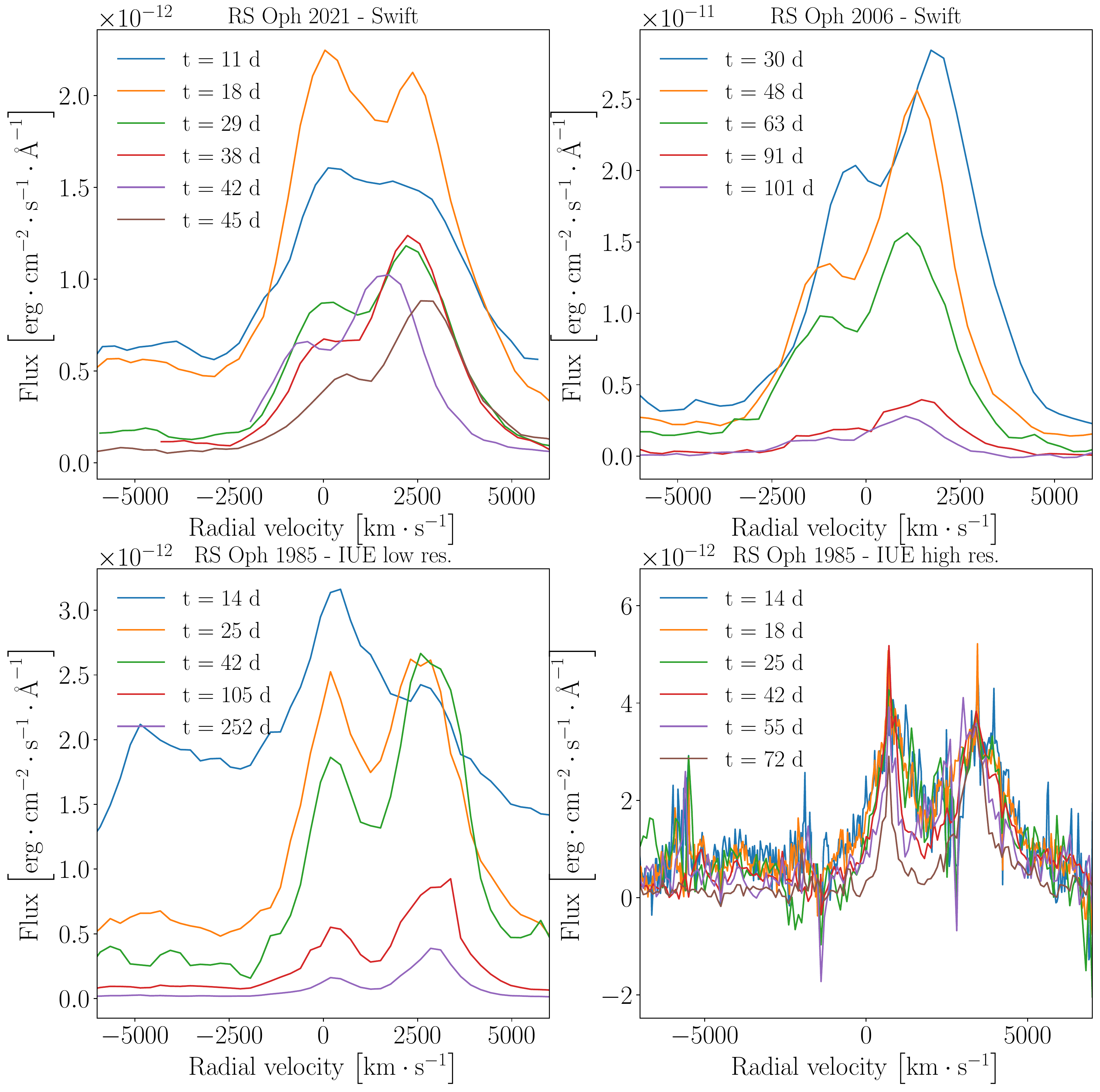}
\caption{\label{fig:SiIIICIIIprofiles} \footnotesize Si III] $1892\;\AA$ (left) and C III] $1892\;\AA$ (right) in various systems: emission profiles of the lines at different stages during the outburst. The centre is on the Si line.}
\end{figure}

This section provides examples of the different components showing up in the profiles of detected emission lines. The UV and optical line profiles show two contributions:  (1) a very broad wing due to the very fast but optically thin material that has been expelled from the white dwarf and expands later on and this dominates the shape of the lines until about 60 days after the explosion, and (2) a narrow feature, whose formation occurs because the explosion generates high-energy UV photons that last until about 100 days after the start of the explosion, beyond which ejecta-RG wind shock break-out occurs and recombination processes of the previously ionised wind become the main source of emission. Fig.\ref{fig:v745scosaao} is a representative illustration of these features: the two components are clearly visible in both lines. Fig. \ref{fig:SiIIICIIIprofiles}, instead, displays the profiles for Si and C lines for different SyRNe in outburst.

\begin{figure}[ht!]
\centering
\includegraphics[width=0.5\textwidth]{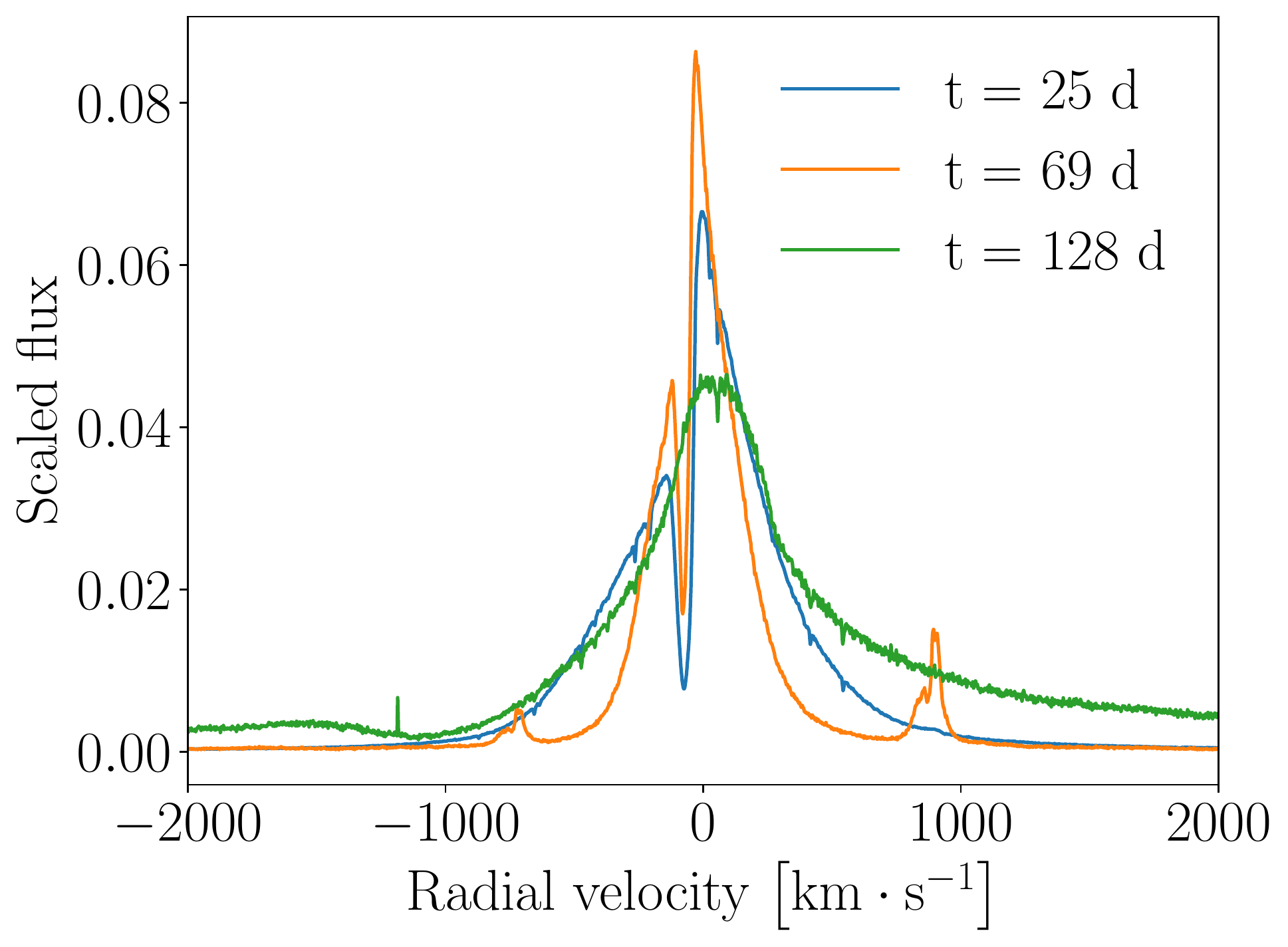}
\caption{\label{fig:Haprofile} \footnotesize H$\alpha$ $6563\;\AA$ in V407 Cyg 2010: emission profiles at different stages during the outburst in NOT spectra.}
\end{figure}

\begin{figure}[ht!]
\centering
\includegraphics[width=0.5\textwidth]{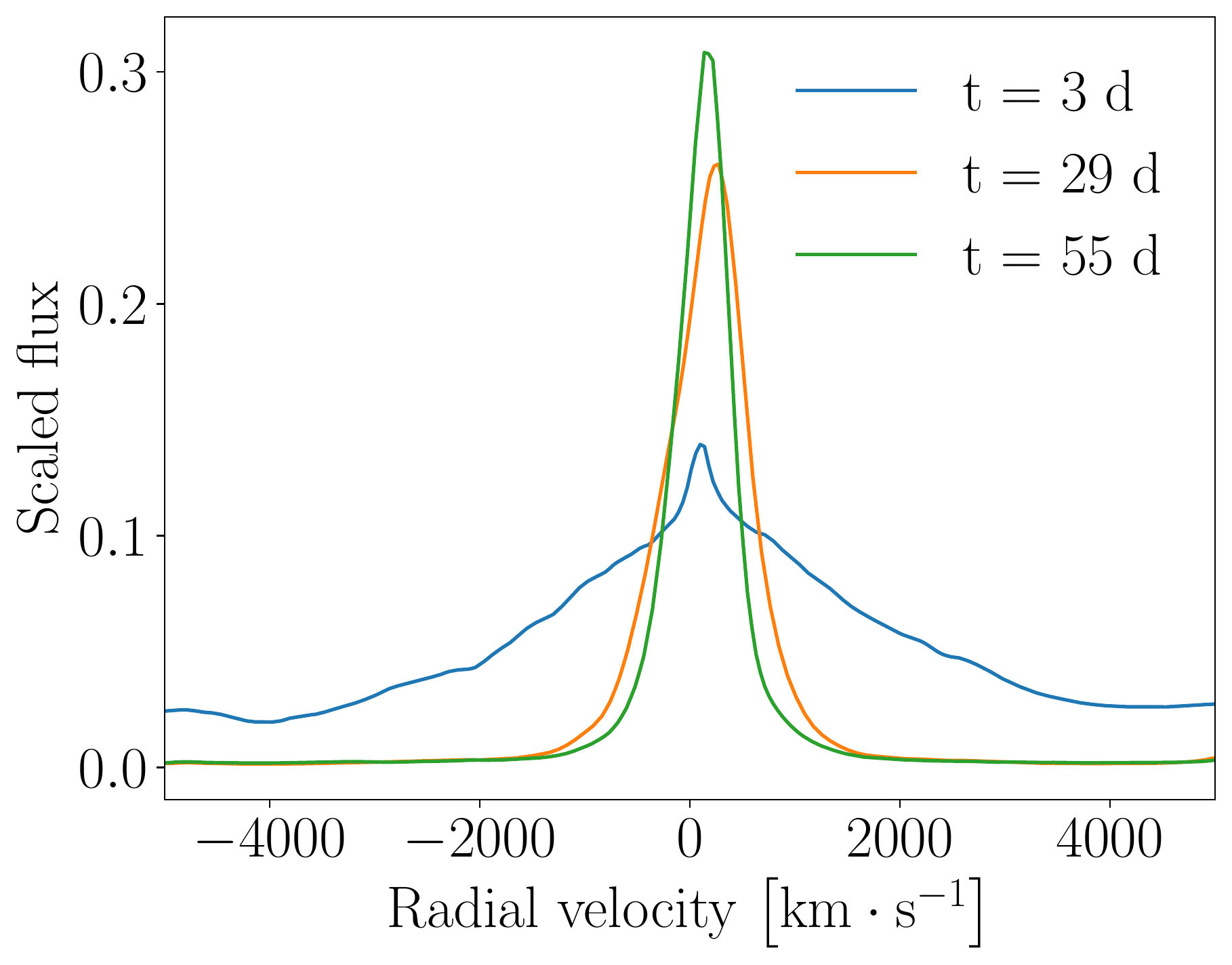}
\caption{\label{fig:Haprofile_rsoph} \footnotesize H$\alpha$ $6563\;\AA$ in RS Oph 2021: emission profiles at different stages during the outburst in ARAS spectra.}
\end{figure}

As explained in Section \ref{section:discussion}, V407 Cyg shows P Cyg components in the spectral profiles (\citet{vcyg1}, \citet{vcyg2}, \citet{iijima}. An example is given by the Balmer H$\alpha$ line in Fig.\ref{fig:Haprofile}. The same emission line appears as in Fig.\ref{fig:Haprofile_rsoph} during the 2021 outburst of RS Oph in ARAS spectra.

More numerous examples of line profiles observed in each dataset of this analysis may be found in \cite{Azz21}.

\section{Electron density}
\label{section:appendix3}
This section provides further explanation about the estimation of electron density from the ratio of Si III] to C III] emission lines. Under the assumption of homogeneous gas:

\begin{equation}
\label{eqn:ratio}
   \rm{r} = \frac{\rm{F}(\rm{Si\;III}]\; \lambda 1892)}{\rm{F}(\rm{C\; III}]\; \rm{\lambda} 1909)} = \frac{\rm{\epsilon(\lambda} 1892)}{\rm{\epsilon(\lambda} 1909)},
\end{equation}

where 

\begin{equation}
\label{eqn:epsilon}
   \rm{\epsilon(\lambda)} = \rm{N(X}_{3}^{2+}) \cdot\rm{h} \cdot\frac{\rm{c}}{\rm{\lambda} A_{ul}}
\end{equation}

is the emissivity of the line, $\rm{N}\left(\rm{X}_{3}^{2+}\right)$ the number density of the ion under examination - C or Si in our case. Eq.\ref{eqn:ratio} can be re-written:

\begin{equation}
\label{eqn:ratio1}
   \rm{r} = \frac{\rm{N(Si)}}{\rm{N(C)}} \frac{\rm{N}(\rm{Si}^{2+})/\rm{N}(\rm{Si})}{\rm{N}(\rm{C}^{2+})/\rm{N}(\rm{C})} \rm{\rho(\lambda} 1892 / \rm{\lambda} 1909)
\end{equation}

if 

\begin{equation}
\label{eqn:ndens}
   \rm{N(X}_{3}^{2+}) = \rm{N(X)} \frac{\rm{N(X}^{2+})}{\rm{N(X)}} \frac{\rm{N(X}_{3}^{2+})}{\rm{N(X}^{2+})}
\end{equation}

and

\begin{equation}
\label{eqn:rho}
\rm{\rho(\lambda} 1892 / \rm{\lambda} 1909) = \frac{\rm{\epsilon(\lambda} 1892) / \rm{N(Si}^{2+})}{\rm{\epsilon(\lambda} 1909) / \rm{N(C}^{2+})} .
\end{equation}

The  electron density dependence explicitly enters as $rate = N_{e} q_{lu}$ because the ratios of number densities depend on the rates of collisional excitation, which can be expressed as:

\begin{equation}
\label{eqn:collrate}
\rm{q_{lu}} = \frac{8.63\times10^{-6}}{\rm{g_l} \sqrt{\rm{T_{e}}}} \rm{\Omega_{lu}}\; exp\left(\frac{- \rm{\Delta E_{lu}}}{k T_{e}} \right) \;[\rm{cm^{3}s^{-1}}],
\end{equation}

with $\rm{g_{l}}=1$ statistical weight of the ground state, $\rm{\Omega}$ collision strength averaged over a Maxwellian electron distribution of temperature $\rm{T_{e}}$ and density $\rm{N_{e}}$, $\rm{\Delta E_{lu}}$ the energy difference between the levels. Using these relations, electron density can be quantitatively determined (\citet{osterbrock}). Transition probabilities and related quantities were taken from \citet{nussbaumerstorey} for C III] and \citet{ojhakeenanhibbert}, \citet{keenan1992improved} for Si III]. The Si III] to C III] ratio is almost unaffected by systematics, and therefore the accuracy in $\rm{n_e}$ estimation depends on how well the profiles of the two lines can be distinguished. The individual measures for the flux ratios are good to a few per cent, the accuracy of the $\rm{n_e}$ at the 1\% level.
\\
Table \ref{table:Ne_rsophSWIFT} reports the electron densities evaluated from Swift spectra of RS Oph, while Tables \ref{table:Ne_rsophIUE} - \ref{table:Ne_v3890sgrSWIFT} contain the correspondent measurements for the other datasets. The estimations in Table \ref{table:Ne_rsophSWIFT} are derived from the spectra, and short-timescales changes are appreciable.

\begin{table}
\caption{RS Oph 2021 - Swift grism}   % title of Table
\label{table:Ne_rsophSWIFT}      % is used to refer this table in the text
\centering                          % used for centering table
\begin{tabular}{cccc}        % centered columns (4 columns)
\hline\hline                 % inserts double horizontal lines
MJD & t $\left[ \rm{days} \right]$ & r & $\rm{n_{e}} \left[ \rm{cm^{-3}} \right]$ \\     % table heading 
\hline                        % inserts single horizontal line
   59445.53 & 11 & 1.09 & $3.489\times 10^{9}$ \\
59446.06 & 12 & 1.17 & $3.86\times 10^{9}$ \\
59449.12 & 15 & 0.98 & $3.05\times 10^{9}$\\
59450.24 & 16 & 0.94 & $2.88\times 10^{9}$\\
59451.17 & 17 & 1.02 & $3.20\times 10^{9}$\\
59452.10 & 18 & 1.08 & $3.47\times 10^{9}$\\
59453.09 & 19 & 1.00 & $3.13\times 10^{9}$\\
59455.02 & 21 & 0.99 & $3.07\times 10^{9}$\\
59456.15 & 22 & 0.78 & $2.25\times 10^{9}$\\
59457.28 & 23 & 0.95 & $2.92\times 10^{9}$\\
59458.07 & 24 & 0.90 & $2.74\times 10^{9}$\\
59463.58 & 29 & 0.73 & $2.091\times 10^{9}$\\
59472.41 & 38 & 0.56 & $1.46\times 10^{9}$\\
59472.48 & 38 & 0.54 & $1.40\times 10^{9}$\\
59472.74 & 38 & 0.60 & $1.60\times 10^{9}$\\
59473.46 & 39 & 0.63 & $1.71\times 10^{9}$\\
59473.73 & 39 & 0.54 & $1.41\times 10^{9}$\\
59474.33 & 40 & 0.48 & $1.20\times 10^{9}$\\
59474.59 & 40 & 0.68 & $1.89\times 10^{9}$\\
59474.85 & 40 & 0.38 & $9.01\times 10^{8}$\\
59475.58 & 41 & 0.52 & $1.34\times 10^{9}$\\
59476.12 & 42 & 0.52 & $1.33\times 10^{9}$\\
59476.38 & 42 & 0.65 & $1.79\times 10^{9}$\\
59476.98 & 42 & 0.5 & $1.45\times 10^{9}$\\
59477.51 & 43 & 0.55 & $1.44\times 10^{9}$\\
59477.97 & 43 & 0.53 & $1.37\times 10^{9}$\\
59478.79 & 44 & 0.46 & $1.15\times 10^{9}$\\
59477.16 & 45 & 0.58 & $1.55\times 10^{9}$\\
59477.37 & 45 & 0.43 & $1.05\times 10^{9}$\\
59479.96 & 45 & 0.39 & $9.02\times 10^{8}$\\
\hline                                   %inserts single line
\end{tabular}
\end{table}

\begin{table*}
\caption{RS Oph 2006 - Swift grism}             
\label{table:Ne_rsophSWIFT}      
\centering          
\begin{tabular}{cccc|cccc}     % 7 columns 
\hline\hline       
                      % To combine 4 columns into a single one 
MJD & t $\left[ \rm{days} \right]$ & r & $\rm{n_{e}} \left[ \rm{cm^{-3}} \right]$ & MJD & t $\left[ \rm{days} \right]$ & r & $\rm{n_{e}} \left[ \rm{cm^{-3}} \right]$ \\ 
\hline                    
   53808.71 &   30   &  0.51  &   $1.30\times10^{9}$   &  53841.66 &   63   &  0.59  &   $1.58\times10^{9}$   \\
53809.84 &   31   &  0.54  &   $1.42\times10^{9}$   &  53842.00 &   63   &  0.40  &   $9.72\times10^{8}$   \\
53810.84 &   32   &  0.52  &   $1.33\times10^{9}$   &  53842.66 &   64   &  0.59  &   $1.57\times10^{9}$   \\
53811.73 &   33   &  0.39  &   $9.15\times10^{8}$   &  53843.00 &   65   &  0.53  &   $1.39\times10^{9}$   \\
53818.02 &   40   &  0.45  &   $1.13\times10^{9}$   &  53844.61 &   66   &  0.48  &   $1.20\times10^{9}$   \\
53819.01 &   41   &  0.30  &   $6.59\times10^{8}$   &  53845.61 &   67   &  0.47  &   $1.19\times10^{9}$   \\
53822.03 &   44   &  0.54  &   $1.41\times10^{9}$   &  53846.00 &   68   &  0.50  &   $1.27\times10^{9}$   \\
53823.03 &   45   &  0.42  &   $1.03\times10^{9}$   &  53847.14 &   70   &  0.47  &   $1.18\times10^{9}$   \\
53824.04 &   46   &  0.35  &   $8.13\times10^{8}$   &  53848.15 &   71   &  0.50  &   $1.27\times10^{9}$   \\
53825.37 &   47   &  0.27  &   $5.80\times10^{8}$   &  53849.08 &   72   &  0.49  &   $1.25\times10^{9}$   \\
53826.31 &   48   &  0.34  &   $7.92\times10^{8}$   &  53850.83 &   73   &  0.36  &   $8.27\times10^{8}$   \\
53827.38 &   49   &  0.42  &   $1.03\times10^{9}$   &  53851.77 &   74   &  0.50  &   $1.29\times10^{9}$   \\
53828.32 &   50   &  0.35  &   $8.17\times10^{8}$   &  53852.71 &   75   &  0.45  &   $1.11\times10^{9}$   \\
53829.32 &   51   &  0.43  &   $1.04\times10^{9}$   &  53853.65 &   77   &  0.40  &   $9.56\times10^{8}$   \\
53831.01 &   53   &  0.50  &   $1.28\times10^{9}$   &  53855.71 &   79   &  0.45  &   $1.12\times10^{9}$   \\
53832.02 &   54   &  0.52  &   $1.36\times10^{9}$   &  53857.65 &   81   &  0.46  &   $1.16\times10^{9}$   \\
53833.02 &   55   &  0.60  &   $1.61\times10^{9}$   &  53859.00 &   85   &  0.35  &   $8.04\times10^{8}$   \\
53836.10 &   58   &  0.53  &   $1.38\times10^{9}$   &  53863.62 &   91   &  0.22  &   $4.34\times10^{8}$   \\
53836.62 &   58   &  0.45  &   $1.13\times10^{9}$   &  53869.51 &   94   &  0.37  &   $8.54\times10^{8}$   \\
53837.10 &   59   &  0.41  &   $1.00\times10^{9}$   &  53872.06 &   98   &  0.42  &   $1.02\times10^{9}$   \\
53838.36 &   60   &  0.47  &   $1.19\times10^{9}$   &  53876.34 &   101  &  0.27  &   $5.70\times10^{8}$   \\
53840.12 &   62   &  0.62  &   $1.69\times10^{9}$   &  53879.34 &   108  &  0.37  &   $8.73\times10^{8}$   \\
53840.65 &   62   &  0.46  &   $1.14\times10^{9}$   &  53886.06 &   112  &  0.53  &   $1.39\times10^{9}$   \\
53841.06 &   63   &  0.51  &   $1.31\times10^{9}$   &  53890.59 &   119  &  0.36  &   $8.47\times10^{8}$   \\
\hline                  
\end{tabular}
\end{table*}

\begin{table*}
\caption{RS Oph 1985 - IUE low resolution}             
\label{table:Ne_rsophIUE}      
\centering          
\begin{tabular}{cccc|cccc}     % 7 columns 
\hline\hline       
                      % To combine 4 columns into a single one 
MJD & t $\left[ \rm{days} \right]$ & r & $\rm{n_{e}} \left[ \rm{cm^{-3}} \right]$ & MJD & t $\left[ \rm{days} \right]$ & r & $\rm{n_{e}} \left[ \rm{cm^{-3}} \right]$ \\ 
\hline                    
   46098.83   &  6     &  0.80  &   $2.35\times10^{9}$   &  46164.09   &  93    &  1.53  &   $5.48\times10^{9}$  \\
46104.54   &  14    &  1.04  &   $3.32\times10^{9}$   &  46164.29   &  93    &  1.64  &   $6.00\times10^{9}$  \\
46106.10   &  18    &  0.97  &   $3.04\times10^{9}$   &  46185.11   &  105   &  0.66  &   $1.84\times10^{9}$  \\
46106.15   &  19    &  0.94  &   $2.90\times10^{9}$   &  46197.53   &  105   &  0.70  &   $1.97\times10^{9}$  \\
46111.02   &  25    &  0.88  &   $2.68\times10^{9}$   &  46197.59   &  110   &  0.44  &   $1.10\times10^{9}$  \\
46117.42   &  25    &  0.97  &   $3.03\times10^{9}$   &  46202.05   &  252   &  6.78  &   $3.81\times10^{10}$ \\
46117.47   &  30    &  0.61  &   $1.65\times10^{9}$   &  46344.72   &  252   &  1.56  &   $5.59\times10^{9}$  \\
46122.98   &  42    &  0.61  &   $1.67\times10^{9}$   &  46631.00   &  540   &  0.71  &   $2.00\times10^{9}$  \\
46134.19   &  42    &  0.45  &   $1.13\times10^{9}$   &  46632.97   &  553   &  0.71  &   $2.03\times10^{9}$  \\
46134.32   &  55    &  0.78  &   $2.30\times10^{9}$   &  46645.99   &  613   &  0.68  &   $1.93\times10^{9}$  \\
46147.23   &  61    &  0.60  &   $1.63\times10^{9}$   &  46705.65   &  764   &  2.50  &   $1.04\times10^{10}$ \\
46147.29   &  72    &  0.88  &   $2.66\times10^{9}$   &  46905.23   &  1194  &  1.07  &   $3.43\times10^{9}$  \\
46152.98   &  72    &  0.95  &   $2.97\times10^{9}$   &  48317.71   &  2225  &  2.73  &   $1.16\times10^{10}$ \\  
\hline                  
\end{tabular}
\end{table*}

\begin{table}
\caption{V745 Sco 1989 - IUE low resolution}             % title of Table
\label{table:Ne_v745scoIUE}      % is used to refer this table in the text
\centering                          % used for centering table
\begin{tabular}{cccc}        % centered columns (4 columns)
\hline\hline                 % inserts double horizontal lines
MJD & t $\left[ \rm{days} \right]$ & r & $\rm{n_{e}} \left[ \rm{cm^{-3}} \right]$ \\     % table heading 
\hline                        % inserts single horizontal line
47741.29 & 4  & 0.84  & $2.48\times10^{9}$  \\
47742.64 & 5  & 0.81  & $2.36\times10^{9}$  \\
47744.39 & 7  & 0.90  & $2.71\times10^{9}$  \\
47751.95 & 14 & 1.12  & $3.63\times10^{9}$  \\ 
\hline                                   %inserts single line
\end{tabular}
\end{table}

\begin{table}
\caption{V3890 Sgr 1990 - IUE low resolution}             % title of Table
\label{table:Ne_v3890sgrIUE}      % is used to refer this table in the text
\centering                          % used for centering table
\begin{tabular}{cccc}        % centered columns (4 columns)
\hline\hline                 % inserts double horizontal lines
MJD & t $\left[ \rm{days} \right]$ & r & $\rm{n_{e}} \left[ \rm{cm^{-3}} \right]$ \\     % table heading 
\hline                        % inserts single horizontal line
48026.17  &  19   &  0.63  & $1.70\times10^{9}$  \\
48026.23  &  19   &  0.65  & $1.79\times10^{9}$  \\
48034.04  &  27   &  0.60  & $1.61\times10^{9}$  \\
48034.16  &  27   &  0.62  & $1.65\times10^{9}$  \\
48150.77  &  143  &  0.30  & $6.54\times10^{8}$  \\   
\hline                                   %inserts single line
\end{tabular}
\end{table}

\begin{table}
\caption{V3890 Sgr 2019 - Swift grism}             % title of Table
\label{table:Ne_v3890sgrSWIFT}      % is used to refer this table in the text
\centering                          % used for centering table
\begin{tabular}{cccc}        % centered columns (4 columns)
\hline\hline                 % inserts double horizontal lines
MJD & t $\left[ \rm{days} \right]$ & r & $\rm{n_{e}} \left[ \rm{cm^{-3}} \right]$ \\     % table heading 
\hline                        % inserts single horizontal line
58743.93 &  5   &  0.61  &  $1.66\times10^{9}$  \\
58744.39 &  6   &  0.62  &  $1.68\times10^{9}$  \\
58745.73 &  6   &  1.12  &  $3.64\times10^{9}$  \\
58746.26 &  6   &  1.40  &  $4.88\times10^{9}$  \\
58728.74 &  20  &  1.01  &  $3.16\times10^{9}$  \\
58729.46 &  21  &  0.79  &  $2.30\times10^{9}$  \\
58729.59 &  22  &  1.20  &  $3.99\times10^{9}$  \\
58731.25 &  23  &  1.07  &  $3.41\times10^{9}$  \\   
\hline                                   %inserts single line
\end{tabular}
\end{table}

\begin{table}
\caption{RS Oph 1985 - IUE high resolution}             % title of Table
\label{table:Ne_rsophIUEHIRES}      % is used to refer this table in the text
\centering                          % used for centering table
\begin{tabular}{cccc}        % centered columns (4 columns)
\hline\hline                 % inserts double horizontal lines
MJD & t $\left[ \rm{days} \right]$ & r & $\rm{n_{e}} \left[ \rm{cm^{-3}} \right]$ \\     % table heading 
\hline                        % inserts single horizontal line
46106.17   &   14   &   1.51  &  $5.30\times10^{9}$  \\
46110.46   &   18   &   0.84  &  $2.50\times10^{9}$  \\
46117.49   &   25   &   0.59  &  $1.58\times10^{9}$  \\
46123.02   &   31   &   0.82  &  $2.40\times10^{9}$  \\
46134.25   &   42   &   0.28  &  $6.00\times10^{8}$  \\
46147.34   &   55   &   0.45  &  $1.10\times10^{9}$  \\
46164.17   &   72   &   0.50  &  $1.27\times10^{9}$  \\
\hline                                   %inserts single line
\end{tabular}
\end{table}

\end{appendix}

\end{document}